\documentclass[12pt,a4paper,twoside]{article}
\usepackage{times}
\usepackage{subscripts}
\usepackage{Martin}
\usepackage{MHenvs}
\usepackage{MHfig}
\usepackage{epsfig}
\usepackage{wrapfig}
\usepackage{array}

\final

\newcommand\F{{\mathfrak F}}
\renewcommand\L{{\cal L}}
\renewcommand\H{{\cal H}}
\newcommand\A{{\cal A}}
\newcommand\g{{g}}

\let\sauf=\backslash

\def\fC{{\mathfrak C}}
\newcommand\B{{\mathfrak B}}
\def\HF{H_{\!S}}
\def\HB{H_{\!B}}
\def\rL{{r_{\!L}}}
\def\rR{{r_{\!R}}}
\def\V#1{V_{#1}}
\def\G{G}
\def\T{{\cal T}}
\def\Power{W}
\newcommand\Exp[1]{\textbf{E}[#1]}
\mydimensions
\mynumbering

\begin{document}

\title{Non-Equilibrium Statistical Mechanics \\
of Strongly Anharmonic Chains of Oscillators}

\begin{author}{J.-P.~Eckmann\thanks{D\'ept.~de Physique Th\'eorique,
Universit\'e de Gen\`eve. Email: \texttt{Jean-Pierre.Eckmann@physics.unige.ch}}
\thanks{Section de Math\'ematiques, Universit\'e de Gen\`eve.} \and
M.~Hairer\thanks{D\'ept.~de Physique Th\'eorique, Universit\'e de Gen\`eve.
Email: \texttt{Martin.Hairer@physics.unige.ch}}}
\end{author}
\date{}

\maketitle

\begin{abstract}
We study the model of a strongly non-linear chain of particles
coupled to two heat baths at different temperatures. Our main result is the
existence and uniqueness of a stationary state at all temperatures. 
This result extends those of Eckmann, Pillet, Rey-Bellet
\cite{EPR,EPR2}
to potentials
with essentially arbitrary growth at infinity.
This extension is possible by introducing a stronger version
of H\"ormander's theorem for Kolmogorov equations to vector fields
with polynomially bounded coefficients on unbounded domains.
\end{abstract}

\thispagestyle{empty}
\pagestyle{myheadings}
\setcounter{page}{0}

\mysection*{Introduction}

In this paper, we study the statistical mechanics of a highly
non-linear chain of oscillators coupled to two heat reservoirs which
are at (arbitrary) different temperatures. We show that such systems
have, under suitable conditions, a \emph{unique} stationary state, in
which heat flows from the hotter reservoir to the cooler one.

These results are an extension of the same statements obtained by Eckmann,
Pillet and Rey-Bellet in \cite{EPR,EPR2}
where it was assumed that the Hamiltonian is essentially ``quadratic
at high energies.'' Since quadratic Hamiltonians have been discussed
much earlier by Lebowitz and Spohn \cite{LS}, there is an issue here
of whether the quadratic nature of the forces at infinite energies is
an essential ingredient of existence and uniqueness of the stationary
state. Our result shows that this is not the case, since we allow for
potentials of arbitrary polynomial growth. 

Our models, which are described in Section~\ref{sec:model}, treat a
Hamiltonian of the form
\equ{
	\HF(p,q) = \sum_{i=0}^N \Bigl( \frac{p_i^2}{2} + \V1(q_i)\Bigr) +
\sum_{i=1}^{N} \V2(q_{i} - q_{i-1})\;,
}
describing a chain of particles with nearest-neighbor interaction
(Figure~\ref{fig}). This chain is linearly coupled to heat baths $B_i$
represented by free fields at temperatures $T_i$. We proceed then, as in
\cite{EPR}, to a reduction to a stochastic differential equation, see
\eref{e:sysStoch}. Associated with it is an ``effective energy'' $G$,
described in \eref{e:defG}, which is equal to $\HF$ with
some quadratic terms from the heat baths added.
The generator corresponding to the stochastic differential equation
above, represented in a space weighted with an exponential of $G$,
will be called $K$ and is the main object of study of this paper.

It is for this generator that we show existence and uniqueness of an
invariant state. This will be done by first showing that $K$ has
compact resolvent (which is really more than needed), and then using
this result to derive the properties of the invariant measure.
Our conditions on $\HF$ are spelled out in Section~\ref{sec:settRes} below. 
They basically say that
the coupling \emph{between} the oscillators must be stronger than the
single particle potential. This condition might be physically
relevant, since it implies that transport is favored over storing of
energy, but we have not found a counterexample when this condition is
violated. Furthermore, the interparticle coupling must be convex.

The main technical insight behind our generalization of the results of
\cite{EPR,EPR2} is a new, and stronger version of the H\"ormander
theorem for Kolmogorov equations. We will develop
this in more generality in Section~\ref{sec:Hormander}, but here we
just indicate how 
we use this result.
The operator $K$ is of the form
\equ[K]{
	K = \sum_{i=1}^n X_i^* X_i^{} + X_0\;,
}
where the $X_i$ are smooth vector fields on $\R^d$.
For example, see Eq.~\eref{e:defXi},
$X_0$ contains terms of the form $p_i\partial_{q_i}$ and
$(\partial_{q_i}V) \partial_{p_i}$, where $V$ is the interaction.
The $X_i$ for $i\ne0$ are first order operators. Here, $\partial V$
is polynomially bounded, whereas, in \cite{EPR}, $\partial V$ was
assumed to be linearly bounded.
Letting $g_0$ be an adequate inverse power of the effective energy
$G$, one successively considers the \emph{finite} sets of
operators
\equ{
\A_{-1} = \{X_1,\ldots,X_n\}\;,\quad \A_0 = \{g_0X_0,X_1,\ldots,X_n\}\;,
}
and then---see Section~\ref{sec:proof2} for the detailed definition---
\equ{
\A_\ell=\A_{\ell-1}\cup [g_0 X_0, \A_{\ell-1}]\;.
}
We stop this iteration after at most $2N$ steps, where $N$ is the
number of particles in the chain, obtaining the set $\A=\A_{2N+1}$.
We now define the operator $\Lambda_{\!\A}$ as the finite sum
\equ{
\Lambda^2_{\!\A}= 1+\sum_{A\in\A} A^* A\;.
}
This is a generalization to our case of an elliptic operator of the
type
$\Lambda ^2 = 1 -\sum_i \partial_{x_i}^2
$ used in \cite{Ho} or $\Lambda ^2 = 1 -\sum_i \partial_{x_i}^2+\sum_i {x_i}^2
$ used in \cite{EPR}.

With these definitions, one then has the bound
\begin{proposition}[Momentum space bound]
There is a constant $C$ such that for all $f\in\cOinf(\R^d)$ one
has in $\Ltwo$:
\equ[momentum]{
\| \Lambda^{16^{-N}}_{\!\A} f\|\,\le\,  C\bigl (\|Kf\|+\|f\|\bigr )~.
}
\end{proposition}

We also derive a similar bound in the conjugate variables:
\begin{proposition}[Position space bound]
There is a constant $C$ such that for all $f\in\cOinf(\R^d)$ one
has in $\Ltwo$:
\equ[position]{
\| G^\eps f\|\,\le\,  C\bigl (\|Kf\|+\|f\|\bigr )~,
}
where $\eps >0$ depends on the asymptotic behavior of the potential
$V$ and on $N$.
\end{proposition}
Combining these two propositions one easily shows that $K$ has compact
resolvent. Then one derives from that result the existence of an
invariant measure. Its properties are then found adapting the
techniques of \cite{EPR,EPR2}.

The remainder of this paper is organized as follows. 
In Section \ref{sec:model} we describe the physical model and in
Section \ref{sec:settRes} we refine the setting and state the results.
Section~\ref{sec:proof1}
will be devoted to the proof of the position space bound
(Proposition~\ref{prop:estG}).
In
Section~\ref{sec:Hormander}, we present in detail the general scheme 
for studying operators of the form of \eref{K}, and show the
inequality corresponding to \eref{momentum}.
This section is as much as possible self-contained as it presents some
independent interest. 
The detailed application of this general scheme to the problem of the chain
allows us to prove the momentum space bound (Proposition
\ref{prop:estDeltaPrime}) in
Section~\ref{sec:proof2}.
In 
Section~\ref{sec:theoChain} we combine these two bounds and prove
Theorem~\ref{theo:Chain} showing that $K$ has compact resolvent and
hence discrete spectrum. In Section~\ref{sec:inv} 
we show existence, uniqueness, and further properties of the invariant
measure (Theorem~\ref{theo:exist}).

Appendix~\ref{App:main} contains some technical estimates used in
Section~\ref{sec:Hormander}. Appendix~\ref{App:ops} contains the proof of a result concerning the domains of $K$ and $K^*$. The method used there probably works
for more general accretive second-order differential operators.
Appendix~\ref{App:core} finally contains the proof of a technical result used
in Section~\ref{sec:inv}.

\subsection*{Acknowledgments}We have profited from helpful discussions
with M. Mantoiu, C.-A. Pillet, L. Rey-Bellet, and  J. Rougemont. This work
was partially supported by the Fonds National Suisse.

\mysection{The model}
\label{sec:model}

We will study the model of a (small) classical $N$-particle Hamiltonian
system coupled to $M$ stochastic heat baths proposed in \cite{EPR}. The small
system without the heat baths is governed by a Hamiltonian
\equ{
	\HF \in \cinf(\R^{2N})\;.
}
(We stay here with $d=1$ dimensional position space for each particle
to simplify notation.)
The heat baths are modeled by classical field theories associated to the wave
equation. The fields will be called $\phi_i$ and their conjugate momenta
$\pi_i$, where the index $i$ ranges from $1$ to $M$.

The Hamiltonian for one heat bath is given by
\equ{
	\HB(\pi, \phi) = \frac12\int_\R \bigl(|\d\phi|^2 + |\pi|^2\bigr)\,dx\;.
}
The couplings allowed for the model are linear in the field variables. The
total Hamiltonian for our model is then given by
\equ[e:totHam]{
	H(p,q,\pi,\phi) = \sum_{i=1}^M \Bigl(\HB(\pi_i,\phi_i) + F_i(p,q)\int_\R
\d\phi_i(x) \rho_i(x)\, dx\Bigr) + \HF(p,q)\;.
}
We assume the initial conditions describe the heat baths at equilibrium at
inverse temperatures $\beta_i$, \ie they are distributed in a sense
according to the measure with ``weight''
\equ{
	e^{-\beta_i \HB(\pi_i,\phi_i)}\;.
}

The paper \cite{EPR} explains in detail how, and under which conditions on the
coupling functions $\rho_i$, one can reduce the resulting ``big'' system to a
``small'' system, where the heat baths are described by a finite number of
variables. The price to pay for that is that we are now dealing with the
following system of stochastic differential equations:
\begin{alignat}{2}\label{e:sysStoch}
	dq_j &= \d_{p_j}\HF\,dt - \sum_{i=1}^M \bigl(\d_{p_j}F_i\bigr) r_i\,dt\;,
&\qquad j&=1,\ldots,N\;,\notag\\
	dp_j &= -\d_{q_j}\HF\,dt + \sum_{i=1}^M \bigl(\d_{q_j}F_i\bigr) r_i\,dt\;,
&&\\
	dr_i &= -\gamma_i r_i\,dt + \lambda_i^2\gamma_i F_i(p,q)\,dt -
\lambda_i\sqrt{2\gamma_i T_i}\,dw_i(t)\;, &\qquad i&=1,\ldots,M\notag\;,
\end{alignat}
where the $w_i$ are independent Wiener processes. The various constants
appearing in \eref{e:sysStoch} have the following meaning. $T_i$ is the
temperature of the $i^{\text{th}}$ heat bath, $\lambda_i$ is the strength of
the coupling between that heat bath and the small system and $1/\gamma_i$ is
the relaxation time of the $i^{\text{th}}$ heat bath. The value of $\gamma_i$
depends on the choice of the coupling function $\rho_i$. If we wanted to be
more general, we would have to introduce for each bath a family of auxiliary
variables $r_{i,m}$ as is done in \cite{EPR}. This would only cause
notational problems and does not change our argument.

If we consider a generic $n$-dimensional system of stochastic differential
equations with additive noise of the form
\equ[e:genStoch]{
	dx_i(t) = b_i(x(t))\,dt + \sum_{j=1}^n \sigma_{i j}\,dw_j(t)\;,
}
we can associate with it the second-order differential operator $\CL$ formally
defined by
\equ[e:genProc]{
	\CL \equiv \frac12 \sum_{i,j=1}^n \d_i (\sigma\sigma^T)_{i j}\d_j +
\sum_{i=1}^n b_i(x)\,\d_i\;.
}
It is a classical result that if the solution of such a system of stochastic
differential equations exists, the probability density of the solution
satisfies the partial differential equation
\equ{
	\d_t p(x,t) = \bigl(\CL p\bigr)(x,t)\;.
}

In our case, the differential operator $\CL$ is given by
\equ[e:genChain]{
	\CL = \sum_{i=1}^M \lambda_i^2\gamma_i T_i \d_{r_i}^2 - \sum_{i=1}^M
\gamma_i\bigl(r_i - \lambda_i^2 F_i(p,q)\bigr)\d_{r_i} + X^{\HF} - \sum_{i=1}^M
r_i X^{F_i}\;,
}
where the symbol $X^F$ denotes the Hamiltonian vector field associated to the
function $F$. It is convenient to introduce the ``effective energy'' given by
\equ[e:defG]{
	G(p,q,r) = \HF(p,q) + \sum_{i=1}^M \Bigl(\frac{r_i^2}{2\lambda_i^2} - F_i(p,q)
r_i\Bigr)\;.
}
At this point, we make the following assumption on the asymptotic behavior of
$G$.

\begin{assum}{0}
There exist constants $\tilde d_i, C > 0$ and $\alpha > 0$, as well as
constants $\tilde c_i > 2/\lambda_i^2$ such that
\sublabels
\equs[1,e:ass0]{
	\HF(p,q) &\ge C(1 + \|p\|^\alpha + \|q\|^\alpha)\;, \sublabel{e:ass01}\\[1mm]
	F_i^2(p,q) &\le \tilde c_i \HF(p,q) + \tilde d_i\;. \sublabel{e:ass02}
}
\end{assum}

\begin{remark}
This assumption essentially means that the effective energy $G$ grows at
infinity at least like $1+\|r\|^2 + \|p\|^\alpha + \|q\|^\alpha$. This implies
the stability of the system, as follows easily from the inequality
\equ{
	|r_i F_i(p,q)| \le s^2 r_i^2 + \frac{F_i^2(p,q)}{s^2}\;,
}
which holds for every $s > 0$.
In particular, this implies that $\exp(-\beta G)$ is integrable for every
$\beta > 0$.
\end{remark}

We also define
\equ{
	\Power \equiv \sum_{i=1}^M \gamma_i T_i\;,
}
which is, in some sense that will be clear in a moment, the maximal power the
heat baths can pull into the chain.
We have the following result.

\begin{proposition}
Assume {\bf A0} holds. Then the solution $\xi(t;x_0,w)$ of \eref{e:sysStoch}
exists and is continuous for all $t>0$ with probability $1$. Moreover, the mean
energy of the system satisfies for all values of $t$ and $x_0$ the estimate
\equ[e:GrowG]{
	\Exp{G(x(t;x_0,w))} - G(x_0) \le \Power t\;,
}
where $\Exp{\cdot}$ denotes the expectation with respect to the $M$-dimensional
Wiener process $w$.
\end{proposition}

\begin{remarque}
The bound \eref{e:GrowG} allows the energy to
grow forever, which would cause the system to
``explode.'' 
But this is not the case for the systems we consider in this paper.
Indeed, we will prove that the process possesses a unique stationary state.
This implies among other features that the mean time needed to reach any
compact region is finite, and so the energy can not grow forever.
\end{remarque}

\begin{proof}
A classical result (see \eg \cite[Thm~4.1]{Ha}) states the following. Assume
that the vector field $b$ of \eref{e:genStoch} is locally Lipshitz and that
there exists a confining ${\cal C}^2$ function $G : \R^n \to \R$ and a constant
$k$ such that
\equ{
	(\CL G)(x) \le k \qquad\text{for all}\qquad x\in \R^n\;.
}
Then there exists a unique stochastic process $\xi(t)$ solving
\eref{e:genStoch}. The process $\xi$ is regular (\ie it does not blow up in a
finite time) and continuous for all $t>0$. It satisfies the statistics of a
Markovian diffusion process with generator $\CL$. Moreover, we have the
estimate
\equ{
	\Exp{G(x(t;x_0,w))} - G(x_0) \le k t\;.
}
This result can be applied to our case, if we take for $G$ the effective energy
defined in \eref{e:defG}. An explicit computation yields indeed
\equ[e:dissip]{
	\CL G = \Power - \sum_{i=1}^M \frac{\gamma_i}{\lambda_i^2} \bigl(r_i - \lambda_i^2
F_i(p,q)\bigr)^2\;.
}
Moreover, $G$ is confining by {\bf A0}. This proves the assertion.
\end{proof}

\subsection{Definition and simple properties of the semi-group}

In this paper, we will mainly be interested in studying under which assumptions
on the chain Hamiltonian $\HF$ it is possible to prove the existence of a
\emph{unique invariant measure} for the stochastic process $\xi(t;x_0,w)$
solving \eref{e:sysStoch}. 
Throughout, we will use
the notation $$
\CX \,=\,\R^{2N+M}
$$ 
for the extended phase space $(p,q,r)$.
This stochastic process defines a semi-group $\T^t$
on $\cOinf[\CX]$ by
\equ[e:defTt]{
	\T^{t} f(x_0) = \Exp{f(\xi(t;x_0,w))}\;.
}
This semi-group satisfies the following
\begin{proposition}
\label{prop:Tt}
Assume {\bf A0} holds. Then $\T^t$ extends to a strongly continuous,
quasi-bounded semi-group of positivity preserving operators on $\Ltwo(\CX)$.
Its generator $L$ is the closure of the operator $\CL$ with domain
$\cOinf[\CX]$. The adjoint $L^*$ is the closure of the formal adjoint $\CL^T$
with domain $\cOinf[\CX]$.
\end{proposition}

\begin{proof}
The proof will be given in Appendix~\ref{App:ops}.
\end{proof}

This in turn defines a dual semi-group $(\T^t)^*$ by
\equ{
\int \bigl(\T^t f\bigr)(x)\,\nu(dx) = \int f(x)\,\bigl((\T^t)^*
\nu\bigr)(dx)\;.
}
The generator of $(\T^t)^*$ is given by the adjoint of $\CL$ in $\Ltwo$ that
will be denoted $\CL^T$. It is possible to check that if the heat baths are all
at the same temperature $T = 1/\beta$, we have
\equ{
	\CL^T \mu_0 = 0\;, \qquad\text{where}\qquad \mu_0(p,q,r) = e^{-\beta
G(p,q,r)}\;.
}
Thus, the generalized Gibbs measure
\equ{
	d\mu_0 = e^{-\beta G(p,q,r)}\,dp\,dq\,dr = \mu_0(p,q,r)\,dp\,dq\,dr\;,
}
is an invariant measure for the Markov process described by \eref{e:sysStoch}.
This confirms our definition of $G$ as the effective energy of our system. We
want to consider the more interesting case where the temperatures of the heat
baths are not the same. The idea is to work in a Hilbert space that is weighted
with a Gibbs measure for some reference temperature.

We will therefore study an extension $\T_0^t$ of $\T^t$ acting on an auxiliary
weighted Hilbert space $\CH_0$, given by
\equ{
	\CH_0 \equiv \Ltwo\bigl(\CX, Z_0^{-1}e^{-2\beta_0
G(p,q,r)}\,dp\,dq\,dr\bigr)\;,
}
where $Z_0$ is a normalization constant and $\beta_0$ is a ``reference''
inverse temperature that we choose such that
\equ[e:condTRefGen]{
	1/\beta_0 \equiv T_0 > \max\{T_i\;|\; i=1,\ldots,M\}\;.
}
We have the following
\begin{proposition}
\label{prop:Tt0}
Assume {\bf A0} holds. Then the semi-group $\T^t$ given by \eref{e:defTt}
extends to a strongly continuous quasi-bounded semi-group $\T_0^t$ on $\CH_0$.
Moreover, $\T_0^t 1 = 1$ and $\T_0^t$ is positivity preserving, \ie
\equ{
	\T_0^t f \ge 0 \qquad\text{if}\qquad f \ge 0\;.
}
Let $L_0$ be the generator of $\T_0^t$. Then $L_0$ coincides on $\cOinf[\CX]$
with $\CL$ of \eref{e:genChain} and $\cOinf[\CX]$ is a core for both $L_0$ and
$L_0^*$.
\end{proposition}

\begin{proof}
The statement can be proven by simply retracing the proof of Lemma~3.1 in
\cite{EPR}. There are only three points that have to be checked. We define the
vector fields $b$ and $b_0$ respectively by
\equs{
 b &= - \sum_{i=1}^M \gamma_i\bigl(r_i - \lambda_i^2 F_i(p,q)\bigr)\d_{r_i} +
X^{\HF} - \sum_{i=1}^M r_i X^{F_i}\;, \\
 b_0 &= 2\beta_0 \sum_{i=1}^M \lambda_i^2 \gamma_i T_i \bigl(\d_{r_i} G\bigr)
\d_{r_i} = 2\beta_0 \sum_{i=1}^M \gamma_i T_i \bigl(r_i - \lambda_i^2
F_i(p,q)\bigr) \d_{r_i}\;.
}
In order to make the proof of \cite{EPR} work, we have to check that
\equ{
\|\!\div b\|_\infty < \infty\;,\quad
\|\!\div b_0\|_\infty < \infty\;,\quad
\sup_{x\in \CX} \bigl(b + {\textstyle\frac{1}{2}}b_0 \bigr) G(x) < \infty\;,
}
where $b$ and $b_0$ are considered as first-order differential operators in the
last inequality. The divergence of any Hamiltonian vector field vanishes, and
so we have
\equ{
	\|\!\div b\|_\infty = -\sum_{i=1}^M \gamma_i < \infty\;.
}
The term involving the divergence of $b_0$ can easily be computed to give
\equ{
\|\!\div b_0\|_\infty = \beta_0 \sum_{i=1}^M \gamma_i T_i < \infty\;.
}
In order to check the last inequality, we compute the expression
\equ{
\bigl(b + {\textstyle\frac{1}{2}}b_0 \bigr) G(p,q,r) = \sum_{i=1}^M
\frac{\gamma_i}{\lambda_i^2} (\beta_0 T_i - 1) \bigl( r_i - \lambda_i^2
F_i(p,q)\bigr)^2\;.
}
We see that condition \eref{e:condTRefGen} on $\beta_0$ obviously implies
$\beta_0 T_i - 1 < 0$, and so the desired inequality holds.

The domains of $L_0$ and $L_0^*$ are controlled by the techniques of
Appendix~\ref{App:ops}.
\end{proof}

We are mainly interested in the case $M=2$. The Hamiltonian $\HF$ will describe
a chain of $N+1$ strongly anharmonic oscillators coupled to two heat baths at
the first and the last particle. In the case in which the Hamiltonian $\HF$ can
be written as a quadratic function plus some bounded terms, the existence and
uniqueness of a stationary state for every temperature difference has been
proved in \cite{EPR,EPR2}. We will extend this result to the case where the
potentials grow faster than quadratically at infinity. Besides some weak
conditions on the derivatives of the one and two-body potentials, we will only
require that they grow algebraically and that the two-body potentials grow
asymptotically faster than the one-body potentials, \ie at large separation the
interaction energy between neighboring particles grows \emph{faster} than the
one-particle energy.

\subsection{Notations}
\label{sec:defs}

Throughout, the domain of an operator $A$ will be denoted by $\CD(A)$. Unless
specified, the domain of any operator will always be the closure in the graph
norm of $\cOinf$. For example, if we write $[A,B]$, we mean in fact
$\overline{(AB - BA)\upharpoonright \cOinf}$, so that the domain of $[A,B]$ can
be larger than that of $A$ or $B$ separately.

\mysection{Setting and results}
\label{sec:settRes}

In order to set up our model, we need to be able to describe precisely the
growth rates of the potentials at infinity. This will be achieved with
the following function spaces.

\begin{definition}
\label{def:func}
Choose $\alpha \in \R$. We call ${\cal F}_\alpha$ the set of all $\cinf$
functions from $\R^n$ to $\R$ such that for every multi-index $k$ there exists
a constant $C_k\;$ for which
\equ{
	\| D^k f(x)\| \le C_k (1 +
\|x\|^2)^{\alpha/2}\;,\quad\text{for all}\quad x \in \R^n\;.
}
\end{definition}

\begin{definition}
\label{def:func2}
Choose $\alpha \in \R$ and $i \in \N \cup \{\infty\}$. We call ${\cal
F}_\alpha^i$ the set of all $\cinf$ functions from $\R^n$ to $\R$ such that for
every multi-index $k$ with $|k| \le i$, we have $D^k f(x) \in \CF_{\alpha -
|k|}$.
\end{definition}

\begin{remarque}
For any $\alpha \in \R$, the function
\equs[e:defPoly]{
	P^\alpha:\R^n &\to \R \\
	x\;\, &\mapsto (1 + \|x\|^2)^{\alpha/2}
}
belongs to $\CF_\alpha^\infty$. Moreover, any polynomial of degree $n$ belongs
to $\CF_n^\infty$.
\end{remarque}


\subsection{The chain}

\begin{MHfig}{ChainFig}[-2mm]
	\vspace{-5mm}
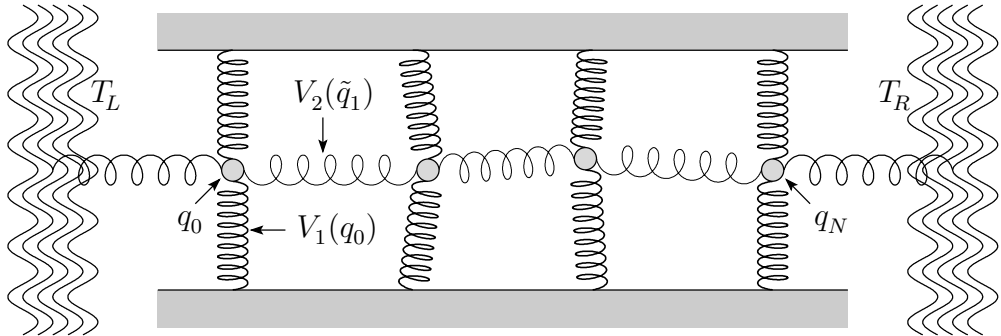
	\caption{Chain of oscillators}
	\label{fig}
\end{MHfig}

We consider the Hamiltonian
\equ[e:defHChain]{
	\HF(p,q) = \sum_{i=0}^N \Bigl( \frac{p_i^2}{2} + \V1(q_i)\Bigr) +
\sum_{i=1}^{N} \V2(q_{i} - q_{i-1})\;,
}
describing a chain of particles with nearest-neighbor interaction
(Figure~\ref{fig}). We slightly modify the notations used so far. Because there
are only two heat baths, we will not use for them the indices $i \in \{1,2\}$,
but rather $i\in\{L,R\}$.
Concerning the coupling between the chain and the baths, we assume that we can
make a dipole approximation, so we set
\equ[e:defCouplChain]{
	F_L = q_0 \qquad\text{and}\qquad  F_R = q_N\;,
}
in equation \eref{e:totHam}. We will make the assumptions {\bf
A1}--{\bf A3} on $\V1$ and $\V2$.
\begin{assum}{1}
The potential $\V1$ is in $\CF_{2n}^2$ for some $n>1$. Moreover, there are
constants $c_i > 0$ such that
\sublabels
\equs[1,ass1]{
	\V1(x) &\ge c_1 P^{2n}(x)\;, \sublabel{e:boundV1}\\
x\V1'(x) &\ge c_2P^{2n}(x) - c_3\;, \sublabel{e:boundxV1}
}
for all $x \in \R$.
\end{assum}
\begin{assum}{2}
The potential $\V2$ is in $\CF_{2m}^2$ for some $m>n$. Moreover, there are
constants $c_i' > 0$ such that
\sublabels
\equs[1,ass2]{
 \V2(x) &\ge c_1' P^{2m}(x)\;, \sublabel{e:boundV2}\\
x\V2'(x) &\ge c_2' P^{2m}(x) - c_3'\;, \sublabel{e:boundxV2}
}
for all $x \in \R$.
\end{assum}
\begin{assum}{3}
The function
\equ{
 x \mapsto \frac{1}{\V2''(x)}
}
belongs to $\CF_\ell$ for some $\ell$.
\end{assum}
\begin{remarque}
It is clear that \eref{e:defCouplChain}, together with {\bf A1} and {\bf A2}
immediately imply {\bf A0}.
Notice that the assumptions $\V1 \in \CF_{2n}^2$ and $\V2 \in \CF_{2m}^2$ give
bounds not only on the asymptotic behavior of $\V1$ and $\V2$, but also of
their derivatives. The numbers $n$, $m$ and $\ell$ need not be integers. The
generalization to a Hamiltonian with $\V1$, $\V2$ depending also on the number
of the particle only creates notational problems and is left to the reader.
\end{remarque}

An example of potentials that satisfy {\bf A1}--{\bf A3} is
\equ{
	\V1(x) = x^4 - x^2 + 2 \quad\text{and}\quad \V2(x) = (1+x^2)^{5/2} -
\cos(x)\;.
}

The effective energy of the system chain+baths is given by
\equ[e:defGChain]{
	\G(p,q,r) = \HF(p,q) + \frac{\rL^2}{2\lambda_L^2} + \frac{\rR^2}{2\lambda_R^2}
- q_0\rL - q_N \rR + \Gamma\;,
}
where we choose the constant $\Gamma$ such that $\G \ge 1$, which is always
possible, because $n>1$. In fact, it is important that the function
$\exp(-\beta G)$ be integrable for any $\beta > 0$. This could also be achieved
with for example only one of the one-body potentials non-vanishing, but would
cause some unimportant notational difficulties.
The case $n=1$ is marginal, the stability of the system depends on the values
of the constants $\lambda_i$ and was treated in \cite{EPR}. We will not treat
this case, but it would not cause any big trouble, as long as $G$ remains
confining.

In the sequel, we will extensively use the notations
\equ{
	\tilde q_i \equiv q_i - q_{i-1} \qquad\text{and}\qquad Q \equiv \sum_{i=0}^N
q_i \;.
}
The system of stochastic differential equations we consider is given by
\equs[e:stochChain]{
	dq_i &= p_i\,dt\;,\\[1mm]
	dp_0 &= -\V1'(q_0)\,dt + \V2'(\tilde q_1)\,dt + \rL\,dt\;, \\[1mm]
	dp_j &= -\V1'(q_j)\,dt - \V2'(\tilde q_{j})\,dt + \V2'(\tilde q_{j+1})\,dt\;,
\\[1mm]
	dp_N &= -\V1'(q_N)\,dt - \V2'(\tilde q_{N})\,dt + \rR\,dt\;, \\[1mm]
	d\rL &= -\gamma_L \rL\,dt + \lambda_L^2\gamma_L q_0\,dt -
\lambda_L\sqrt{2\gamma_L T_L}\,dw_L(t)\;, \\[1mm]
	d\rR &= -\gamma_R \rR\,dt + \lambda_R^2\gamma_R q_N\,dt -
\lambda_R\sqrt{2\gamma_R T_R}\,dw_R(t)\;,
}
where $i = 1,\ldots,N$ and $j=1,\ldots,N-1$. Since {\bf A0} holds, the results
of the preceding section apply. Therefore, there exists for any initial
condition $x_0$ a unique stochastic process $\xi(t;x_0,w)$ solving
\eref{e:stochChain}. It obeys the statistics of a Markov diffusion process with
generator
\equs[e:defL]{
	\CL =&\; \lambda_L^2\gamma_L T_L \d_{\rL}^2 + \lambda_R^2\gamma_R T_R
\d_{\rR}^2 - \gamma_L(\rL - \lambda_L^2 q_0)\d_{\rL} - \gamma_R(\rR -
\lambda_R^2 q_N)\d_{\rR} \\
	&+ \rL \d_{p_0} + \rR \d_{p_N} + \sum_{i=0}^N \bigl(p_i\d_{q_i} -
\V1'(q_i)\d_{p_i}\bigr) - \sum_{i=1}^{N} \V2'(\tilde
q_i)\bigl(\d_{p_i}-\d_{p_{i-1}}\bigr)\;.
}
We want to prove the existence of a smooth invariant measure with
density $\mu(p,q,r)$.
It is the solution of
$(\T^t)^*\mu = 0$, where $(\T^t)^*$ is the dual semi-group of
$\T^t$. To achieve this, we introduce, as above, the Hilbert space 
\equ{
	\CH_0 \equiv \Ltwo\bigl(\R^{2N+4}, Z_0^{-1}e^{-2\beta_0
G(p,q,r)}\,dp\,dq\,dr\bigr)\;,
}
where $Z_0$ is a normalization constant and $\beta_0$ is a ``reference''
inverse temperature that we choose such that
\equ[e:condTRef]{
	1/\beta_0 \equiv T_0 > \max\{T_L,\,T_R\}\;.
}
Proposition~\ref{prop:Tt} holds, so the dynamics of our system is described by
a semi-group $\T_0^t$ acting in $\CH_0$ with generator $L_0$, formally given by
$\CL$. The extended phase space of our system will again be denoted by $\CX
\equiv \R^{2N+4}$.

For convenience, we would like to work in $\CH = \Ltwo(\CX)$, so we define the
unitary transformation $U:\CH \to \CH_0$ by
\equ{
	\bigl(U f\bigr)(x) = e^{\beta_0 G(x)}f(x)\;.
}
So $L_0$ is unitarily equivalent to the operator $L_{\CH} : \CD(L_{\CH}) \to
\CH$ defined by
\equ{
	L_{\CH} = U^{-1} L_0 U = e^{-\beta_0 G}L_0 e^{\beta_0 G}\;.
}
An explicit computation shows that $L_{\CH}$ is given by
\equ{
	L_{\CH} = \alpha - K\;,
}
where the formal expression for the differential operator $K$ is
\equs[e:defKchain]{
K =&\; \alpha_K - c_L^2 \d_{\rL}^2 + a_L^2(\rL - \lambda_L^2 q_0)^2  - c_R^2
\d_{\rR}^2 + a_R^2(\rR - \lambda_R^2 q_N)^2 \\
	&\;- \rL\d_{p_0}  + b_L(\rL - \lambda_L^2 q_0)\d_\rL - \rR\d_{p_N}  + b_R(\rR
- \lambda_R^2 q_N)\d_\rR \\
	&\;- \sum_{i=0}^N \bigl(p_i\d_{q_i} - \V1'(q_i)\d_{p_i}\bigr) + \sum_{i=1}^{N}
\V2'(\tilde q_i)\bigl(\d_{p_i}-\d_{p_{i-1}}\bigr)\;.
}
Since $\cOinf[\CX]$ is invariant under the unitary transformation
$U$, it remains a core for both $K$ and $K^*$. The various constants appearing
in \eref{e:defKchain} are given by
\begin{alignat*}{2}
a_i^2 &= \gamma_i(\beta_0 T_i - 1)\;, \quad&\quad &\\
b_i &= \frac{\gamma_i \beta_0}{\lambda_i^2}\bigl(\beta_0 T_i - 1\bigr)\;,
\quad&\quad i&\in\{L,R\}\;,\\
c_i &= \lambda_i\sqrt{\gamma_i T_i}\;,\quad&\quad&\\
\alpha_K &= -\frac{b_L}{2} - \frac{b_R}{2}\;,\quad&\quad&\\
\alpha &= \alpha_K + {\beta_0} \sum_{i\in\{L,R\}} \gamma_i T_i\;.\quad&\quad &
\end{alignat*}
We see that condition \eref{e:condTRef} ensures the positivity of the constants
$a_L^2$ and $a_R^2$, which in turn implies that the closure of $\Re K = (K +
K^*)/2$ is a strictly positive self-adjoint operator.

The first feature we notice about $K$ is that {\bf A3} implies the
hypoellipticity of the operators $K$, $K^*$, $\d_t + K$ and $\d_t + K^*$. We
recall that a differential operator $L$ acting on functions in a
finite-dimensional differentiable manifold $\CM$ is called hypoelliptic if
\equ{
	\text{sing supp } f = \text{sing supp } L f \;,\qquad \text{for all}\quad f
\in \CD'(\CM)\;,
}
where $\CD'(\CM)$ is the space of distributions on $\cOinf[\CM]$. In
particular, the eigenfunctions of a hypoelliptic operator are $\CC^\infty$.

The hypoellipticity of the above operators is a consequence of a theorem by
H\"ormander \cite{H1,Ho}: given a second-order differential operator
\equ{
	L = \sum_{i=1}^n L_i^* L_i + L_0 + c\;,
}
where $c : \CM \to \C$ is a smooth function and the $L_i$ are smooth vector
fields. Then a sufficient condition for $L$ to be hypoelliptic is that the Lie
algebra generated by $\{L_i\;|\; i=0,\ldots,n\}$ has maximal rank everywhere.
It is not hard to verify that {\bf A3} ensures that this condition is verified
for $K$, $K^*$, $\d_t + K$ and $\d_t + K^*$.

\begin{proposition}
If {\bf A0} and {\bf A3} are satisfied, the transition probabilities of the
Markov process solving \eref{e:stochChain} have a smooth density
\equ{
	P(t,x,y) \in \CC^\infty\bigl((0,\infty) \times \CX \times \CX\bigr)\;.
}
\end{proposition}

\begin{proof}
This is an immediate consequence of the Kolmogorov equations which state that
\equ{
	\d_t P = \CL P\qquad\Rightarrow\qquad (\d_t + K - \alpha)U^{-1}P = 0\;,
}
so $U^{-1}P$ is an eigenfunction of the operator $\d_t + K - \alpha$, which is
hypoelliptic.
\end{proof}

\subsection{Main results}

Our main technical result is

\begin{theorem}
\label{theo:Chain}
If Assumptions {\bf A1}--{\bf A3} are satisfied, then the operator
$K$ defined in \eref{e:defKchain} has compact resolvent.
\end{theorem}

In order to prepare the proof of Theorem \ref{theo:Chain}, we will prove the
following two propositions.

\begin{proposition}
\label{prop:estG}
If Assumptions {\bf A1} and {\bf A2} are satisfied, there exist constants $C$
and $\eps > 0$ such that
\sublabels
\equs[2,e:estG]{
	\|\G^{\eps}f\| &\le C(\|K f\| + \|f\|)\;, \quad&\text{for all}\quad f&\in
\CD(K)\;,\sublabel{e:estG1}\\
	\|\G^{\eps}f\| &\le C(\|K^* f\| + \|f\|)\;, \quad&\text{for all}\quad f&\in
\CD(K^*)\;.\sublabel{e:estG2}
}
\end{proposition}

\begin{proposition}
\label{prop:estDeltaPrime}
If Assumptions {\bf A1}--{\bf A3} are satisfied, there exist
constants $C$, $\eps > 0$, a positive function $a_0:\CX \to \R$ and a finite
number $\bar N$ of smooth vector fields $L_i$ with bounded coefficients such
that, for every function $f \in \cOinf[\CX]$, we have
\equs[1,e:defDeltaPrime]{
	\|\tilde\Delta^{\eps}f\| &\le C(\|K f\| + \|f\|)\;,\\
\intertext{where}
	\tilde \Delta &= \sum_{i=1}^{\bar N}L_i^* L_i + a_0\;.\notag
}
Moreover, the $L_i$ span the whole of $\R^{2N+4}$ at every point.
\end{proposition}

Given Theorem~\ref{theo:Chain}, we can state and prove the main result of this
paper, namely the existence and uniqueness of an invariant measure for our
Markov process. More precisely, we have the following result.

\begin{theorem}
\label{theo:exist}
If Assumptions {\bf A1}--{\bf A3} are satisfied, then the
stochastic process $\xi(t)$ solving \eref{e:sysStoch} possesses a unique and
strictly positive invariant measure $\mu$. Its density $h$ is ${\cal
C}^\infty$ and satisfies for any $\beta_0 <
\min\{\beta_L,\beta_R\}$,
\equ{
	h(x) = \tilde h(x) e^{-\beta_0 G(x)}\;,
}
where $\tilde h$ decays at infinity faster than any polynomial.
\end{theorem}

\begin{MHwrap}{r}{7cm}{SpectrFig}[-5mm]
	\vspace{-5mm}
	\caption{Spectrum of $K$.}
	\label{fig:spec}
\end{MHwrap}
The above results say that the spectrum of $K$ looks roughly like the one
schematically depicted in Figure~\ref{fig:spec}. We see that it is discrete
(compactness of the resolvent) and located in the right half of the complex
plane ({\it m}-accretivity). Moreover, it is symmetric along the real axis,
because $K$ is a differential
operator with real coefficients.

Most of the remainder of this paper is devoted to the proofs of
Theorems~\ref{theo:Chain} and \ref{theo:exist}.
In the sequel, we will always use the notation
\equ{
	K = \sum_{i=1}^4 X_i^* X_i + X_0\;,
}
where we define
\sublabels
\equs[2,e:defXi]{
	X_1 &= c_L \d_{\rL}\;, \quad&\quad X_2 &= a_L(\rL - \lambda_L^2 q_0)\;,
\sublabel{e:defLeft}\\
	X_3 &= c_R \d_{\rR}\;, \quad&\quad X_4 &= a_R(\rR - \lambda_R^2 q_N)\;,
\sublabel{e:defRight}
}
\vspace{-5mm}
\sublabels*
\equs[e:defX0]{
X_0 =&\; - \rL\d_{p_0} + b_L(\rL - \lambda_L^2 q_0)\d_\rL  - \rR\d_{p_N} +
b_R(\rR - \lambda_R^2 q_N)\d_\rR \\
 &\;- \sum_{i=0}^N \bigl(p_i\d_{q_i} - \V1'(q_i)\d_{p_i}\bigr) + \sum_{i=1}^{N}
\V2'(\tilde q_i)\bigl(\d_{p_i}-\d_{p_{i-1}}\bigr) - \alpha_K\;.
}
The operator $X_0$ is antisymmetric, \ie
\equ[e:propX0star]{
	X_0^* = -X_0\;.
}
This implies that
\equ[e:defReK]{
	\Re K = \sum_{i=1}^4 X_i^* X_i \quad\text{and}\quad X_0 = K-\Re K\;,
}
and thus $\Re K$ is a positive self-adjoint operator. We have one more estimate
that will be extensively used in the sequel. If $f$ is some function in
$\cOinf[\CX]$ and $i\in\{1,\ldots,4\}$ we have
\equ[e:estXchain]{
	\|X_i f\|^2 = \scal{f,X_i^* X_i f} \le \scal{f, \Re K f} = \Re\scal{f,K f} \le
\|f\|\|K f\| \le (\|Kf\| + \|f\|)^2\;,
}
and by a similar argument also
\equ[e:estXstarchain]{
	\|X_i^* f\|^2 \le (\|Kf\| + \|f\|)^2\;.
}

\mysection{Proof of the bound in position space (Proposition~\ref{prop:estG})}
\label{sec:proof1}

First of all, we need a collection of functions belonging to
$\CF_0$, as defined in Definition~\ref{def:func}. We have the following result.
\begin{proposition} \label{prop:Degs}
Let $r$, $p$, $q$ and $\tilde q$ designate the vectors
\equs[2]{
	r &= (\rL, \rR)\;,\quad&\quad q &= (q_0, \ldots, q_N)\;,\\
	p &= (p_0, \ldots, p_N)\;,\quad&\quad\tilde q &= (\tilde q_1, \ldots, \tilde
q_N)\;.
}
Choose $\alpha \ge 0$ and let $h_k : \R^k \to \R$ be functions in $\CF_\alpha$.
Then the functions
\equ{
	G^{-\alpha/2}h_2(r)\;,\quad G^{-\alpha/2}h_{N+1}(p)\;,\quad
G^{-\alpha/(2n)}h_{N+1}(q)\;,\quad\text{and}\quad \;G^{-\alpha/(2m)}h_N(\tilde
q)
}
belong to $\CF_0$.
\end{proposition}

\begin{proof}
We will only sketch the proof of the statement for $G^{-\alpha/(2m)}h_N (\tilde
q)$. The other expressions can easily be treated in a similar way.

We first notice that $G^{-1}(D^k G)$ is bounded for every multi-index $k$. This
is a straightforward consequence of two observations. The first one is that
because of the lower bounds \eref{e:boundV1} and \eref{e:boundV2} of {\bf A1}
and {\bf A2} and the expression \eref{e:defGChain} of $G$, there exists a
constant $C >0$ for which
\equ[e:bound1]{
G(p,q,r) \ge C\bigl(r^2 + p^2 + P^{2n}(q) + P^{2m}(\tilde q)\bigr)\;,
}
where $P^k$ was defined in \eref{e:defPoly}. The second observation is that,
because $\V1 \in \CF_{2n}$ and $\V2 \in \CF_{2m}$, we have for every
multi-index $k$ some constant $C_k$ for which
\equ[e:bound2]{
|D^k G(p,q,r)| \le C_k \bigl(r^2 + p^2 + P^{2n}(q) + P^{2m}(\tilde q)\bigr)\;.
}
Notice that $G^{-\alpha/(2m)}D^k h_N(\tilde q)$ is bounded by a similar
argument, in particular because $h_N \in \CF_\alpha$.

We set $\alpha = -\alpha/(2m)$ and write
\equ{
	\d_i\bigl(G^\alpha h_N(\tilde q)\bigr) = \alpha\bigl(G^{-1}\d_i G\bigr)
G^\alpha h_N(\tilde q) + G^\alpha \d_i h_N(\tilde q)\;.
}
Both terms are bounded by \eref{e:bound1}, \eref{e:bound2} and the fact that
$h_N \in \CF_\alpha$. It is easy to see that all the derivatives can be bounded
similarly. The proof of Proposition~\ref{prop:Degs} is complete.
\end{proof}

Let us define
\equ{
	\Lambda_1 \equiv G^{1/2}\;.
}
The symbol $\Lambda_1$ was chosen in order to emphasize the similarity between
the proof of Proposition~\ref{prop:estG} and the proof of the main result of
Section~\ref{sec:Hormander}, Theorem~\ref{theo:princ}.

Before we start the proof of Proposition~\ref{prop:estG}, we notice two more facts.
Let us choose $\alpha$,$\beta \in \R$ with $0 \le \beta \le 1$, and let $A$,
$B$ be two operators of multiplication by positive functions $A \le B$. We then
have
\equ[e:estPosFun]{
	\scal{\Lambda_1^\alpha A f, f} \le \scal{\Lambda_1^\alpha B f, f}\;,
}
as well as the implication
\equ[e:estExpos]{
	\|\Lambda_1^\alpha A f\| \le C(\|K f\| + \|f\|) \quad\Rightarrow\quad
\|\Lambda_1^{\alpha \beta} A^\beta f\| \le C(\|K f\| + \|f\|)\;.
}
Both inequalities are trivial consequences of the fact that $\Lambda_1$ is an
operator of multiplication by a positive function and the estimate $x^s \le
1+x$ if $x \ge 0$ and $s \le 1$.

\subsection{The main tool of the proof}

The main tool in the proof of Proposition~\ref{prop:estG} is the following lemma.

\begin{lemma}
\label{lem:estG}
Let $\Lambda_1$ and $K$ be defined as above. Let $A$ and $B$ be multiplication
operators represented by functions of the form
\equ{
	h(p,q,r) = c_L \rL + c_R \rR + \tilde h(p,q)\;,\qquad \tilde h \in
\cinf(\R^{2N+2})\;.
}
Assume moreover that there are exponents $\alpha_i$ and $\beta_i$ and positive
constants $C_i$ such that the following estimates are true for every $f \in
\cOinf[\CX]$.
\begin{alignat*}{2}
\|\Lambda_1^{-\alpha_1}A f\| &\le C_1(\|K f\| + \|f\|)\;, \qquad&\qquad
\|\Lambda_1^{-\beta_1}B f\| &\le C_2(\|K f\| + \|f\|)\;, \\
\|\Lambda_1^{-\alpha_2}A f\| &\le C_3\|f\|\;, \qquad&\qquad
\|\Lambda_1^{-\beta_2}B f\| &\le C_4\|f\|\;, \\
\|\Lambda_1^{-\alpha_3}[X_0,A]f\| &\le C_5(\|K f\| + \|f\|)\;, \qquad&\qquad
\|\Lambda_1^{-\beta_3}[X_0,B]f\| &\le C_6(\|K f\| + \|f\|)\;.
\end{alignat*}
If $\gamma$ satisfies the conditions
\begin{eqnarray}
\gamma &\ge& \alpha_3 + \beta_1 \label{e:cond1}\;,\\
\gamma &\ge& \alpha_2 + \frac{\beta_1 + \max\{\beta_2,\beta_3\}}2
\label{e:cond2}\;,\\
\gamma &\ge& \min\{\alpha_1 + \beta_2\,,\, \alpha_2 + \beta_1\}\;,
\label{e:cond4}
\end{eqnarray}
then there exists a constant $C$ such that
\equ[e:estWork]{
	|\scal{[X_0,B]f,\Lambda_1^{-\gamma}A f}| \le C(\|K f\| +
\|f\|)^2\;,\quad\text{for all}\quad f\in
\cOinf[\CX]\;.
}
\end{lemma}

\begin{proof}
The proof of this lemma involves some of the commutation techniques developed
by H\"ormander \cite{Ho}, but it uses the fact that most operators involved are
multiplication operators, \ie they commute. An explicit computation, using
\eref{e:defGChain} and \eref{e:defXi} yields
\sublabels
\equs[2,e:comms]{
[X_0, G] &= \sum_{j\in\{R,L\}} \frac{b_j}{\lambda_j^2}\bigl(r_j - \lambda_j^2
F_j\bigr)^2 \;,&\quad [X_1,G] &= c_L\bigl(\rL/\lambda_L^2 - q_0\bigr)\;,
\sublabel{e:comms1}\\
[X_2,G] &= [X_4,G] = 0\;,&\quad [X_3,G] &= c_R\bigl(\rR/\lambda_R^2 -
q_N\bigr)\;. \sublabel{e:comms2}
}
We therefore see that, by Proposition~\ref{prop:Degs}, we have for
$i=0,\ldots,4$
\equ[e:boundG]{
	G^{-1}[X_i,G] \in \CF_0\;.
}
Since the $X_i$ are either differentiation operators or multiplicative
operators, we have, for any $\alpha \in \R$, the relation
\equ{
	G^{-\alpha}[X_i, G^\alpha] = \alpha G^{-1}[X_i,G] \in \CF_0\;,
}
and so, since $\Lambda_1^2 = G$,
\equ[e:boundComm]{
	\|\Lambda_1^{\alpha}[X_i,\Lambda_1^{-\alpha}]\| < \infty\;.
}

We can now start to bound \eref{e:estWork}. Since $[X_0, B] = - X_0^* B - B
X_0$, we can write \eref{e:estWork} as
\equs{
|\scal{[X_0,B]f,\Lambda_1^{-\gamma}A f}| &\le |\scal{B
X_0f,\Lambda_1^{-\gamma}A f}| + |\scal{B f, X_0 \Lambda_1^{-\gamma}A f}| \\
& \equiv T_1 + T_2\;.
}
Both terms will be estimated separately.

\proclaim{Term $\boldsymbol{T_1}$.} Since we know by \eref{e:defReK} that $X_0
= K - \Re K$, we can write it as
\equ{
	T_1 \le |\scal{B(\Re K)f, \Lambda_1^{-\gamma}A f}| + |\scal{B K f,
\Lambda_1^{-\gamma}A f}|
	\equiv T_{11} + T_{12}\;.
}
The term $T_{12}$ can be estimated by using \eref{e:cond4}. We indeed have
either $\gamma \ge \alpha_1 + \beta_2$, or $\gamma \ge \alpha_2 + \beta_1$. In
the former case, we write
\equ{
T_{12} \le \|\Lambda_1^{-\beta_2} B\| \|K f\| \|\Lambda_1^{-\gamma + \beta_2}A
f\| \le C(\|K f\| + \|f\|)^2\;.
}
In the latter case, we use the fact that $A$, $B$ and $\Lambda_1$ commute and
are self-adjoint to write similarly
\equ{
T_{12} = |\scal{A K f, \Lambda_1^{-\gamma}B f}|
	\le \|\Lambda_1^{-\alpha_2} A\| \|K f\| \|\Lambda_1^{-\gamma + \alpha_2}B f\|
\le C(\|K f\| + \|f\|)^2\;.
}
Let us now focus on the term $T_{11}$. Using the positivity of $\Re K$, it can
be written as
\equs{
T_{11} &= \scal{(\Re K)^{1/2}\Lambda_1^{-\gamma_1}B f,(\Re
K)^{1/2}\Lambda_1^{-\gamma_2}A f} + \scal{[\Lambda_1^{-\gamma_1}B,\Re K]f,
\Lambda_1^{-\gamma_2}A f}\\
&\equiv T_{13} + T_{14}\;,
}
where
\equ{
\gamma_1, \gamma_2 > 0\;,\qquad\gamma_1 + \gamma_2 = \gamma\;,
}
are to be chosen later. We estimate both terms separately. The commutator in
$T_{14}$ can be expanded to give
\equ{
T_{14} = \scal{\Lambda_1^{-\gamma_1}[B,\Re K]f,\Lambda_1^{-\gamma_2}A f} +
\scal{[\Lambda_1^{-\gamma_1},\Re K]B f,\Lambda_1^{-\gamma_2}A f}\;.
}
In order to estimate these terms, we recall that $\Re K = \sum_{i=1}^4 X_i^*
X_i$. We therefore have
\equs{
T_{14} =&\; \sum_{i=1}^4 \Bigl( \scal{\Lambda_1^{-\gamma_1}[B,X_i^*]X_i
f,\Lambda_1^{-\gamma_2}A f} + \scal{\Lambda_1^{-\gamma_1}X_i^*[B,X_i]
f,\Lambda_1^{-\gamma_2}A f} \\[0mm]
	&\qquad+\scal{[\Lambda_1^{-\gamma_1},X_i^*]X_i B f,\Lambda_1^{-\gamma_2}A
f}+\scal{X_i^*[\Lambda_1^{-\gamma_1},X_i] B f,\Lambda_1^{-\gamma_2}A f}\Bigr)
\\[0mm]
	\equiv&\; \sum_{i=1}^4 \bigl(T_{i}^{(1)} +
T_{i}^{(2)}+T_{i}^{(3)}+T_{i}^{(4)}\bigr)\;.
}
Noticing that $[B, X_i^*]$ is a multiple of the identity operator and that
$\Lambda_1$ is self-adjoint, we have
\equ{
	|T_i^{(1)}| \le C\scal{X_i f, \Lambda_1^{-\gamma} A f} \le \|X_i f\|
\|\Lambda_1^{-\gamma} A f\| \le C(\|K f\| + \|f\|)^2\;,
}
where we used \eref{e:estXchain} and the fact that $\gamma > \alpha_2$ to get
the last inequality.
The term $T_{i}^{(2)}$ is bounded by $C(\|K f\| + \|f\|)^2$ in a similar way.
The term $T_{i}^{(3)}$ is written as
\equs{
|T_{i}^{(3)}| &= |\scal{\Lambda_1^{\gamma_1}[\Lambda_1^{-\gamma_1},X_i^*]X_i f,
\Lambda_1^{-\gamma}A B f} +
\scal{\Lambda_1^{\gamma_1}[\Lambda_1^{-\gamma_1},X_i^*][X_i,B] f,
\Lambda_1^{-\gamma}A f}| \\[1mm]
&\le C\|X_i f\|\|\Lambda_1^{-\gamma} A B f\| + C\|f\|\|\Lambda_1^{-\gamma} A
f\|\;,
}
where we used \eref{e:boundComm} and the fact that $[X_i, B]$ is bounded.
Now we can bound $T_i^{(3)}$ by $C(\|Kf\| + \|f\|)^2$, using \eref{e:estXchain}
to estimate $\|X_i f\|$ and \eref{e:cond4} to estimate $\|\Lambda_1^{-\gamma} A
B f\|$ and $\|\Lambda_1^{-\gamma} A f\|$. The term $T_{i}^{(4)}$ can be
estimated in a similar way.

Let us now focus on the term $T_{13}$. We can write
\equ{
	|T_{13}| \le |\Re\scal{K \Lambda_1^{-\gamma_1}B f,\Lambda_1^{-\gamma_1}B
f}|^{1/2}\sqrt{\sum_{i=1}^4\|X_i\Lambda_1^{-\gamma_2}A f\|}\;.
}
If we choose
\equ[e:hyp1]{
\gamma_2 = \alpha_2\;,
}
the terms under the square root are easily estimated by writing them as
\equs{
\|X_i\Lambda_1^{-\gamma_2}A f\| &\le \|\Lambda_1^{-\gamma_2}A\| \|X_i f\| +
\|[X_i,\Lambda_1^{-\gamma_2}]\Lambda_1^{\gamma_2}\| \|\Lambda_1^{-\gamma_2}A
f\| \\
	&\quad + \|\Lambda_1^{-\gamma_2}[X_i,A] f\|\;,
}
and estimating the two commutators by \eref{e:boundComm} and \eref{e:comms}
respectively.

The term preceding the square root can be written as
\equs{
\scal{K \Lambda_1^{-\gamma_1}B f,\Lambda_1^{-\gamma_1}B f} &=
\scal{\Lambda_1^{-\gamma_1}B K f,\Lambda_1^{-\gamma_1}B f} +
\scal{[K,\Lambda_1^{-\gamma_1}B]f,\Lambda_1^{-\gamma_1}B f} \\
	&\equiv T_{15} + T_{16}\;.
}
The term $T_{15}$ can be bounded if we choose
\equ[e:hyp2]{
	2\gamma_1 \ge \beta_1 + \beta_2\;,
}
because we have then
\equ{
	T_{15} \le \|K f\|\|\Lambda_1^{-\beta_2}B\| \|\Lambda_1^{-\beta_1}B f\| \le
C(\|K f\| + \|f\|)^2\;.
}
In order to estimate the term $T_{16}$, we use $K = \Re K + X_0$ to write
\equs{
T_{16} &= \scal{[X_0,\Lambda_1^{-\gamma_1}B]f,\Lambda_1^{-\gamma_1}B f} +
\scal{[\Re K,\Lambda_1^{-\gamma_1}B]f,\Lambda_1^{-\gamma_1}B f}\\[1mm]
	&\equiv T_{16}^{(1)} + T_{16}^{(2)}\;.
}
The term $T_{16}^{(1)}$ can be estimated by writing it as
\equ{
 T_{16}^{(1)} = \scal{\Lambda_1^{-\gamma_1}[X_0,B]f,\Lambda_1^{-\gamma_1}B f} +
\scal{[X_0,\Lambda_1^{-\gamma_1}]\Lambda_1^{\gamma_1} \Lambda_1^{-\gamma_1}B
f,\Lambda_1^{-\gamma_1}B f}\;.
}
The first term can be bounded by $C(\|K f\| + \|f\|)^2$ if we choose
\equ[e:hyp3]{
2\gamma_1 \ge \beta_1 + \beta_3\;.
}
In order to bound the second term, it suffices to have $\gamma_1 \ge \beta_1$,
which is the case because of \eref{e:hyp2} and the fact that $\beta_2 \ge
\beta_1$.

The term $T_{16}^{(2)}$ can be bounded by $C(\|K f\| + \|f\|)^2$, by treating
it in a similar way than the term $T_{14}$. We leave to the reader the
verification that no additional conditions on $\gamma_1$ have to be made. This
completes the estimate of $T_1$, because \eref{e:hyp1}, \eref{e:hyp2} and
\eref{e:hyp3} can be satisfied simultaneously by \eref{e:cond2}.

\proclaim{Term $\boldsymbol{T_2}$.} We decompose this term as
\equs{
	T_2 &\le |\scal{B f, \Lambda_1^{-\gamma}A X_0 f}| + |\scal{B f,
\Lambda_1^{-\gamma}[X_0,A] f}| + |\scal{B f, [X_0, \Lambda_1^{-\gamma}]A f}|
\\[1mm]
	& \equiv T_{21} + T_{22} + T_{23}\;.
}

Since $\gamma \ge \alpha_3 + \beta_1$ the term $T_{22}$ is easily estimated by
\equ{
	T_{22} \le \|\Lambda_1^{-\beta_1} B f\|\|\Lambda_1^{-\alpha_3} [X_0,A] f\| \le
C(\|K f\| + \|f\|)^2\;.
}

Noticing that we can assume $\alpha_1 \le \alpha_2$ and $\beta_1 \le \beta_2$,
condition \eref{e:cond4} implies $\gamma \ge \alpha_1 + \beta_1$. Since $[X_0,
\Lambda_1^{-\gamma}]$ is a function, it commutes with $\Lambda_1$, and so
$T_{23}$ can be estimated writing
\equs{
	T_{23} &\le |\scal{\Lambda_1^{-\beta_1}B f, \Lambda_1^\gamma [X_0,
\Lambda_1^{-\gamma}]\Lambda_1^{-\alpha_1}A f}| \\
	& \le \|\Lambda_1^{-\beta_1}B f\| \|[X_0,
\Lambda_1^{-\gamma}]\Lambda_1^\gamma\|\|\Lambda_1^{-\alpha_1}A f\| \le C(\|Kf\|
+ \|f\|)^2\;,
}
where we used \eref{e:boundComm} to get the last bound.

We finally bound $T_{21}$. Since $X_0 = K - \Re K$, it can be expanded as
\equ{
T_{21} \le |\scal{B f, \Lambda_1^{-\gamma}A K f}| + |\scal{B f,
\Lambda_1^{-\gamma}A(\Re K) f}| \equiv T_{21}^{(1)} + T_{21}^{(2)}\;.
}
The term $T_{21}^{(1)}$ can be estimated by writing
\equ{
 T_{21}^{(1)} \le \|K f\| \|\Lambda_1^{-\gamma}A B f\|\;,
}
and using \eref{e:cond4}. The term $T_{21}^{(2)}$ can be written as
\equ{
T_{21}^{(2)} = \scal{B f, \Lambda_1^{-\gamma}A(\Re K) f} = T_{13} +
\scal{\Lambda_1^{-\gamma_1}B f, [\Lambda_1^{-\gamma_2}A, \Re K] f}\;.
}
The term $T_{13}$ has already been estimated. The other term can be treated
like the term $T_{14}$. We leave to the reader the verification that one can
indeed bound it by $C(\|K f\| + \|f\|)^2$ without any further restriction on
$\gamma_1$ and $\gamma_2$.

This completes the proof of the lemma.
\end{proof}

\newcommand\expos[7][]{
\begin{alignat*}{2}
	\alpha_1 &= #2\;,\qquad&\qquad \beta_1 &= #5\;, \\
	\alpha_2 &= #3\;,\qquad&\qquad \beta_2 &= #6\;, \\
	\alpha_3 &= #4\;,\qquad&\qquad \beta_3 &= #7\;#1
\end{alignat*}
}

\subsection{The main step of the proof of Proposition~\ref{prop:estG}}

By an elementary approximation argument, it is sufficient to prove the
inequalities \eref{e:estG} for $f \in \cOinf[\CX]$, since this is a core for
both $K$ and $K^*$. Moreover, we will prove only \eref{e:estG1}. The interested
reader may verify that the same arguments also apply for \eref{e:estG2}.

We want to show that we can find constants $\eps$ and $C$ such that
\equ{
	\|\Lambda_1^{\eps}f\| \le C(\|K f\| + \|f\|)\;,\quad\text{for all}\quad f\in
\cOinf[\CX]\;.
}
In order to show this, we notice that there is a constant $C$ such that
\equ{
	\Lambda_1^2 \le C\Bigl(1 + (\rL - \lambda_L^2 q_0)^2 + (\rR - \lambda_r^2
q_N)^2 + \sum_{i=0}^N p_i^2 + P^{2n}(Q) + \sum_{i=1}^N P^{2m}(\tilde q_i)\Bigr)
\equiv \tilde G\;.
}
The immediate consequence is that
\equ{
	\|\Lambda_1^{\eps}f\|^2 = \scal{f, \Lambda_1^{2\eps}f} \le
\scal{f,\Lambda_1^{2\eps-2}\tilde G f}\;.
}
It is therefore enough to show that there exists a (small) constant $\eps$ such
that the terms
\equ{
	\|\Lambda_1^{\eps-1}P^n(Q)f\|\,,\;\|\Lambda_1^{\eps-1}p_i
f\|\,,\;\|\Lambda_1^{\eps-1}P^m(\tilde q_i)f\|\,,\ldots
}
are bounded by $C(\|K f\| + \|f\|)$.

We are first going to bound the terms involving variables near the boundary of
the chain. Then, we will proceed by induction towards the middle of the chain.

\proclaim{The term $\boldsymbol{\|\Lambda_1^{\eps-1}(\rL - \lambda_L^2
q_0)f\|}$.} We have
\equs[e:estR]{
	\|(\rL - \lambda_L^2 q_0)f\|^2 &= |\scal{(\rL-\lambda_L^2 q_0)^2f,f}| \le
C|\scal{(\Re K)f,f}| \\
&= C|\Re\scal{K f,f}| \le C\|K f\|\|f\| \le C(\|K f\| + \|f\|)^2\;,
}
where we used the fact that $a_L \neq 0$ to obtain the first inequality.
Since $\Lambda_1 \ge 1$, we thus have the estimate
\equ{
\|\Lambda_1^{\eps-1}(\rL - \lambda_L^2 q_0)f\| \le C(\|K f\| + \|f\|)^2
}
if we take $\eps \le 1$.

%
%

\proclaim{The term $\boldsymbol{\|\Lambda_1^{\eps-1}p_0 f\|}$.} We will prove
the estimate
\equ[e:estp1]{
	\|\Lambda_1^{\eps_0-1}p_0 f \| \le C(\|K f\| + \|f\|)\;,
}
for $\eps_0 \le 1/(2m)$. An explicit computation yields the relation
\equ[e:estp02]{
	[X_0, \rL - \lambda_L^2 q_0] = b_L (\rL - \lambda_L^2 q_0) - \lambda_L^2
p_0\;.
}
Solving \eref{e:estp02} for $p_0$, we get
\equ{
\|\Lambda_1^{\eps_0-1}p_0 f \|^2 = \scal[b]{\lambda_L^{-2} \bigl(b_L (\rL -
\lambda_L^2 q_0) - \lambda_L^{-2}[X_0,\rL - \lambda_L^2 q_0]\bigr)f,
\Lambda_1^{2\eps_0-2}p_0f} \equiv X_0^{(1)} - X_0^{(2)}\;.
}
The term $X_0^{(1)}$ can be estimated as
\equ{
	|X_0^{(1)}| \le \lambda_L^{-2} \|b_L (\rL - \lambda_L^2 q_0)
f\|\|\Lambda_1^{2\eps_0-2}p_0 f\| \le C(\|K f\| + \|f\|)^2\;,
}
where the last inequality holds because $\eps_0 \le 1/2$.

In order to estimate $X_0^{(2)}$, we apply Lemma~\ref{lem:estG} with $A = p_0$
and $B = \rL - \lambda_L^2 q_0$. An explicit computation yields $[X_0, A] =
\V1'(q_0) + \V2'(\tilde q_1) - \rL$. The term $[X_0,B]$ has already been
computed in \eref{e:estp02}. Because of Proposition~\ref{prop:Degs} and of
\eref{e:estR}, we can choose
\expos[.]{1}{1}{2-1/m}{0}{1}{1}
The hypotheses of Lemma~\ref{lem:estG} are thus fulfilled if we choose $\gamma
= 2-1/m$. We therefore have the estimate \eref{e:estp1} with $\eps_0 = 1/(2m)$.
We have a similar estimate for the symmetric term at the other end
of the chain.

%
%

\proclaim{The term $\boldsymbol{\|\Lambda_1^{\eps-1} P^m(\tilde q_1) f\|}$.} We
will prove the estimate
\equ{
	\|\Lambda_1^{\eps_0'-1}P^m(\tilde q_1) f \| \le C(\|K f\| + \|f\|)\;,
}
for some $\eps_0' < \eps_0$. Because of the bound \eref{e:boundxV2} of {\bf
A2}, we can find some constants $c_1$ and $c_2$ such that
\equ[e:estq1q0]{
	\scal[b]{\Lambda_1^{2\eps_0'-2}P^{2m}(\tilde q_1) f,f} \le
c_1\bigl|\scal[b]{\Lambda_1^{2\eps_0'-2} \V2'(\tilde q_1)f, \tilde q_1 f}\bigr|
+ c_2\bigl|\scal[b]{\Lambda_1^{2\eps_0'-2}f, f}\bigr|\;,
}
where we also used \eref{e:estPosFun}. The second term is easily estimated
because $\Lambda_1^{2\eps_0'-2}$ is bounded if $\eps_0' \le 1$.
We use once again the fact that $[X_0, p_0] = \V1'(q_0) + \V2'(\tilde q_1) -
\rL$ to write the first term as
\equ{
\bigl|\scal[b]{\Lambda_1^{2\eps_0'-2} \V2'(\tilde q_1)f, \tilde q_1 f}\bigr| =
\bigl|\scal[b]{\Lambda_1^{2\eps_0'-2} \bigl([X_0, p_0] - \V1'(q_0) +
\rL\bigr)f, \tilde q_1 f}\bigr| \equiv |Y_1^{(1)} + Y_1^{(2)} + Y_1^{(3)}|\;.
}
The term $Y_1^{(2)}$ can be written as
\equ{
|Y_1^{(2)}| = \bigl|\scal[b]{\Lambda_1^{2\eps_0'-2+1/m}\V1'(q_0)f,
\Lambda_1^{-1/m}\tilde q_1 f}\bigr| \le
\bigl\|\Lambda_1^{2\eps_0'-2+1/m}\V1'(q_0)f\bigr\| \|\Lambda_1^{-1/m}\tilde q_1
f\|\;.
}
By Proposition~\ref{prop:Degs} and the fact that $\V1' \in \CF_{2n-1}$, this
term is bounded by $C\|f\|^2$ if we take $\eps_0'$ so small that 
\equ[e:eps01]{
	2\eps_0' \le 1/n - 1/m\;.
}
The term $Y_1^{(3)}$ is bounded similarly by writing
\equ{
|Y_1^{(3)}| \le \bigl\|\Lambda_1^{2\eps_0'-2+1/m}\rL f\bigr\|
\|\Lambda_1^{-1/m}\tilde q_1 f\|\;,
}
if we impose
\equ[e:eps02]{
	2\eps_0' \le 1 - 1/m\;.
}
Both conditions can be satisfied because
we assumed that $1 < n < m$. In order to estimate $Y_1^{(1)}$, we apply once
again Lemma~\ref{lem:estG}. This time we have $A = \tilde q_1$ and $B = p_0$.
Using \eref{e:estp1} and Proposition~\ref{prop:Degs}, we see that we can choose
\expos[.]{1/m}{1/m}{1}{1-\eps_0}{1}{2-1/m}
By using $m > 1$, we see that the hypotheses of Lemma~\ref{lem:estG} are
fulfilled if \eref{e:eps01} and \eref{e:eps02} hold, together with $\eps_0' < \eps_0/2$.
Once again, we have the same estimate at the other end of the chain.

We can now go along the chain by induction. At each step, we go one particle
closer towards the middle of the chain. We present here only the terms arising
when we go from the left to the right of the chain.

%
%

\proclaim{The term $\boldsymbol{\|\Lambda_1^{\eps-1}p_i f\|}$.} We already
treated the case $i=1$. Let us therefore assume $i>1$. We moreover assume that
there exist constants $\eps_{i-1}, \eps_{i-1}'>0$ such that we have the
estimates
\equ[e:indp]{
\bigl\|\Lambda_1^{\eps_{i-1}-1}p_{i-1} f\bigr\| \le C(\|K f\| + \|f\|)
\quad\text{and}\quad \bigl\|\Lambda_1^{\eps_{i-1}'-1}P^m(\tilde q_i) f\bigr\|
\le C(\|K f\| + \|f\|)\;.
}
We will show that this implies the existence of a constant $\eps_i > 0$ such
that
\equ[e:estpi]{
\bigl\|\Lambda_1^{\eps_i-1}p_i f\bigr\| \le C(\|K f\| + \|f\|)\;.
}
We use $p_i = p_{i-1} + [X_0, \tilde q_i]$ to write
\equ{
\bigl\|\Lambda_1^{\eps_i-1}p_i f\bigr\|^2 =
\scal{\Lambda_1^{2\eps_i-1}p_{i-1}f, \Lambda_1^{-1}p_i f} +  \scal{[X_0, \tilde
q_i]f, \Lambda_1^{2\eps_i-2}p_i f} \equiv X_i^{(1)} + X_i^{(2)}\;.
}
The term $X_i^{(1)}$ is easily bounded if we write
\equ{
	|X_i^{(1)}| \le \|\Lambda_1^{2\eps_i-1}p_{i-1}f\| \|\Lambda_1^{-1}p_i f\| \le
C(\|K f\| + \|f\|)^2\;,
}
where the last inequality is obtained by using Proposition~\ref{prop:Degs} and
\eref{e:indp}. We only have to make the assumption $2\eps_i \le \eps_{i-1}$.

In order to estimate the term $X_i^{(2)}$, we apply Lemma~\ref{lem:estG} with
$A = p_i$ and $B = \tilde q_i$. Explicit computation yields $[X_0,p_i] =
\V1'(q_i) - \V2'(\tilde q_{i+1}) - \V2'(\tilde q_i)$. Using the induction
hypothesis \eref{e:indp} and Proposition~\ref{prop:Degs}, we see that we can
choose
\expos[.]{1}{1}{2-1/m}{(1-\eps_{i-1}')/m}{1/m}{1}
If we take $\eps_i \le \eps_{i-1}'/(2m)$, we see that the hypotheses of
Lemma~\ref{lem:estG} are satisfied. We thus have the desired bound
\eref{e:estpi}.

%
%

\proclaim{The term $\boldsymbol{\|\Lambda_1^{\eps-1}P^m(\tilde q_{i+1}) f\|}$.}
We assume that there exist strictly positive constants $\eps_i$ and
$\eps_{i-1}'$ such that
\equ{
\bigl\|\Lambda_1^{\eps_i-1}p_i f\bigr\| \le C(\|K f\| + \|f\|)
\quad\text{and}\quad \bigl\|\Lambda_1^{\eps_{i-1}'-1}P^m(\tilde q_i) f\bigr\|
\le C(\|K f\| + \|f\|)\;.
}
We will show that this implies the existence of a constant $\eps_i' > 0$ for
which
\equ[e:estqtilde]{
\bigl\|\Lambda_1^{\eps_i'-1}P^m(\tilde q_{i+1}) f\bigr\| \le C(\|K f\| +
\|f\|)\;.
}
Expression \eref{e:estq1q0} with $\tilde q_1$ replaced by $\tilde q_{i+1}$
holds. In order to prove \eref{e:estqtilde}, it suffices therefore to show that
\equ{
	|\scal{\Lambda_1^{2\eps_i'-2}\V2'(\tilde q_{i+1})f, \tilde q_{i+1} f}| \le
C(\|Kf\| + \|f\|)^2\;.
}
Since, for $i>1$ we have $[X_0, p_i] = \V1'(q_i) - \V2'(\tilde q_{i+1}) -
\V2'(\tilde q_i)$, the preceding term can be written as
\equ{
\bigl|\scal[b]{\Lambda_1^{2\eps_i'-2}\bigl([X_0, p_i] + \V1'(q_i) + \V2'(\tilde
q_i)\bigr)f, \tilde q_{i+1}f}\bigr| \equiv |Y_i^{(1)} + Y_i^{(2)} +
Y_i^{(3)}|\;.
}
We impose $2\eps_i' \le 1/n - 1/m$. The term $Y_i^{(2)}$ is then estimated as
\equ{
|Y_i^{(2)}| \le \|\Lambda_1^{-1/m}\tilde
q_{i+1}f\|\bigl\|\Lambda_1^{2\eps_i'-2+1/m}\V1'(q_i)f\bigr\| \le C(\|K f\| +
\|f\|)^2\,,
}
where the last step uses Proposition~\ref{prop:Degs} and $\V1' \in \CF_{2n-1}$.
In order to estimate the term $Y_i^{(3)}$, we notice that by the Cauchy-Schwarz
inequality and assumption {\bf A2}, we have
\equs{
	|Y_i^{(3)}| &\le C\|\Lambda_1^{-1/m}\tilde q_{i+1} f\|
\bigl\|\Lambda_1^{2\eps_i'-2+1/m}P^{2m-1}(\tilde q_i) f\bigr\|\\
	&\le C\|f\|\bigl\|\Lambda_1^{1/m-1}P^{m-1}(\tilde
q_i)\Lambda_1^{2\eps_i'-1}P^{m}(\tilde q_i) f\bigr\|\\
&\le C\|f\|\bigl\|\Lambda_1^{2\eps_i'-1} P^{m}(\tilde q_i) f\bigr\|
}
We can choose $2\eps_i' < \eps_{i-1}'$, so this term can be estimated by the
induction hypothesis. The term $Y_i^{(1)}$ is once again estimated by using
Lemma~\ref{lem:estG}, this time with $A = \tilde q_{i+1}$ and $B=p_i$. Using
Proposition~\ref{prop:Degs}, it is easy to verify that one can take
\expos[.]{1/m}{1/m}{1}{1-\eps_i}{1}{2-1/m}
It suffices then to choose $2\eps_i' < \eps_i$ to satisfy the assumptions of
Lemma~\ref{lem:estG} and get the desired estimate.

It is obvious that this induction also works in the other direction, starting
from the other end of the chain. It also accommodates to a little bit more
complicated topologies, as long as the chain does not contain any closed loop.
In order to complete the proof of the lemma, we have to estimate the last term
corresponding to the motion of the center of mass.

%
%

\proclaim{The term $\boldsymbol{\|\Lambda_1^{\eps-1}P^n(Q) f\|}$.} Finally, we
want to show the estimate
\equ[e:estQ]{
\|\Lambda_1^{\eps - 1}P^n(Q) f\| \le C(\|K f\| + \|f\|),
}
for some $\eps$. We start with a little computation. We write
\equ{
	(N+1) q_0 = Q + (q_{N-1}-q_N) + 2(q_{N-2}-q_{N-1}) + \ldots + N(q_0 - q_1)\;.
}
Moreover, we have $q_i = q_0 + (q_1-q_0) + \ldots + (q_i - q_{i-1})$. We can
thus write
\equ{
	\frac{Q}{N+1} - q_i =  \sum_{j=1}^N b_{i j} \tilde q_j\;,\quad\text{with}\quad b_{i j} \in
\R\;.
}
This, together with the mean-value theorem, implies the useful relation
\equs[e:relSec]{
	(N+1)Q\V1'\bigl(Q/(N+1)\bigr) &= Q\sum_{i=0}^N \V1'(q_i) + Q\sum_{i=0}^N
\Bigl(\V1'\bigl(Q/(N+1)\bigr) - \V1'(q_i)\Bigr) \\
	&= Q\sum_{i=0}^N \V1'(q_i) + Q\sum_{i=0}^N \V1''(\xi_i) \sum_{j=1}^N b_{i j}
\tilde q_j \;,
}
where $\xi_i$ is located somewhere on the  $Q/(N+1)$ and $q_i$.

In the case of $d$-dimensional particles, the expression corresponding to
\eref{e:relSec} is
\equs{
	|(N+1)Q\V1'\bigl(Q/(N+1)\bigr)| \le&\; (N+1)|Q||\nabla\V1(q_i)| \\
	&+ |Q|\sum_{i=0}^N \sup_{t \in (0,1)} \bigl|\nabla^2\V1\bigl(tQ/(N+1) +
(1-t)q_i\bigr)\bigr| \sum_{j=1}^N b_{i j} |\tilde q_j| \;.
}
The subsequent expressions can be rewritten accordingly.

We  use {\bf A1} and \eref{e:relSec} to write the left-hand side of
\eref{e:estQ} as
\equs{
	\|\Lambda_1^{\eps-1} P^n(Q)f\|^2 &= |\scal{\Lambda_1^{2\eps-2}P^{2n}(Q) f,f}|
\le C(N+1)\bigl|\scal[b]{\Lambda_1^{2\eps-2}\V1'\bigl(Q/(N+1)\bigr)f, Q
f}\bigr| + C\|f\|^2 \\
	& \le C\Bigl|\Bigl\langle\Lambda_1^{2\eps-2}\Bigl(\sum_{i=0}^N \V1'(q_i)\Bigr)
f, Q f \Bigr\rangle\Bigr| + C \sum_{i,j=1}^N b_{i
j}|\scal{\Lambda_1^{2\eps-2}\tilde q_j \V1''(\xi_i) f, Q f}| + C\|f\|^2\\
	&\equiv Y^{(1)} + Y^{(2)} + C\|f\|^2\;.
}
The term $Y^{(2)}$ can be bounded because $\V1'' \in \CF_{2n-2}$, and so
\equ{
|\V1''(\xi_i)| \le C(1+\xi_i^2)^{n-1} \le C P^{2n-2}(Q) + C P^{2n-2}(q_i) \le
C\sum_{k=0}^N P^{2n-2}(q_k) \;.
}
Thus, $Y^{(2)}$ can be split in terms of the form
\equ{
|\scal{\Lambda_1^{2\eps-2}\tilde q_j P^{2n-2}(q_k) f, Q f}| \le
\|\Lambda_1^{1/n-2} P^{2n-2}(q_k) Q f\| \|\Lambda^{2\eps-1/n} \tilde q_j f\|\;.
}
The first factor clearly can be bounded by $C\|f\|$ if we notice that $q
\mapsto P^{2n-2}(q_k) Q$ belongs to $\CF_{2n-1}$ and then apply Proposition
\ref{prop:Degs}. The second factor can also be bounded by $C\|f\|$ if we impose
\equ{
	0 < \eps \le \frac{1}{2n} - \frac{1}{2m}\;,
}
which can be done because we assumed $n < m$.

It remains to estimate $Y^{(1)}$. We define $P = \sum_{i=0}^N p_i$. Since it
may easily be verified that $[X_0, P] = \sum_{i=0}^N \V1'(q_i) - \rL - \rR$, we
can write $Y_1$ as
\equ{
Y_1 = \scal[b]{\Lambda_1^{2\eps-2} \bigl([X_0, P] + \rL + \rR\bigr)f, Q f}
\equiv Y^{(3)} + Y^{(4)} + Y^{(5)}\;.
}
We leave to the reader the verification that the terms $Y^{(4)}$ and $Y^{(5)}$
can be bounded by $C\|f\|^2$ without introducing any stronger condition on
$\eps$. The term $Y^{(3)}$ can be estimated by using Lemma~\ref{lem:estG} with
$A = Q$ and $B = P$. We have already verified that \eref{e:estpi} holds for
every $i$, so we can define
\equ{
	\eps_P \equiv \min\{\eps_i \;|\; i=0,\ldots,N\}\;.
}
This, together with Proposition \ref{prop:Degs}, allows us to choose,
\expos[,]{1/n}{1/n}{1}{1-\eps_P}{1}{2-1/n}
and thus \eref{e:estQ} is fulfilled if we choose $2\eps \le \eps_P$.
This completes the proof of the lemma.\endproof

\mysection{Generalization of H\"ormander's theorem}
\label{sec:Hormander}

\def\CoK{C_{\!{\cal K}}}

In a celebrated paper \cite{H1}, H\"ormander studied second-order differential
operators of the form
\equ[e:defP]{
	P = \sum_{j=1}^r L_j^* L_j + L_0\;,
}
where the $L_j$ are some smooth vector fields acting in $\R^d$. He showed that
a sufficient condition for the operator $P$ to be hypoelliptic is that the Lie
algebra generated by $\{L_0,\ldots,L_r\}$ has maximal rank
everywhere. 
The main
step in his proof
is to show that there exists a
constant $\eps>0$ and, for every compact domain $\CK \subset \R^d$, a constant
$\CoK$ such that
\equ[e:resHor]{
	\|u\|_{(\eps)} \le \CoK(\|P u\| + \|u\|)\;,\quad\forall\;u\in\cOinf[\CK]\;.
}
In this expression, the norm $\|\cdot\|_{(\eps)}$ is the natural norm
associated to the Sobolev space $H^{\eps}(\R^d)$, \ie
\equ{
\|u\|_{(\eps)}^2 = \int_{\R^d} |\hat u(k)|^2 (1+k^2)^{\eps}\, d^d \!k\; \equiv
\|(1 + \Delta)^{\eps/2} u\|.
}
We base our discussion on the proof presented in \cite{Ho}. H\"ormander first
defines $Q_1$ as the set of all properly supported symmetric first-order
differential operators $q$ such that for every compact domain $\CK$, there
exist constants $\CoK'$ and $\CoK''$ with
\equ[e:constQ1]{
	\|q u\|^2 \le \CoK' \Re \scal{Pu,u} + \CoK'' \|u\|^2\;,\quad u\in
\cOinf[\CK]\;.
}
In particular, if we write $L_j^* = -L_j + c_j$, where $c_j$ is some function,
$Q_1$ contains all the operators of the form
\equ{
(L_j - c_j/2)/i\;,\quad j \ge 1\;,
}
as well as their linear combinations. It also contains every operator of order
$0$. H\"ormander then defines $Q_2$ as consisting of the operator $(P-P^*)/i$,
as well as all the commutators of the form $[q,q']/i$ with $q,q' \in Q_1$. For
$k>2$, he defines $Q_k$ as the set of all commutators $[q,q']/i$ with $q\in
Q_{k-1}$ and $q' \in Q_{k-2}$. One feature of this construction is that a
finite number of steps suffices to catch every symmetric first-order
differential operator. This is a consequence of the maximal rank hypothesis.
The main point of H\"ormander's proof is then the following result.
\begin{lemma}[H\"ormander]
If $q_k\in Q_k$ and $\eps \le 2^{1-k}$, we have for every $\CK \subset \R^d$
\equ[e:estHor]{
	\|q_k u\|_{(\eps-1)} \le C(\|Pu\| +\|u\|)\;,\qquad u \in \cOinf[\CK]\;.
}
\end{lemma}
The proof can be found in \cite[p.~355]{Ho}. The result \eref{e:resHor} then
follows almost immediately, because the operators $i\d_j$ all belong to some
$Q_k$. Thus there exists some $\eps>0$ such that
\equ{
	\sum_{j=1}^d \|\d_j u\|^2_{(\eps-1)} \le \CoK(\|Pu\|
+\|u\|)^2\;,\qquad u \in 
\cOinf[\CK]\;,
}
which implies \eref{e:resHor}. One of the major problems encountered in this
paper is to find a \emph{global} estimate analogous to
\eref{e:resHor}, \ie to find constants $C$ and $\eps$
such 
that
\equ{
	\|\tilde \Delta^\eps u \| \le C(\|P u\| +
\|u\|)\;,\quad\text{for all}\quad u\in \cOinf(\R^d)\;,
}
where $\tilde \Delta$ is some modified Laplacean. There are two major
difficulties:
\begin{list}{$\bullet$}{\setlength{\leftmargin}{5mm}}

\item If we were to construct the sets $Q_k$ as in \cite{Ho}, they would not
necessarily ``close'' in the sense that the successive commutators could blow
up, and the whole proof would break down. To avoid this we do not necessarily
put $(P-P^*)/i$ into $Q_2$, but rather $g_0(P-P^*)/i$, where $g_0$ is some
bounded function. This allows to get decreasing bounds on the successive
commutators.

This problem does \emph{not} appear in \cite{EPR}, where the successive
commutators are all first-order differential operators with constant (or
bounded) coefficients. On the other hand, the commutator technique is
essentially the same as in \cite{EPR}.

\item The above construction does not allow to deal with arbitrary symmetric
first-order differential operators. The reason is that if we want a global
equivalent of \eref{e:constQ1}, the set $Q_1$ is no longer allowed to contain
products of the $L_j$ and unbounded functions. We thus work with fewer
operators, which means that we track much more closely
the expressions which appear
in the constructions.
\end{list}

\subsection{General setting}

Let us consider the Hilbert space $\H = \Ltwo(\R^d, dx)$ for some integer $d
\ge 1$. We define the set $\fC(\H)$ as the set of closed operators on $\H$ and
the algebra $\B(\H)$ as the everywhere defined bounded operators on $\H$.

We define $\CD \equiv \cOinf[\R^d]$, which is dense in $\CH$. Let us fix some
sub-algebra $\F \subset \B(\H)$ that is closed under conjugation and such that
$F\CD \subset \CD$ for all $F \in \F$ (typically $\F$ is some algebra of
bounded functions). The advantage of considering $\cOinf[\R^d]$ is that
\emph{every} differential operator with sufficiently smooth coefficients is
closable on it (see \cite{Yo} for a justification). Moreover, every
differential operator with smooth coefficients maps $\CD$ into itself. This
allows us to make a formal calculus, \ie every relationship between operators
appearing in this section is supposed to hold on $\CD$. The actual operators
are then the closures of the operators defined on $\CD$.

We define ${\mathfrak L}$ as the set of all formal expressions of the form
\equ{
	\sum_{|\ell| \le k} a_\ell(x) D^\ell\;,\qquad k \ge 0\;,\quad a \in
\CC^\infty(\R^d)\;,
}
where $D^\ell$ denotes the $|\ell|^{\text{th}}$ derivative with respect to the
multi-index $\ell$. By the above remark, any element of ${\mathfrak L}$ can
naturally be identified with a differential operator in $\fC(\CH)$.

Consider a differential operator $K$ that can be written as
\equ[e:defK]{
	K = \sum_{i=1}^n X_i^* X_i^{} + X_0^{}\;, \qquad X_j \in {\mathfrak
L}\;,\qquad j=1,\ldots,n\;,
}
where $X_0$ is such that
\equ[e:propX0]{
	X_0^* = -X_0 + \g\;, \qquad \g \in \F\;.
}
We introduce now a definition that will be very useful in the sequel.

\begin{definition}
Let $\CS \subset {\mathfrak L}$ be a finite set of differential operators and
$i\ge 0$ a natural number. We define the set $\CY_{\F}^i(\CS)$ as the
module on $\F$ generated by the terms
\equ{
	S_1 S_2 \cdots S_i\;, \qquad S_k \in \CS \cup \{1\}\;, \qquad k=1,\ldots,i\;.
}
The elements of $\CY_{\F}^i(\CS)$ are naturally identified with densely defined
closed operators on $\CH$. If $i=0$, we use the convention $\CY_\F^0(\CS)
\equiv \F$.
\end{definition}

The subscript $\F$ will be dropped in the sequel when the algebra $\F$ is clear
from the context. We construct the sets
\equ[e:defA0]{
\A_{-1} = \{X_1,\ldots,X_n\}\;,\quad \A_0 = \{g_0X_0,X_1,\ldots,X_n\}\;, \qquad
g_0 \in \F\;,
}
where the operator $g_0$ is assumed to be self-adjoint, positive and such that
\equ[e:propg0]{
	[g_0, X_0] \in \F\;.
}
Let us now construct recursively up to a level $R < \infty$ some finite sets
$\CB_i, \A_i \subset {\mathfrak L}$ by the following procedure. Assume
$\A_{i-1}$ is known. 
Consider next the set $\CB_i^{(0)}$ of all $A$ of the form
\equ[e:defA]{
	A = \sum_{B \in \A_{i-1}} \Bigl(f_B B + \sum_{X\in \A_0} f_{XB}[X,
B]\Bigr)\;, \qquad f_B,\, f_{XB} \in \F\;.
}
We then select a finite subset $\CB_i\subset \CB_i^{(0)}$.
The set $\A_i$ is then defined as
\equ{
	\A_i \equiv \A_{i-1} \cup \CB_i\;.
}
\begin{remark}It is here that our construction differs from similar
ones where \emph{all} elements of $\CB_i^{(0)}$ would have been
selected. This makes the set of operators which we study much smaller,
but then we of course have to verify that the operators of interest
are really covered by our construction.
\end{remark}
We will make some working hypotheses on the sets $\A_i$.
\begin{hypo}{1}
	The pair $(\A_R , \F)$ satisfies the following. If $A,B \in \A_R$ and $f \in
\F$, then
\equ{ [A,B] \in \CY^1(\A_R)\;,\quad A^* \in \CY^1(\A_R)\;,\quad [A,f] \in \F\;.
	}
\end{hypo}
\begin{hypo}{2}
If $A \in \A_i$ with $i \ge -1$, we have $A^* \in \CY^1(\A_i)$.
\end{hypo}
\begin{remark}
Hypothesis \hyp{1} implies that if $X \in \CY^j(\A_R)$ and $Y \in \CY^k(\A_R)$,
then $[X,Y] \in \CY^{k+j-1}(\A_R)$. This will be very useful in the sequel.
Hypothesis \hyp{2} implies that the classes $\CY^k(\A_i)$ are closed under
conjugation.
\end{remark}

We define now the operator $\Lambda^2$ by
\equ[e:defLambda]{
	\Lambda^2 = 1 + \sum_{A\in\A_R} A^*A\;.
}
This is, in some sense that will immediately be clear from
Lemma~\ref{lem:power}, the ``biggest'' operator contained in $\CY^2(\A_R)$. The
operator $\Lambda^2$ is symmetric, densely defined and positive. We will
moreover assume that
\begin{hypo}{3}
$\Lambda^2$ is essentially self-adjoint on $\CD$.
\end{hypo}
The powers $\Lambda^\alpha$ thus exist and are also essentially self-adjoint on
$\CD$ for $\alpha \le 2$.

\subsection{Results and a preliminary lemma}

The following theorem is the main result of this section.
\begin{theorem}
\label{theo:princ}
Let $K$ and $\Lambda$ be defined as above and assume \hyp{1}--\hyp{3} are
satisfied for some $R$. Then there exist some constants $C, \eps > 0$ such that
for every $f \in \CD$, we have
\equ[e:estPrinc]{
	\|\Lambda^\eps f\| \le C(\|K f\| + \|f\|)\;.
}
\end{theorem}

In the sequel, we will write $\A$ instead of $\A_R$ to simplify the notation.

In order to prove Theorem~\ref{theo:princ}, we need the following lemma, which
will be extensively used in the sequel.

\begin{lemma}
\label{lem:power}
Let $\Lambda$, $\F$ and $\A$ be as above and assume \hyp{1} and \hyp{3} hold.
If $X \in \CY_\F^j(\A)$, then the operators
\equ{
	\Lambda^\beta X \Lambda^\gamma \quad\text{with}\quad \beta+\gamma \le -j
}
are bounded.

If $Y \in {\mathfrak L}$ is such that $[Y,\Lambda^2] \in \CY_\F^j(\A)$, then
the operators
\equ{
	\Lambda^\beta[\Lambda^\alpha,Y]\Lambda^\gamma \quad\text{with}\quad \alpha +
\beta + \gamma \le 2-j
}
are bounded.

If $X,Y \in {\mathfrak L}$ are such that
\equ{
[X,\Lambda^2] \in \CY_\F^j(\A)\;\;,\quad [Y,\Lambda^2] \in
\CY_\F^k(\A)\quad\text{and}\quad\bigl[[\Lambda^2,X],Y\bigr] \in
\CY_\F^{j+k-2}(\A)\;,
}
then the operators
\equ{
	\Lambda^\beta\bigl[[\Lambda^\alpha,X],Y\bigr]\Lambda^\gamma
\quad\text{with}\quad \alpha + \beta + \gamma \le 4-j-k
}
are bounded.
\end{lemma}

\begin{proof}
The proof of this lemma is postponed to Appendix \ref{App:main}.
\end{proof}

\begin{remark}
Lemma~\ref{lem:power} allows us to count powers in the following sense. Each
time we see an operator that is a monomial containing fractional powers of
$\Lambda$ and some operators of $\CY^j(\A)$, we know that the operator is
bounded if its ``degree'' is less or equal to $0$. The rule is that if $Y \in
\CY^j(\A)$, its degree is $j$ and the degree of $\Lambda^\alpha$ is $\alpha$.
Moreover, every time we encounter a commutator, we can lower the degree by one unit.
\end{remark}

Lemma~\ref{lem:power} also shows that if $f\in \CD$, $A \in\CA$ and $\alpha \le
2$, expressions such as $A\Lambda^\alpha f$ can be well defined by
\equ{
	A\Lambda^\alpha f \equiv \Lambda^\alpha A f +
[A,\Lambda^\alpha]\Lambda^{-2}\Lambda^2 f\;,
}
where $[A,\Lambda^\alpha]\Lambda^{-2}$ is bounded and can therefore be defined
on all of $\H$. Similar expressions hold to show that any expression of this
section can be well defined.

We are now ready to prove the theorem.

\subsection{Proof of Theorem~\ref{theo:princ}}

The proof uses the commutation techniques developed by H\"ormander \cite{Ho}
and improved by Eckmann, Pillet, Rey-Bellet \cite{EPR}. Large parts of this
proof are inspired from this latter work.

Before we start the proof itself, let us make a few computations, the results
of which will be used repeatedly in the sequel. We first show that we can
assume $\Re K$ positive. An explicit computation, using \eref{e:defK} and
\eref{e:propX0}, shows that
\equ[e:ReK]{
	\Re K = \sum_{i=1}^n X_i^* X_i^{} + \frac{\g}{2}\;,\quad\text{and thus
also}\quad X_0 = K - \Re K + g/2\;.
}
Because $\g \in \F$, we can add a sufficiently big constant to $X_0$ to make
$\Re K$ positive. This will change neither the commutation relations, nor the
estimate \eref{e:estPrinc}.

Another useful equality is
\equ[e:g0ReK]{
	g_0\Re K = \Re(g_0K + K_1) + K_2\qquad K_1, K_2 \in \CY^1(\A_{-1})\;,
}
where $K_1$ is a self-adjoint operator such that $\Re(g_0K + K_1)$ is a
positive self-adjoint operator.
This is a consequence of the following two equalities, which are easily
verified by inspection
\equs{
	g_0 \Re K &= \sum_{i=1}^n X_i^* g_0 X_i^{} + K_2 \quad K_2 \in
\CY^1(\A_{-1})\;, \\
	\Re (g_0 K) &= \sum_{i=1}^n X_i^* g_0 X_i^{} - K_1 \quad K_1 \in
\CY^1(\A_{-1})\;.
}
We therefore have
\equ{
\Re(g_0K + K_1) = \sum_{i=1}^n X_i^* g_0 X_i^{}\;.
}
This proves \eref{e:g0ReK}.

Another useful identity will be
\equs[e:explX0]{
	(g_0 X_0)^* &= -X_0 g_0 + \g g_0 = -g_0 X_0 + [g_0, X_0] + \g g_0 \\
	&= -g_0 X_0 + g_0'\;, \qquad g_0' \in \F\;,
}
where the last equality is a consequence of \eref{e:propg0}.

We will now verify the estimate \eref{e:estPrinc} for some vector $f\in \CD$.
In the sequel, the symbol $C$ will be used to denote some constant depending
only on the operator $K$. This constant can change from one line to the other.
We will first prove that $A \in \CY^1(\A_i)$ with $0 \le i \le R$ implies
\equ[e:est]{
	\|\Lambda^{1/4^{i+1} - 1}A f\| \le C(\|K f\| + \|f\|)\;.
}
In fact, an immediate consequence of the first part of Lemma~\ref{lem:power} is
that we only have to prove this assertion for $A \in \A_i$. The proof will
proceed by induction on $i$.

\subsubsection{Verification for $\boldsymbol{i=0}$}

We want to verify the estimate
\equ{
\|\Lambda^{-3/4}A f\| \le C(\|K f\|+\|f\|)\;,\quad\text{for all}\quad A \in \A_0\;.
}
The cases $A=g_0 X_0$ and $A = X_j$ with $j \neq 0$ will be treated separately.

%
%

\proclaim{The case $\boldsymbol{A = X_j}$.} We write
\equs{
	\|\Lambda^{-3/4}X_jf\|^2 \le C\|X_j f \|^2 &\le C\scal{f,X_j^* X_j^{}f} \le C
\scal{f,(K + K^* - \g)f} \\
	&\le C\Re\scal{f,K f} + C\|f\|^2 \le C\|f\|(\|K f\| + \|f\|)\;.
}
This implies the desired estimate. Because $X_j^* \in \CY^1(\A_{-1})$ by
hypothesis, this computation immediately implies the estimates
\sublabels
\equs[1,e:estX]{
 \|X_j f \| &\le C(\|K f\| + \|f\|)\;, \sublabel{e:myestX} \\
 \|X_j^* f \| &\le C(\|K f\| + \|f\|)\;,\sublabel{e:estXstar}
}
which hold for every $j \ge 1$.

%
%

\proclaim{The case $\boldsymbol{A = g_0X_0}$.} We write, using expression
\eref{e:ReK},
\equs{
	\|\Lambda^{-3/4}A f\|^2 &= \scal{g_0X_0f, \Lambda^{-3/2}A f} \\
	&= \scal{K f, g_0\Lambda^{-3/2}A f} + \scal{g_0\g f, \Lambda^{-3/2}A f}/2 -
\scal{(\Re K)f,g_0 \Lambda^{-3/2}A f} \\
	&\equiv S_1 + S_2 - S_3\;.
}
The terms $S_1$ and $S_2$ are easily bounded by $C(\|K f\| + \|f\|)^2$,  using
the Cauchy-Schwarz inequality and the first part of Lemma~\ref{lem:power}.
Using the positivity of $\Re K$ and the explicit form of $K$, the term $S_3$
can be bounded as
\equs{
	|S_3| &= \scal{(\Re K)^{1/2}f,(\Re K)^{1/2}g_0 \Lambda^{-3/2}A f} \\[2mm]
	&\le |\Re\scal{K f,f}|^{1/2}|\scal{(\Re K)g_0 \Lambda^{-3/2}A f,g_0
\Lambda^{-3/2}A f}|^{1/2} \\
	&\le \sqrt{\|K f\|\|f\|}\,\Bigl|\scal{gg_0 \Lambda^{-3/2}A
f,g_0\Lambda^{-3/2}A f}/2+\sum_{i=1}^n\|X_ig_0 \Lambda^{-3/2}A
f\|^2\Bigr|^{1/2} \\
	&\equiv \sqrt{\|K f\|\|f\|}\,\sqrt{S_0 + \sum_{i=1}^n S_{0,i}^2}\;.
}
The term $S_0$ is estimated by simple power counting (the $\Lambda$'s contribute
for $-3$ and the $A$'s for $2$ in the total degree of the expression, hence
$|S_0| \le C\|f\|^2$). The terms $S_{0,i}$ are estimated by writing
\equ{
	|S_{0,i}| \le \|g_0 \Lambda^{-3/2}AX_if\| + \|[X_i,g_0 \Lambda^{-3/2}A]f\|\;.
}
The first term is estimated by using \eref{e:estX} and power counting. The
second term is estimated by expanding the commutator as
\equ{
 [X_i,g_0 \Lambda^{-3/2}A] = [X_i,g_0] \Lambda^{-3/2}A +
g_0[X_i,\Lambda^{-3/2}]A + g_0 \Lambda^{-3/2}[X_i,A]\;,
}
and estimating separately the resulting terms.

\subsubsection{The induction hypothesis}

We shall proceed by induction. Let us fix $j>0$, take $A \in \A_j$ and assume
\eref{e:est} holds for $i<j$. Let us moreover define $\eps \equiv 1/4^{j+1}$ in
order to simplify the notation. Our assumption is therefore that
\equ[e:indHyp]{
\|\Lambda^{4\eps-1}B f\| \le C(\|K f\| + \|f\|)\qquad \forall\; B\in
\CY^1(\A_{j-1})\;.
}
We will now prove that this assumption implies the desired estimate, \ie
\equ[e:mainEst]{
\|\Lambda^{\eps-1}A f\| \le C(\|K f\| + \|f\|)\qquad \forall\; A\in
\CY^1(\A_{j})\;.
}
This, together with the preceding paragraph, will imply the estimate
\eref{e:est}.

\subsubsection{Proof of the main estimate}

Because of the induction hypothesis, we only have to check \eref{e:mainEst} for
$A \in \A_j \backslash \A_{j-1}$. By \eref{e:defA}, we can write
\equ{
	A = \sum_{B \in \A_{j-1}} \Bigl(f_B B + f^0_B [g_0X_0,B] + \sum_{i=1}^n f^i_B
[X_i,B]\Bigr)\;,
}
with all the $f$ belonging to $\F$. 
We have
\equs{
	\|\Lambda^{\eps - 1}A f\|^2 &= \sum_{B \in \A_{j-1}} \scal[B]{\Bigl(f_B B +
f^0_B [g_0X_0,B] + \sum_{i=1}^n f^i_B [X_i,B]\Bigr)f, \Lambda^{2\eps-2} A f} \\
	 &\equiv \sum_{B \in \A_{j-1}} \Bigl(T_{B} + T^0_B + \sum_{i=1}^n
T_{B}^i\Bigr)\;.
}
We are going to bound each term of this sum separately by $C(\|K f\| +
\|f\|)^2$.

%
%

\proclaim{Term $\boldsymbol{T_{B}}$.} We have
\equ{
	|T_{B}| = |\scal{\Lambda^{2\eps-1} f_B \Lambda^{1-2\eps}\Lambda^{2\eps-1}B f,
\Lambda^{-1}A f}|\;.
}
The operators $\Lambda^{2\eps-1} f_B \Lambda^{1-2\eps}$ and $\Lambda^{-1}A$ are
bounded by Lemma~\ref{lem:power}. Using the induction hypothesis
\eref{e:indHyp}, we thus get the bound $|T_{B}| \le C(\|K f\|+\|f\|)^2$.

%
%

\proclaim{Term $\boldsymbol{T_B^i}$ with $\boldsymbol{i \neq 0}$.} We define $h
\equiv f^i_B$. The term  $T_B^i$ is then written as
\equ{
	T_B^i = \scal{B f, X_i^*h^*\Lambda^{2\eps-2}A f} - \scal{X_if, h^* B^*
\Lambda^{2\eps-2}A f} \equiv Q_1 - Q_2\;.
}

%
%

\proclaim{Term $\boldsymbol{Q_1}$.} It can be estimated by writing
\equ{
	Q_1 = \scal{B f, h^*\Lambda^{2\eps-2}AX_i^*f} + \scal{B f,
[X_i^*,h^*\Lambda^{2\eps-2}A]f}\;.
}
The first term is estimated by rewriting it as
\equs{
|\scal{B f, h^*\Lambda^{2\eps-2}AX_i^*f}| &= |\scal{\Lambda^{2\eps-1}B f,
\Lambda^{1-2\eps}h^*\Lambda^{2\eps-2}AX_i^*f}| \\
&\le \|\Lambda^{2\eps-1}B f\|\|\Lambda^{1-2\eps}h^*\Lambda^{2\eps-2}AX_i^*f\|
\le C(\|K f\|+\|f\|)^2\;.
}
The last inequality has been obtained by using the induction hypothesis
\eref{e:indHyp}, the estimate \eref{e:estXstar} and the fact that the operator
$\Lambda^{1-2\eps}h^*\Lambda^{2\eps-2}A$ is bounded by Lemma~\ref{lem:power}.

The second term is estimated as
\equs{
|\scal{B f, [X_i^*,h^*\Lambda^{2\eps-2}A]f}| &= |\scal{\Lambda^{2\eps-1}B
f,\Lambda^{1-2\eps} [X_i^*,h^*\Lambda^{2\eps-2}A]f}| \\
&\le \|\Lambda^{2\eps-1}B f\| \|\Lambda^{1-2\eps}
[X_i^*,h^*\Lambda^{2\eps-2}A]f\|\;.
}
The term $\|\Lambda^{2\eps-1}B f\|$ is bounded by the induction hypothesis
\eref{e:indHyp}. The other term can be estimated by writing the commutator as
\equ{
[X_i^*,h^*\Lambda^{2\eps-2}A] = [X_i^*,h^*]\Lambda^{2\eps-2}A +
h^*[X_i^*,\Lambda^{2\eps-2}]A + h^*\Lambda^{2\eps-2}[X_i^*,A]\;.
}
The resulting terms are estimated by power counting, using the fact that $X_i^*
\in \CY^1(\A)$.

%
%

\proclaim{Term $\boldsymbol{Q_2}$.} We bound this term as
\equs{
|Q_2| &= \bigl| \scal{X_i f, h^* \Lambda^{2\eps-2}AB^*f} + \scal{X_if,
h^*[B^*,\Lambda^{2\eps-2}A]f}\bigr| \\[1.5mm]
	&\le \|X_i f\|\bigl(\|h^* \Lambda^{2\eps-2}AB^*f\| + \|
h^*[B^*,\Lambda^{2\eps-2}A]f\|\bigr) \\[1.5mm]
	&\le \|X_i f\|\bigl(\|h^*
\Lambda^{2\eps-2}A\Lambda^{1-2\eps}\|\|\Lambda^{2\eps-1}B^*f\| + \|
h^*[B^*,\Lambda^{2\eps-2}A]f\|\bigr)\;.
}
We leave to the reader the not too hard task to verify that it is indeed
possible to get the bound $|Q_2| \le C(\|K f\| + \|f\|)^2$ by similar estimates
as for the term $Q_1$.

%
%

\proclaim{Term $\boldsymbol{T_{B}^0}$.} We define $h \equiv f^0_B$. The term
$T_B^0$ is thus equal to
\equ{
	T_B^0 = \scal{[g_0 X_0,B]f,h^*\Lambda^{2\eps-2}A f} = \scal{g_0 X_0 B
f,h^*\Lambda^{2\eps-2}A f} - \scal{B g_0 X_0 f,h^*\Lambda^{2\eps-2}A f}\;.
}
We use \eref{e:explX0} to write this as
\equs{
T_B^0 =&\, -\scal{B f, h^*\Lambda^{2\eps-2}A g_0 X_0 f} + \scal{B f,
g_0'h^*\Lambda^{2\eps-2}A f} \\[1mm]
	&\, -\scal{B f, [g_0 X_0, h^*\Lambda^{2\eps-2}A]f} - \scal{B g_0
X_0,h^*\Lambda^{2\eps-2}A f} \\[1mm]
	\equiv&\, - U_1 + U_2 - U_3 - U_4\;,
}
where $g_0' \in \F$.
The term $U_2$ can easily be estimated by
\equs{
 |U_2| &= |\scal{\Lambda^{2\eps-1}B f,
\Lambda^{1-2\eps}g_0'h^*\Lambda^{2\eps-2}A f}| \le \|\Lambda^{2\eps-1}B
f\|\|\Lambda^{1-2\eps}g_0'h^*\Lambda^{2\eps-2}A f\| \\[1mm]
	&\le C(\|K f\| + \|f\|)\|f\|\;,
}
using the induction hypothesis. In order to estimate the term $U_3$, we notice
that $g_0 X_0 \in \A$, and thus $[g_0X_0,\Lambda^2] \in \CY^2(\A)$. We can
therefore write
\equ{
 |U_3| = |\scal{\Lambda^{2\eps-1}B f, \Lambda^{1-2\eps}[g_0 X_0,
h^*\Lambda^{2\eps-2}A]f}| \le \|\Lambda^{2\eps-1}B f\|\|\Lambda^{1-2\eps}[g_0
X_0, h^*\Lambda^{2\eps-2}A]f\|\;,
}
expand the commutator and estimate the resulting terms separately by
power counting. We use the equality
\equ{
	X_0  = K - \Re K + \g/2\;,
}
to write the terms $U_1$ and $U_4$ as
\equs{
	U_1 &= \bigl\langle B f, h^*\Lambda^{2\eps-2}A g_0 \bigl(K - (\Re K) +
\g/2\bigr) f\bigr\rangle \equiv T_{B,1} - T_{B,2} + T_{B,3}\;,\\
	U_4 &= \bigl\langle B g_0 \bigl(K - (\Re K) + \g/2\bigr)f,
h^*\Lambda^{2\eps-2}A f\bigr\rangle \equiv T_{B,4} - T_{B,5} + T_{B,6}\;.
}
Each of these terms will now be estimated separately.

%
%

\proclaim{Terms $\boldsymbol{T_{B,3}}$ and $\boldsymbol{T_{B,6}}$.} They are
easily bounded like the term $U_2$ by power counting and using the induction
hypothesis to bound $\|\Lambda^{2\eps-1}B f\|$. In the case of $T_{B,6}$, we
first have to commute $B$ with $g_0 g/2$, but this does not cause any problem.

%
%

\proclaim{Term $\boldsymbol{T_{B,1}}$.} This term can be estimated by
\equ{
	|T_{B,1}| \le \|K f\|\|g_0^* A^* \Lambda^{2\eps-2}h B f\| \le  \|K f\|\|g_0^*
A^* \Lambda^{2\eps-2}h\Lambda^{2-2\eps}\|\|\Lambda^{2\eps-2} B f\|\;.
}
The norm of $g_0^* A^* \Lambda^{2\eps-2}h\Lambda^{2-2\eps}$ is bounded by
power counting. Using the induction hypothesis \eref{e:indHyp}, we thus have
$|T_{B,1}| \le C(\|K f\| + \|f\|)^2$.

%
%

\proclaim{Term $\boldsymbol{T_{B,4}}$.} We have the estimate
\equ{
	|T_{B,4}| = |\scal{K f,g_0^*B^*h^*\Lambda^{2\eps-2}A f}| \le \|K
f\|\|g_0^*B^*h^*\Lambda^{2\eps-2}A f\|\;.
}
The second norm can be estimated by writing
\equ{
\|g_0^*B^*h^*\Lambda^{2\eps-2}A f\| \le
\|g_0^*h^*\Lambda^{2\eps-2}A\Lambda^{1-2\eps}\|\|\Lambda^{2\eps-1}B^*f\| +
\|g_0^*[B^*,h^*\Lambda^{2\eps-2}A]f\|\;.
}
Here, the first term can be bounded by $C(\|K f\| + \|f\|)$ because, by
\hyp{2}, we have $B^* \in \CY^1(\A_{j-1})$ and so we can use the induction
hypothesis. The commutator can be expanded and bounded by power counting.

%
%

\proclaim{Term $\boldsymbol{T_{B,2}}$.} We can write this term as
\equs{
	T_{B,2} =&\, \scal{\Lambda^{2\eps-1}h B f,(g_0 \Re K)\Lambda^{-1}A f} +
\scal{\Lambda^{2\eps-1}h B f,[\Lambda^{-1}A, g_0 \Re K]f} \\
	=&\, \scal{\Lambda^{2\eps-1}h B f,K_2 \Lambda^{-1}A f} +
\scal{\Lambda^{2\eps-1}h B f,[\Lambda^{-1}A, g_0 \Re K]f} \\
	&\,+ \scal{\Lambda^{2\eps-1}h B f,\Re(g_0 K + K_1)\Lambda^{-1}A f} \\
	\equiv&\, M_1 + M_2 + M_3\;,
}
where the second equality has been obtained using \eref{e:g0ReK}.
These terms can now be estimated separately.

%
%

\proclaim{Term $\boldsymbol{M_1}$.} We write this term as
\equ{
	M_1 = \scal{\Lambda^{2\eps-1}h B f,\Lambda^{-1}A K_2 f} +
\scal{\Lambda^{2\eps-1}h B f,[K_2,\Lambda^{-1}A]f}\;.
}
The first term is estimated by using
\equ{
	K_2 \in \CY^1(\A_{-1}) \quad\Rightarrow\quad \|K_2f\| \le C(\|K f\| +
\|f\|)\;,
}
where the implication is a straightforward consequence of \eref{e:myestX}. The
second term can be estimated by power counting and the induction hypothesis,
using the fact that $K_2 \in \CY^1(\A_{-1})$, so that $[K_2,\Lambda^{-1}A]$ is
bounded.

%
%

\proclaim{Term $\boldsymbol{M_2}$.} We use the explicit form of $\Re K$ to
write this term as
\equs{
	M_2 =&\; \scal{\Lambda^{2\eps-1}h B f,[\Lambda^{-1}A, g_0] (\Re K) f} \\
	& + \sum_{i=1}^n\bigl(\scal{\Lambda^{2\eps-1}h B f,g_0 X_i^* [\Lambda^{-1}A,
X_i]f} + \scal{\Lambda^{2\eps-1}h B f,g_0[\Lambda^{-1}A,X_i^*]X_i^{}f}) \\
	& + \scal{\Lambda^{2\eps-1}h B f,g_0[\Lambda^{-1}A, \g] f}/2\\
	\equiv&\; M_{20} + \sum_{i=1}^n (M_{i1}+ M_{i2}) + M_{21}\;.
}
The term $M_{20}$ is estimated by using the explicit form of $\Re K$ to
decompose it in terms of the form
\equ{
|\scal{\Lambda^{2\eps-1}h B f,[\Lambda^{-1}A, g_0]X_i^*X_i^{} f}| \le
\|[\Lambda^{-1}A, g_0]X_i^*\|\|\Lambda^{2\eps-1}h B f\|\|X_i f\|\;.
}
The norm $\|[\Lambda^{-1}A, g_0]X_i^*\|$ is finite by Lemma~\ref{lem:power}.
The terms $\|\Lambda^{2\eps-1}h B f\|$ and $\|X_i f\|$ are bounded by $C(\|K
f\| + \|f\|)$, using the induction hypothesis \eref{e:indHyp} and the estimate
\eref{e:myestX} respectively.
The terms $M_{21}$ and $M_{i2}$ are estimated by power counting and the
induction hypothesis. In order to estimate the term $M_{i1}$, we have to
commute once more to find
\equ{
M_{i1} = \scal{\Lambda^{2\eps-1}h B f,g_0 [\Lambda^{-1}A, X_i^{}]X_i^*f} +
\bigl\langle{\Lambda^{2\eps-1}h B f,g_0 \bigl[X_i^*,[\Lambda^{-1}A,
X_i^{}]\bigr]f}\bigr\rangle\;.
}
The first term is estimated by using \eref{e:estXstar}. The second term is
estimated by expanding the double commutator and power counting.

%
%

\proclaim{Term $\boldsymbol{M_3}$.} We use the positivity of $\Re (g_0 K +
K_1)$ to write
\equs{
 |M_3| =&\; \scal[b]{\bigl(\Re (g_0K + K_1)\bigr)^{1/2}\Lambda^{2\eps-1}h B
f,\bigl(\Re (g_0K + K_1)\bigr)^{1/2}\Lambda^{-1}A f} \\
	\le&\; |\Re\scal{(g_0K + K_1)\Lambda^{2\eps-1}h B f,\Lambda^{2\eps-1}h B
f}|^{1/2}|\scal{\Re(g_0K + K_1)\Lambda^{-1}A f,\Lambda^{-1}A f}|^{1/2} \\
	\le&\; \sqrt{|\Re M_4| + |\Re M_5|}\sqrt{|M_6|}\;.
}
We will now estimate $M_4$, $M_5$ and $M_6$ separately.

%
%

\proclaim{Term $\boldsymbol{M_4}$.} We want to put the operator $g_0K$ to the
left of $f$. So we write
\equ{
M_4 = \scal{\Lambda^{-1}h Bg_0K f,\Lambda^{4\eps-1}h B f} +
\scal{[g_0K,\Lambda^{2\eps-1}h B]f,\Lambda^{2\eps-1}h B f}\equiv M_{41} +
M_{42}\;.
}
The term $M_{41}$ is estimated easily by using the induction hypothesis and the
fact that $\Lambda^{-1}h Bg_0$ is bounded. In order to estimate $M_{42}$, we
use the explicit form of $K$ to write
\equs{
	M_{42} =&\; \scal{\Lambda^{-2\eps}[g_0X_0,\Lambda^{2\eps-1}h
B]f,\Lambda^{4\eps-1}h B f} \\
	&+ \sum_{i=1}^n \bigl(\scal{g_0X_i^*[X_i^{},\Lambda^{2\eps-1}h
B]f,\Lambda^{2\eps-1}h B f} + \scal{g_0[X_i^*,\Lambda^{2\eps-1}h B]X_i
f,\Lambda^{2\eps-1}h B f}\bigr) \\
	&+ \scal{\Lambda^{-2\eps}[g_0,\Lambda^{2\eps-1}h B]K f,\Lambda^{4\eps-1}h B f}
\\
	\equiv&\; M_{40} + \sum_{i=1}^n (M_{i3} + M_{i4}) + M_{4K}\;.
}
The terms $M_{40}$ and $M_{4K}$ are estimated by expanding the commutator and
power counting. The term $M_{i4}$ can be written as
\equs{
|M_{i4}| &=  |\scal{\Lambda^{-2\eps}g_0[X_i^*,\Lambda^{2\eps-1}h B]X_i
f,\Lambda^{4\eps-1}h B f}| \\
	&\le \|\Lambda^{1-4\eps}h^*\Lambda^{2\eps-1}g_0[X_i^*,\Lambda^{2\eps-1}h
B]\|\|X_i f\|\|\Lambda^{4\eps-1}B f\|\;.
}
It is then estimated by power counting, using moreover the induction hypothesis
and the estimate \eref{e:estX}. In order to estimate the term $M_{i3}$, we have
to commute once more to write
\equ{
M_{i3} = \scal{\Lambda^{-2\eps}g_0[X_i^{},\Lambda^{2\eps-1}h
B]X_i^*f,\Lambda^{4\eps-1}h B f} +
\bigl\langle\Lambda^{-2\eps}g_0\bigl[X_i^*[X_i^{},\Lambda^{2\eps-1}h
B]\bigr]f,\Lambda^{4\eps-1}h B f\bigr\rangle\;.
}
The first term is estimated exactly like $M_{i4}$. The second term can then be
estimated by expanding the double commutator and power counting.

%
%

\proclaim{Term $\boldsymbol{M_5}$.} We write this term as
\equ{
	M_5 = \scal{\Lambda^{-1}h B K_1f, \Lambda^{4\eps-1}h B f} +
\scal{\Lambda^{-2\eps}[K_1,\Lambda^{2\eps-1}h B]f, \Lambda^{4\eps-1}h B f}\;.
}
The first term is estimated using the induction hypothesis and the fact that
\eref{e:g0ReK} and \eref{e:estX} imply
\equ[e:estK1]{
K_1 \in \CY^1(\A_{-1}) \quad\Rightarrow\quad \|K_1 f\| \le C(\|K f\| +
\|f\|)\;.
}
The other term is estimated by using the fact that $[K_1, \Lambda^2] \in
\CY^2(\A)$ and $[K_1, h B] \in \CY^1(\A)$, which follows from $\A_{-1} \subset
\A$ and thus $K_1 \in \CY^1(\A)$.

%
%

\proclaim{Term $\boldsymbol{M_6}$.} We use the explicit expression for $\Re(g_0
K + K_1)$ to write this term as
\equ{
	M_6 = \sum_{i=1}^n\|g_0^{1/2}X_i \Lambda^{-1}A f\|^2 \le C\sum_{i=1}^n\|X_i
\Lambda^{-1}A f\|^2\;.
}
These terms are easily estimated by putting the $X_i$ to the left of $f$, using
\eref{e:estX} and estimating the commutators.

%
%

\proclaim{Term $\boldsymbol{T_{B,5}}$.} This is the last term we have to
estimate. Using the expression \eref{e:g0ReK} and the positivity of $\Re(g_0 K
+ K_1)$, it can be written in the form
\equs{
	T_{B,5} =&\; \bigl\langle\bigl(\Re(g_0 K + K_1)\bigr)^{1/2}f, \bigl(\Re(g_0 K
+ K_1)\bigr)^{1/2}B^*h^*\Lambda^{2\eps-2}A f\bigr\rangle \\
 &+ \scal{K_2f, B^*h^*\Lambda^{2\eps-2}A f} \\
	\equiv&\; N_1 + N_2\;.
}
These terms are now estimated separately.

%
%

\proclaim{Term $\boldsymbol{N_2}$.} We use the Cauchy-Schwarz inequality to
write
\equ{
	|N_2| \le \|K_2 f\|\|B^*h^*\Lambda^{2\eps-2}A f\| \equiv \|K_2 f\|\|N_3\|\;.
}
We can estimate $N_3$ by writing
\equ{
B^*h^*\Lambda^{2\eps-2}A =
h^*\Lambda^{2\eps-2}A\Lambda^{1-2\eps}\Lambda^{2\eps-1} B^* +
[B^*,h^*\Lambda^{2\eps-2}A]\;,
}
and estimating the resulting terms using the induction hypothesis. We already
noticed that we have the desired estimate for $\|K_2 f\|$.

%
%

\proclaim{Term $\boldsymbol{N_1}$.} Using the Cauchy-Schwarz inequality, we
write it as
\equs{
	N_1 & \le \scal{f,\Re (g_0 K + K_1)f}^{1/2}\scal{\Re(g_0 K +
K_1)B^*h^*\Lambda^{2\eps-2}A f,B^*h^* \Lambda^{2\eps-2}A f}^{1/2} \\
& \le C(\|K f\| + \|f\|)|\scal{\Lambda^{-2\eps}(g_0K +
K_1)B^*h^*\Lambda^{2\eps-2}A f,\Lambda^{2\eps}B^*h^* \Lambda^{2\eps-2}A
f}|^{1/2}\\
& \equiv C(\|K f\| + \|f\|)\sqrt{|\scal{f_1+f_2,f_3}|} \le C(\|K f\| +
\|f\|)\sqrt{(\|f_1\|+\|f_2\|)\|f_3\|}\;.
}

%
%

\proclaim{Estimate of $\boldsymbol{\|f_3\|}$.} We write it as
\equ{
	f_3 =
\Lambda^{2\eps}h^*\Lambda^{2\eps-2}A\Lambda^{1-4\eps}\Lambda^{4\eps-1}B^*f +
\Lambda^{2\eps}[B^*, h^*\Lambda^{2\eps-2}A]f \;.
}
The first term is estimated by using the recurrence hypothesis and the fact
that \hyp{2} implies $B^* \in \CY^1(\A_{j-1})$. The second term is estimated by
power counting and by using the fact that $\eps < 1/4$.

%
%

\proclaim{Estimate of $\boldsymbol{\|f_2\|}$.} We write it as
\equ{
	f_2 = \Lambda^{-2\eps} B^*h^*\Lambda^{2\eps-2}AK_1f +
\Lambda^{-2\eps}[K_1,B^*h^*\Lambda^{2\eps-2}A]f\;.
}
The first term is estimated using the fact that $\|K_1f\| \le C(\|K f\| +
\|f\|)$ and power counting. The second term is simply estimated by
power counting, and the fact that $K_1\in \CY^1(\A)$.

%
%

\proclaim{Estimate of $\boldsymbol{\|f_1\|}$.} We use the explicit form of $K$
to write $f_1$ as
\equs{
	f_1 =&\; \Lambda^{-2\eps}B^*h^*\Lambda^{2\eps-2}Ag_0K f +
	\Lambda^{-2\eps}[g_0X_0,B^*h^*\Lambda^{2\eps-2}A]f \\
	&+ \sum_{i=1}^n\bigl(\Lambda^{-2\eps}g_0
X_i^*[X_i^{},B^*h^*\Lambda^{2\eps-2}A]f + \Lambda^{-2\eps}[g_0
X_i^{*},B^*h^*\Lambda^{2\eps-2}A]X_i f\bigr) \\
	\equiv&\; Q_K + Q_0 + \sum_{i=1}^n(Q_{i,1}+Q_{i,2})\;.
}
These terms will now be estimated separately.

%
%

\proclaim{Term $\boldsymbol{Q_K}$.} We notice that the operator
\equ{
\Lambda^{-2\eps}B^*h^*\Lambda^{2\eps-2}Ag_0
}
is bounded by power counting. This yields the desired estimate.

%
%

\proclaim{Term $\boldsymbol{Q_0}$.} This term is bounded by $C\|f\|$ by
power counting, noticing that $g_0 X_0 \in \A$.

%
%

\proclaim{Term $\boldsymbol{Q_{i,2}}$.} This term can be estimated by
power counting if we expand the commutator and use the estimate \eref{e:estX}.

%
%

\proclaim{Term $\boldsymbol{Q_{i,1}}$.} We use once more the trick that
consists of putting the $X_i^*$ to the left of $f$. We write therefore
\equ{
	Q_{i,1} = \Lambda^{-2\eps}g_0[X_i^{},B^*h^*\Lambda^{2\eps-2}A]X_i^*f +
\Lambda^{-2\eps}g_0\bigl[X_i^*[X_i^{},B^*h^*\Lambda^{2\eps-2}A]\bigr]f\;.
}
The first term is estimated by using \eref{e:estXstar} and expanding the
commutator. The second term is estimated in a similar way by expanding the
double commutator. We don't write the resulting terms here, because there are
too much of them. They are all bounded by simple power counting and by using
Lemma~\ref{lem:power}. This completes the proof of estimate \eref{e:mainEst}.

It is now straightforward to prove the theorem. Recall that $R$ is the level up
to which the $\CA_i$ are defined. We put $\eps = 1/4^{R+1}$, and we write:
\equs{
	\|\Lambda^\eps f\| &= \scal{f, \Lambda^{2\eps-2}\Lambda^2f} = \sum_{A \in \A}
\scal{f, \Lambda^{2\eps-2}A^*A f} \\
	&= \sum_{A \in \A} \bigl(\|\Lambda^{\eps-1}A f\|^2 + \scal{f, [A^*,
\Lambda^{2\eps-2}]A f}\bigr)\;.
}
The first term in the sum is bounded by using \eref{e:est}, the second term by
simple power counting. This finally completes the proof of
Theorem~\ref{theo:princ}.\phantom{a}\nobreak\hfill\qed

We next note a consequence of this theorem, namely a simple criterion
to see if a 
quadratic differential operator has compact resolvent. It is an easy
illustration of the technique that will be used in the sequel to show that $K$
has compact resolvent.

\subsection{Quadratic differential operators}

\begin{definition}
An operator $A : \CD(A) \to \CH$ is called \emph{accretive} if it satisfies
\equ{
	\Re \scal{f, A f} \ge 0 \;,\quad\text{for all}\quad f \in \CD(A)\;.
}
An operator $A$ is called \emph{quasi accretive} if there exists $\lambda \in
\R$ such that $A + \lambda$ is accretive. It is called \emph{strictly
accretive} if there exists $\lambda > 0$ such that $A - \lambda$ is still
accretive.
\end{definition}

If $-A$ is accretive, $A$ is called \emph{dissipative}. An operator $A$ is
called \emph{{\it m}-accretive} if it is accretive and if $(A + \lambda)^{-1}$
exists for all $\lambda > 0$ and satisfies $\|(A + \lambda)^{-1}\| \le
\lambda^{-1}$. The expressions {\it m}-dissipative, quasi dissipative, etc.~are
defined similarly in an obvious way. An equivalent characterization of {\it
m}-accretive operators is that they are accretive with no proper accretive
extension.

It is a classical result (see \eg \cite{Da}) that the quasi {\it m}-dissipative
operators are precisely the generators of quasi-bounded semi-groups. An
immediate consequence is that if an operator $A$ is (quasi) {\it m}-accretive
({\it m}-dissipative), its adjoint $A^*$ is also (quasi) {\it m}-accretive
({\it m}-dissipative).

\begin{proposition} \label{prop:Comp}
Let $\CH$ be a Hilbert space and $\CC$ be a dense subset of $\CH$.
Let $K : \CD(K) \to \CH$ be a quasi {\it m}-accretive (or quasi {\it
m}-dissipative) operator and let $\Lambda^2 : \CD(\Lambda^2) \to \CH$ be a
self-adjoint positive operator such that $\CC \subset \CD(\Lambda^2)$. Assume
moreover that $\CC$ is a core for $K$, that $\Lambda^2$ has compact resolvent
and that there are constants $C>0$ and $0 < \eps < 2$ such that
\equ[e:estComp]{
	\|\Lambda^\eps f\| \le C(\|K f\| + \|f\|)\;,\quad\text{for all}\quad f \in
\CC\;.
}
Then $K$ has compact resolvent too.
\end{proposition}

\begin{proof}
By assumption, there exists a constant $\lambda > 0$ such that $K + \lambda$ is
strictly {\it m}-accretive. Moreover, \eref{e:estComp} with $K$ replaced by
$K+\lambda$ holds if we change the constant $C$. Since $\CC$ is a core for $K$,
a simple approximation argument shows that $\CD(K) \subset \CD(\Lambda^\eps)$
and that \eref{e:estComp} holds for every $f \in \CD(K)$.

This immediately implies that $(K+\lambda)^*(K+\lambda)$ has compact resolvent.
Since $(K+\lambda)$ is strictly {\it m}-accretive, it is invertible and the
operator
\equ{
\bigl((K+\lambda)^*(K+\lambda)\bigr)^{-1} =
(K+\lambda)^{-1}\bigl((K+\lambda)^{-1}\bigr)^{*}\;,
}
is compact. Moreover, we know that $(K+\lambda)^{-1}$ is closed, so we can make
the polar decomposition
\equ{
(K+\lambda)^{-1} = P J\;,
}
with $P$ self-adjoint and $J$ unitary. Thus $P^2$ is compact. By the spectral
theorem and the characterization of compact operators, this immediately implies
$P$ compact, and thus also $P J$ compact. Thus $K$ has compact resolvent.
\end{proof}

We now consider $\H=\Ltwo(\R^d)$ and $\F = \{\lambda I\;|\; \lambda \in \R\}$,
where $I$ is the identity operator in $\H$. We define the formal expressions
\equs{
	x^T &= (x_1,\ldots,x_d)\;,\\
	\qquad\d_x^T &= (\d_{x_1},\ldots,\d_{x_d})\;.
}
Let $A:\R^d \to \R^d$ be a linear map and
\equ{
	\CB = \{b_i \in \R^d \;|\; i=1,\ldots,s\}\;,\quad \CC = \{c_i \in \R^d \;|\;
i=1,\ldots,t\}\;,
}
two vector families. Let us consider the differential
operator $K$ defined as the closure on $\cOinf[\R^d]$ of
\equ[e:defKquad]{
	K = -\sum_{i=1}^s \d_x^T b_i b_i^T \d_x + \sum_{j=1}^t x^T c_j c_j^T x + x^T A
\d_x\;.
}
We are interested in giving a geometrical condition on $A$, $\CB$ and $\CC$
that implies the compactness of the resolvent of $K$, and therefore the
discreteness of its spectrum. It is possible to prove that $K$ is quasi {\it
m}-accretive. Just follow the proof of Proposition~\ref{prop:maccr}, replacing
$G(x)$ by $x^T x$.

We have the following result.

\begin{proposition}
\label{theo:quad}
A sufficient condition for the resolvent of the operator $K$ defined in
\eref{e:defKquad} to be compact is that the vector families
\equ[e:condGeo]{
	\bigcup_{N\ge 0} (A^T)^N\CB \qquad\hbox{and}\qquad \bigcup_{N\ge 0} A^N \CC
}
span the whole space $\R^n$.
\end{proposition}

\begin{remark}
The intuitive meaning of this theorem is that we can apply H\"ormander's
criterion in both direct and Fourier space to obtain an estimate of the form
\equ[e:estHarm]{
	\|H^\eps f\|\le C(\|K f\| + \|f\|)\;,\quad H = -\d_x^T \d_x^{} + x^T x\;.
}
It is well known that $H$ has compact resolvent. By
Proposition~\ref{prop:Comp}, \eref{e:estHarm} implies that $K$ has compact
resolvent.
\end{remark}

\begin{proof}
We have the following relations
\equs{
	[x^T A\d_x, b^T\d_x] &\equiv \sum_{i,j,k}[x_ia_{i j}\d_{x_j}, b_k\d_{x_k}] =
\sum_{i,j,k}b_k[x_i, \d_{x_k}]a_{i j}\d_{x_j} \\
&= -\sum_{i,j,k}b_k\delta_{ki}a_{i j}\d_{x_j} = -b^TA\d_x\;,\\
	[x^TA\d_x, c^Tx] &\equiv \sum_{i,j,k}[x_ia_{i j}\d_{x_j}, c_k x_k] =
\sum_{i,j,k}x_i a_{i j}[\d_{x_j}, x_k]c_k \\
&= \sum_{i,j,k}x_i a_{i j}\delta_{j k}c_k = c^T A^T x\;.
}
We take $g_0 = 1$, so we have $\A_0 = \A_{-1} \cup \{x^T A \d_x\}$. We
construct the remaining $\A_i$ by
\equ{
\CB_i \equiv [x^T A \d_x, \CB_{i-1}].
}
It is very easy to verify \hyp{1} and \hyp{2}, because the assumptions we made
on $A$, $\CB$ and $\CC$ imply that $\CY^1(\A)$ contains every operator of the
form $b^T\d_x$ or $c^Tx$. We have moreover
\equ{
(b^T\d_x)^* = - b^T \d_x \qquad\text{and}\qquad (c^Tx)^* = c^Tx\;.
}
It is well-known that \hyp{3} concerning the essential self-adjointness of the
$\Lambda^2$ constructed in Theorem~\ref{theo:princ} holds. Finally, it is
straightforward that $\Lambda^2$ satisfies $\Lambda^2 \ge C H$, where $H$ is
the ``harmonic oscillator'' defined in \eref{e:estHarm}.

This proves the validity of \eref{e:estHarm}, and hence of the assertion.
\end{proof}

The interested reader may verify that Proposition~\ref{theo:quad} is quite
stable under perturbations. A similar result indeed still holds when the
coefficients $b_i$ and $c_i$ are not constants, but functions in $\CF_0$. This
is precisely what was proved in \cite{EPR}.
\mysection{Proof of the bound in momentum space
(Proposition~\ref{prop:estDeltaPrime})}
\label{sec:proof2}

This proposition is an application of Theorem
\ref{theo:princ}. It is just a little bit cumbersome to verify the hypotheses
of the theorem. In this section, the symbol $K$ will again denote the operator
defined in \eref{e:defKchain}.

We choose\equ{
	\F \equiv \CF_0 \;,
}
which is simply the set of bounded smooth functions with all their derivatives
bounded. It is trivial to check that $\F$ is an algebra of closed operators.
Moreover, they are all self-adjoint. We also define $\CD \equiv \cOinf[\CX]$.

In this section, we will first construct a set $\CA$ according to the rules
explained in Section \ref{sec:Hormander}. Then we will check that
\hyp{1}--\hyp{3} are indeed satisfied, so we will be able to apply Theorem
\ref{theo:princ}. This will prove Proposition~\ref{prop:estDeltaPrime} almost
immediately.

Before we start this program, we write down once again the definition of $X_0$,
as it will be used repeatedly throughout this section:
\equs{
X_0 =&\; - \rL\d_{p_0} + b_L(\rL - \lambda_L^2 q_0)\d_\rL  - \rR\d_{p_N} +
b_R(\rR - \lambda_R^2 q_N)\d_\rR\\
 &\;- \sum_{i=0}^N \bigl(p_i\d_{q_i} - \V1'(q_i)\d_{p_i}\bigr) + \sum_{i=1}^{N}
\V2'(\tilde q_i)\bigl(\d_{p_i}-\d_{p_{i-1}}\bigr) - \alpha_K\;.
}

\subsection{Definition of $\boldsymbol{\CA}$}

We choose an exponent $\alpha < -3/2-\ell/(2m)$ and we let $g_0$ be the
operator of multiplication by $G^\alpha$. It is clear that $g_0$ is
self-adjoint and positive. Moreover, we recall that
\equ{
	[X_0,G^\alpha] = \alpha G^{\alpha} G^{-1}[X_0,G] \in \CF_0\;,
}
and so we have $[X_0, g_0] \in \F$. The set $\A_0$ is defined as
\equ{
	\A_0 = \{ c_L \d_{\rL},\, c_R \d_{\rR},\, G^\alpha X_0\} \cup \bar\A
\;,\qquad\text{with}\qquad \bar \A = \bigl\{a_L (\rL - \lambda_L^2 q_0),\,
a_R(\rR - \lambda_R^2 q_N)\bigr\}\;.
}

Before we define the sets $\A_i$, we need a few functions.
Let $i>0$ be a natural number. The functions $V_L^{(i)}$ and $V_R^{(i)}$ are
defined respectively by
\equs{
	V_L^{(i)}(\tilde q) &= \V2''(\tilde q_{i}) \V2''(\tilde
q_{i-1})\cdot\ldots\cdot\V2''(\tilde q_{1}) \;,\\
	V_R^{(i)}(\tilde q) &= \V2''(\tilde q_{N+1-i}) \cdot\ldots\cdot \V2''(\tilde
q_{N-1}) \V2''(\tilde q_{N})\;.
}
It is useful to notice that
\equ[e:derV]{
	\d_{q_j}V_L^{(i)}(\tilde q) = \left\{\begin{array}{clr@{\,\,}c@{\,\,}l@{}l}
	0\;, &\qquad \text{if} &j &> &i&\;, \\[2mm]
	\bigl(\V2'''(\tilde q_1)\V2''(\tilde q_1)^{-1}\bigr) V_L^{(i)}(\tilde q) \;,
&\qquad \text{if} &j &= &0&\;, \\[2mm]
	\bigl(\V2'''(\tilde q_i)\V2''(\tilde q_i)^{-1}\bigr) V_L^{(i)}(\tilde q) \;,
&\qquad \text{if} &j &= &i&\;, \\[2mm]
	\bigl(\V2'''(\tilde q_i)\V2''(\tilde q_i)^{-1} - \V2'''(\tilde
q_{i+1})\V2''(\tilde q_{i+1})^{-1}\bigr) V_L^{(i)}(\tilde q) \;,
&\multicolumn{4}{c}{\qquad \text{otherwise}}&\;.
\end{array}\right.
}
There are symmetric relations for the derivatives of $V_R^{(i)}$. At
this point, we use assumption {\bf A3} to write
\equ[e:estVi1]{
\d_{q_j}V_L^{(i)}(\tilde q) = f_{ij}(\tilde q) V_L^{(i)}(\tilde q)\;, \qquad
f_{ij} \in \CF_{2m-2+\ell}\;.
}
This implies
\equ[e:estVi2]{
[G^\alpha X_0,V_L^{(i)}(\tilde q)] = G^\alpha\sum_{j=0}^N p_j f_{ij}
V_L^{(i)}(\tilde q) = f_i V_L^{(i)}(\tilde q)\;,\qquad f_i \in \F\;,
}
because of Proposition \ref{prop:Degs} and by the choice $\alpha < -3/2 -
\ell/(2m)$. Moreover, we notice that
\equ{
	G^{2i\alpha}V_R^{(i)} \in \F\;,
}
still because of Proposition \ref{prop:Degs}. One more thing we have to
remember is \eref{e:boundG}, which implies for example that there exists a
function $f_0 \in \F$ such that
\equ{
	[G^\alpha X_0, G^\beta] = \beta f_0 G^\beta\;,\quad\text{for any}\quad \beta
\in \R\;.
}
We are now ready to complete the construction of $\A$.

\subsubsection{Definition of $\boldsymbol{\CA_1}$ and $\boldsymbol{\CA_2}$}

We verify that in the case of our model, we can find functions $f_B$ and
$f_{XB}$ in \eref{e:defA} such that
\equs{
	\A_1 \sauf \A_0 &= \{G^\alpha\d_{p_0},\, G^\alpha \d_{p_N}\}\;, \\
	\A_2 \sauf \A_1 &= \{G^{2\alpha}\d_{q_0},\, G^{2\alpha} \d_{q_N}\} \;.
}
Considering the elements of $\CA_1$, we see that it is indeed possible to write
\equ{
	G^\alpha \d_{p_0} = c_L^{-1} [G^\alpha X_0, c_L \d_{\rL}] - G^{-1}(\d_{\rL}G)
G^\alpha X_0 + G^\alpha b_L \d_\rL\;,
}
and a similar relation concerning $G^\alpha \d_{p_N}$. The operators
$G^{-1}(\d_{\rL}G)$ and $G^\alpha b_L c_L^{-1}$ belong to $\F$, so we
succeeded to construct $\A_1$ 
according to \eref{e:defA}.

Let us now focus on the elements of $\A_2$. We can write
\equ{
	G^{2\alpha}\d_{q_0} = [G^\alpha X_0, G^\alpha \d_{p_0}] -
G^{\alpha-1}(\d_{p_0}G) G^\alpha X_0 - \alpha f_0 G^\alpha \d_{p_0}\;,
}
and an equivalent expression at the other end of the chain. Since
$G^{\alpha-1}(\d_{p_0}G) \in \F$ and $f_0 \in \F$, we succeeded to
construct $\A_2$ 
according to \eref{e:defA}.

\subsubsection{Definition of $\boldsymbol{\CA_{2i-1}}$ and
$\boldsymbol{\CA_{2i}}$}

For $i \ge 1$, these sets are defined by
\equs{
	\A_{2i-1} \sauf \A_{2i-2} &= \{G^{(2i+1)\alpha}V_L^{(i)}\d_{p_i},\,
G^{(2i+1)\alpha}V_R^{(i)}\d_{p_{N-i}}\}\;, \\
	\A_{2i} \sauf \A_{2i-1} &= \{G^{(2i+2)\alpha}V_L^{(i)}\d_{q_{i}},\,
G^{(2i+2)\alpha} V_R^{(i)}\d_{q_{N-i}}\} \;.
}
We repeat this construction until $i=N-1$, \ie we do not stop at the middle of
the chain, but we go on until we reach the other end.
We want to check that these sets were constructed according to \eref{e:defA}.
In fact, we will see that any element $A$ of $\A_j \sauf \A_{j-1}$ with $j \ge
2$ can be written as
\equ[e:construct]{
	A = [G^\alpha X_0, B] + D\;, \quad B \in \A_{j-1}\;,\quad D \in
\CY^1(\A_{j-1})\;.
}
We will verify this only for $2 \le i \le N-2$. We let the reader verify that
\eref{e:construct} is also valid for the remaining sets.

Let us first take $j = 2i-1$ and $A = G^{(2i+1)\alpha}V_L^{(i)}\d_{p_i}$. We
choose $B = G^{2i\alpha}V_L^{(i-1)} \d_{q_{i-1}} \in \A_{j-1}$ and write
\equs{
[G^\alpha X_0, B] =&\; f_{i-1}G^{2i\alpha}V_L^{(i-1)} \d_{q_{i-1}} -
G^{2i\alpha}V_L^{(i-1)} G^{-1}(\d_{q_{i-1}}G) G^{\alpha}X_0 \\
	&+  2i\alpha f_0 B + G^{(2i+1)\alpha}V_L^{(i-1)} [X_0, \d_{q_{i-1}}]\;.
}
The first three terms belong to $\CY^1(\A_{2i-2})$ and can thus be absorbed
into $D$. The last term can be written as
\equs{
G^{(2i+1)\alpha}V_L^{(i-1)}[X_0, \d_{q_{i-1}}] =&\;
G^{2\alpha}\bigl(\V1''(q_{i-1}) + \V2''(\tilde q_i) + \V2''(\tilde
q_{i-1})\bigr) G^{(2i-1)\alpha}V_L^{(i-1)}\d_{p_{i-1}} \\
	&+ G^{4\alpha} \V2''(\tilde q) \V2''(\tilde q)
G^{(2i-3)\alpha}V_L^{(i-2)}\d_{p_{i-2}} \\
	&+ G^{(2i+1)\alpha}V_L^{(i)}\d_{p_i}\;.
}
The first two terms also belong to $\CY^1(\A_{2i-2})$, so they can be absorbed
into $D$ as well. The remaining term is
\equ{
	G^{(2i+1)\alpha}V_L^{(i)}\d_{p_i} = A\;,
}
thus we have verified that $A$ can be written as in \eref{e:construct}. The
procedure to get the symmetric term from the other end of the chain is similar.

We take now $j = 2i$ and $A = G^{(2i+2)\alpha} V_L^{(i)} \d_{q_i}$. We choose
$B = G^{(2i+1)\alpha}V_L^{(i)}\d_{p_i} \in \CY^1(\CA_{j-1})$ and write
\equs{
[G^\alpha X_0, B] =&\; f_i G^{(2i+1)\alpha}V_L^{(i)}\d_{p_i} +
G^{(2i+1)\alpha}V_L^{(i)} G^{-1}(\d_{p_i} G) G^{\alpha} X_0 \\
	&+ (2i+1)\alpha f_0 B + G^{(2i+2)\alpha} V_L^{(i)} \d_{q_i}\;.
}
The first three terms belong to $\CY^1(\A_{2i-1})$ and can be absorbed into
$D$, so we verified that every element of $\A$ can indeed be written as in
\eref{e:defA}.

\subsection{Verification of the hypotheses and proof}

In order to be able to apply Theorem \ref{theo:princ}, we verify the hypotheses
\hyp{1}--\hyp{3}.

\proclaim{Verification of \hyp{2}.} We want to check that $A\in \CA_j$ implies
$A^* \in \CY^1(\CA_j)$. By Proposition \ref{prop:Degs}, we can easily verify
that $\A \sauf \bar \CA \subset \L_0$. But we know that
\equ{
A \in \CL_0 \Rightarrow A^* = -A + g\;,
}
and so \hyp{2} holds for $\CA \sauf \bar \CA$. The elements of $\bar \CA$ being
self-adjoint, \hyp{2} holds trivially.

\proclaim{Verification of \hyp{3}.} The operator $\Lambda^2$ can be written as
\equ{
	\Lambda^2 = -\sum_{i,j} \d_i a_{ij}(x) \d_j + V(x)\;.
}
It is well-known that if $a_{ij}$ and $V$ are sufficiently nice, such operators
are essentially self-adjoint on $\cOinf[\CX]$ (see \eg \cite[Thm.~3.2]{Ag}).

\proclaim{Verification of \hyp{1}.} Let us define $\fL_0 \subset \fL$ as the
set of first-order differential operators with coefficients in $\CF_0$.
We first verify that
\equ{
	A \in \A\,,\; f \in \F\quad\Rightarrow\quad [A,f] \in \F\;.
}
This is trivial, noticing that $\A \subset \fL_0 \cup \bar \A$ and $[\fL_0, \F]
= [\bar \A, \F] = \{0\}$.

We now verify that
\equ{
	A \in \A \quad\Rightarrow\quad A^* \in \CY^1(\A)\;.
}
This is also trivial, because $A \in \fL_0 \Rightarrow A^* = -A + g$, with $g
\in \CF_0$. Moreover, the elements of $\bar \CA$ are self-adjoint.

Finally, we want to verify that
\equ{
A,B \in \A \quad\Rightarrow\quad [A,B] \in \CY^1(\A)\;.
}
This is a little bit longer to verify.

Concerning the commutators of the elements of $\bar \A$ with the other elements
of $\A$, the statement follows easily from the fact that if $F:\R^n \to R$ is
linear and $A \in \fL_0$, then $[A,F] \in \CF_0 \equiv \F$. Moreover, the
com\-mutator between two multiplication operators vanishes.

Concerning the commutators between the $\d_r$ and the other elements, we notice
that they commute with the functions $V_L^{(i)}(\tilde q)$ and
$V_R^{(i)}(\tilde q)$. Moreover, we have for example
\equ{
	[\d_\rL, G^\gamma] = \gamma \bigl(G^{-1} [\d_\rL, G]\bigr)G^\gamma
\;,\quad\text{if}\quad \gamma \in \R\;,
}
and $G^{-1} [\d_\rL, G]$ belongs to $\F$. It is straightforward to verify that
this implies the desired statement.

Concerning the commutators of $G^\alpha X_0$ with the other elements of $\A$,
the statement has already been verified by the construction of $\A$ for every
operator, but those in $\A_{2N-2}\sauf \A_{2N-3}$. These operators are of the
form
\equ{
	A = G^{2N\alpha}V_R^{(N-1)}\d_{q_1}\;,
}
and a similar term at the other end of the chain. We can make a computation
very similar to the one we made when we constructed $\A_{2i-1}$, to show that
\equ{
	[G^\alpha X_0, A] = G^{(2N+1)\alpha}V_R^{(N)}\d_{p_{0}} + C\;, \qquad C\in
\CY^1(\A)\;.
}
But $G^{2N\alpha}V_R^{(N)} \in \F$, so $[G^\alpha X_0, A] \in \CY^1(\A)$.
It remains therefore only to verify the statement for commutators between
elements of $\A \sauf \A_0$. We can divide these commutators in three classes.

\proclaim{Both operators contain a $\boldsymbol{\d_p}$.} We notice that these
operators can all be written in the form $G^{\alpha_i} W_i(q) \d_{p_i}$. The
commutator between two such elements is given by
\equs{
[G^{\alpha_i} W_i(q) \d_{p_i},G^{\alpha_j} W_j(q) \d_{p_j}] =&\;
G^{-1}(\d_{p_i}G)G^{\alpha_i}W_i(q)G^{\alpha_j}W_j(q)\d_{p_j} \\
	& - G^{-1}(\d_{p_j}G)G^{\alpha_j}W_j(q)G^{\alpha_i}W_i(q)\d_{p_i}\;.
}
Both terms belong to $\CY^1(\A)$, because $G^{-1}(\d_p G) \in \F$.

\proclaim{One operator contains a $\boldsymbol{\d_p}$, one contains a
$\boldsymbol{\d_q}$.} Let us compute the commutator between
$G^{(2i+2)\alpha}V_L^{(i)}\d_{q_i}$ and $G^{(2j+1)\alpha}V_L^{(j)}\d_{p_j}$. We
have
\equs{
[G^{(2i+2)\alpha}V_L^{(i)}\d_{q_i},G^{(2j+1)\alpha}V_L^{(j)}\d_{p_j}] =&\;
G^{(2i+2)\alpha}V_L^{(i)}(\d_{q_i} G)G^{-1}G^{(2j+1)\alpha}V_L^{(j)}\d_{p_j} \\
&+ G^{(2i+1)\alpha}V_L^{(i)}(\d_{p_i} G)G^{-1}G^{(2i+2)\alpha}V_L^{(i)}\d_{q_i}
\\
&+  G^{2i\alpha}V_L^{(i)}G^{2\alpha}f_{ij}G^{(2j+1)\alpha}V_L^{(j)}\d_{p_j}\;.
}
All those terms belong to $\CY^1(\A)$. The computation is similar if we take
for example
\equ{
	G^{(2j+1)\alpha}V_R^{(j)}\d_{p_{N-j}}
} instead of $G^{(2j+1)\alpha}V_L^{(j)}\d_{p_j}$.

\proclaim{Both operators contain a $\boldsymbol{\d_q}$.} The computation is
similar to the preceding case and is left to the reader.

It is now easy to give the

\begin{proof}[of Proposition~\ref{prop:estDeltaPrime}]
We have just verified that the hypotheses of Theorem \ref{theo:princ} are
satisfied. We apply it, so we have the estimate
\equ{
\|\tilde\Delta^\eps f\| \le C(\|Kf\| + \|f\|)\;,
}
where $\tilde\Delta$ is given by
\equ{
	\tilde\Delta = 1 + \sum_{A \in \A} A^*A\;.
}
It is easy to see that $\tilde\Delta$ has exactly the form
\eref{e:defDeltaPrime}. This completes the proof of
Proposition~\ref{prop:estDeltaPrime}.
\end{proof}

\mysection{Proof of Theorem \ref{theo:Chain}}
\label{sec:theoChain}

It is now possible to prove that the operator $K$ has compact resolvent, which
is one of the main results of this paper. Before we start the proof itself,
we need two preliminary results. The first one states
\begin{lemma}
\label{prop:sumEps}
Let $\tilde\Delta$ be the closure in $\Ltwo(\R^n)$ of the operator acting on
$\cOinf[\R^n]$ as
\equ{
 \tilde\Delta  = \sum_{i=1}^{\bar N} L_i^*L_i + a_0\;,
}
where the $L_i$ are smooth vector fields with bounded coefficients spanning
$\R^n$ at every point and $a_0$ is a smooth positive function.

Let $V : \R^n \to \R^n$ be a continuous function such that for every constant
$C>0$, there exists a compact $K_C \subset \R^n$ with the property that $V(x) >
C$ for every $x \in \R^n \backslash K_C$. We moreover assume that $V(x) \ge 1$.
Define the operator $H$ as the closure in $\Ltwo(\R^n)$ of the operator acting
on $f \in \cOinf[\R^n]$ as
\equ{
\bigl(H f\bigr)(x) = \bigl(\tilde\Delta f\bigr)(x) + V(x) f(x)\;.
}
Then the operator $H$ is self-adjoint.

Suppose $V$ and the $L_i$ are such that the function
\equ[e:defBound]{
	2a_0 V + \sum_{i=1}^{\bar N}\Bigl((L_i^* + L_i)[L_i,V] -
\bigl[L_i,[L_i,V]\bigr]\Bigr)
}
is bounded. We then have the estimate
\equ[e:sumexp]{
	\scal{f,H ^\eps f} \le \scal{f,\tilde\Delta^\eps f} +\scal{f,V^\eps f} +
C\scal{f,H^{\eps-1}f}\;,\qquad 0<\eps < 1\;,
}
which holds for any $f \in \cOinf[\R^n]$.
\end{lemma}

\begin{proof}
The result concerning the self-adjointness of $H$ and of $\tilde \Delta$ is
classical, we will not prove it here. The interested reader can find a proof in
\cite[Thm.~3.2]{Ag}.

We use the fact that if $T$ is a strictly positive self-adjoint operator and
$\alpha= 1-\eps \in (0,1)$, we can write
\equ{
T^{-\alpha} = C_{\alpha} \int_0^\infty z^{-\alpha} (z+T)^{-1}\,dz\;,\qquad
C_\alpha = \frac{\sin(\pi \alpha)}{\pi}\;,
}
and thus
\equ{
T^{\eps} = C_{\alpha} \int_0^\infty z^{\eps-1} \frac{T}{z+T}\,dz\;.
}
Moreover, a core of $T$ is again a core of $T^\eps$, so \eref{e:sumexp} makes
sense. For a proof of these statements, see \cite[\S V.3]{Ka}.
This allows us to write inequality \eref{e:sumexp} as
\equs[e:inequWanted]{
\int_0^\infty z^{\eps-1} \Bigl\langle f,\frac{H}{z+H}f\Bigr\rangle\,dz \le &\;
\int_0^\infty z^{\eps-1} \Bigl\langle
f,\frac{\tilde\Delta}{z+\tilde\Delta}f\Bigr\rangle\,dz
	+ \int_0^\infty z^{\eps-1} \Bigl\langle f,\frac{V}{z+V}f\Bigr\rangle\,dz
\\[2mm]
	& + C\int_0^\infty z^{\eps-1} \Bigl\langle f,\frac{1}{z+H}f\Bigr\rangle\,dz\;.
}
\def\Del{\tilde\Delta}
In order to prove \eref{e:inequWanted}, let us first show that the operator
$\Del V + V \Del$ is lower bounded. This is an immediate consequence of
\eref{e:defBound} and the equality
\equs{
	L_i^* L_i V + V L_i^* L_i &= 2L_i^* V L_i + (L_i + L_i^*)[L_i,V] -
\bigl[L_i,[L_i,V]\bigr]\;,
}
which is easily verified, using the fact that $L_i + L_i^*$ is simply a
function.
Therefore, there exists a constant $C> 0$ such that
\equ{
\scal[b]{g, (\Del V + V \Del) g} + C\scal{g,g} \ge 0\;,\quad \forall g\in
\cOinf[\R^n]\;.
}
Since $H \ge 1$ in the sense of quadratic forms, we find
\equ{
	\scal[b]{g, (\Del V + V \Del) g} + C\scal{g,(z+H)g} \ge 0\;,
}
which holds for every $z \ge 0$.
Since $\Del$ and $V$ are positive self-adjoint operators, this immediately
implies
\equ[e:inter1]{
	\scal[B]{g, V\frac{\tilde\Delta}{z+\tilde\Delta}V g} + \scal[B]{g,
\tilde\Delta\frac{V}{z+V}\tilde\Delta g} + \scal[b]{g, (\tilde\Delta V + V
\tilde\Delta) g}+ C\scal{g,(z+H)g} \ge 0\;.
}
We can easily check the following identities
\equs{
	V \Del(z+\Del)^{-1} V + \Del V &= (z+H)\Del(z+\Del)^{-1}V\;, \\[1mm]
	\Del V(z+ V)^{-1} \Del + V \Del &= (z+H)\Del(z+V)^{-1}\Del \;.
}
Inserting this in \eref{e:inter1}, we get
\equ{
	\scal[b]{g,(z+H)\bigl(\tilde\Delta(z+\tilde\Delta)^{-1}V  +
V(z+V)^{-1}\tilde\Delta\bigr)g}	 + C\scal{g,(z+H)g} \ge 0\;,
}
and thus
\equ{
	\scal[b]{g,(z+H)Hg} \le
\scal[b]{g,(z+H)\bigl(H+\tilde\Delta(z+\tilde\Delta)^{-1}V  +
V(z+V)^{-1}\tilde\Delta\bigr)g} + C\scal{g,(z+H)g}\;.
}
We can check the equalities
\equs{
	\Del(z+\Del)^{-1}V + \Del &= \Del(z+\Del)^{-1}(z+H)\;,\\[1mm]
	V(z+V)^{-1}\Del + V &= V(z+V)^{-1}(z+H)\;,
}
which allow us to write
\equ{
	\scal[b]{g,(z+H)Hg} \le \scal[b]{(z+H)g,\bigl(\Del(z+\tilde\Delta)^{-1}  +
V(z+V)^{-1}\bigr)(z+H)g} + C\scal{g,(z+H)g}\;.
}
Let us define $f \equiv (z + H)g$. This immediately yields the estimate
\equ[e:rel]{
\scal[B]{f,\frac{H}{z+H}f} \le \scal[B]{f,\frac{\tilde\Delta}{z+\tilde\Delta}f}
+ \scal[B]{f,\frac{V}{z+V}f} + C \scal[B]{f,\frac{1}{z+H}f}\;,
}
which holds for any $f$ in $\CW \equiv (z+H)\cOinf[\R^n]$. But we know that
$\cOinf[\R^n]$ is a core for $H$, therefore $\CW$ is dense in $\Ltwo(\R^n)$.
Since the operators appearing in \eref{e:rel} are all bounded, the inequality
\eref{e:rel} holds for every $f \in \Ltwo(\R^n)$ and thus in particular also
for $f \in \cOinf[\R^n]$.
This implies the wanted estimate \eref{e:inequWanted}.
\end{proof}

The second result we want to use is
\begin{proposition}
\label{prop:compRes}
Let $\tilde\Delta$, $V$ and $H$ be as in Lemma~\ref{prop:sumEps}.
Then $H$ has compact resolvent.
\end{proposition}

\begin{proof}
We know that $\tilde \Delta$ is a positive self-adjoint operator, so
\equ{
	T = (\tilde\Delta + 1)^{-1}
}
exists and $\|T\| \le 1$. The proof of compactness is a modification of the
standard proof of the same theorem with $\tilde \Delta$ replaced by the true
Laplacean $\Delta$, which can be found \eg in \cite{Ag}. It is based on the
fact that if $\chi$ is a function with compact support, then the multiplication
operator $\chi$ is relatively compact with respect to $\tilde\Delta$. We want
to prove that $\chi T$ is a compact operator, \ie that the closure of
\[
Y = \{ \chi T f \; | \; f \in \cOinf[\R^n] \quad \hbox{and} \quad \|f\| \le 1\}
\]
is compact.

Let us define $\CK = \supp \chi$. By hypothesis, $\CK$ is compact. Moreover, we
have $Y \subset \cOinf(\CK)$. It is well-known that if $\CK$ is a
compact domain of $\R^n$, then the set
\[
\{ u\in \cOinf(\CK) \; | \; \|u\| \le 1;\; \scal{u,\Delta u} \le 1\}
\]
is compact (see \eg \cite[Thm.~XIII.73]{RS}). This implies that $Y$ is compact
if we are able to prove that there are strictly positive constants $\eps$,
$c_1$ and $c_2$ such that $u\in Y$ implies
\[
	\|u\| \le c_1 \qquad\hbox{and}\qquad \scal{u,\Delta u} \le c_2	\;.
\]
We take any element $u$ in $Y$ and write it as $u=\chi T f$. We have
\[
\|u\| \le \|\chi\|_\infty \, \|T\| \, \|f\| \le c_1\;.
\]
Recall that we assumed the vector fields $L_i$ appearing in the construction of
$\tilde \Delta$ span $\R^n$ at any point and that $a_0$ is a strictly positive
function. Together with the compactness of the support of $u$, this implies
that there are constants $C$ and $k_1$ such that
\equs[equ:est1]{
	|\scal{u,\Delta u}| &\le C|\scal{u,\tilde\Delta u}| =  C|\scal{u,\tilde\Delta
\chi T f}| \le C\|u\|\|\tilde\Delta \chi T f\| \\
	& \le C\|\chi \tilde\Delta T f\| + C\|[\tilde\Delta,\chi] T f\| \le k_1 +
C\|[\tilde\Delta,\chi] T f\|\;,
}
where the last inequality is a consequence of $T = (1 +\tilde \Delta)^{-1}$.
We therefore only need to bound the term containing the commutator of
$\tilde\Delta$ and $\chi$. Explicit calculation yields
\equ{
[\tilde\Delta,\chi] = \sum_{i=1}^{\bar N} \Bigl( -2[L_i, \chi]L_i +
\bigl[L_i,[L_i,\chi]\bigr] + (L_i^* + L_i)[L_i,\chi]\Bigr)
	\equiv \sum_{i=1}^{\bar N} \eta_i L_i + \eta_0\;,
}
where the $\eta_i$ are bounded functions with $\supp \eta_i \subset \CK$. So
the only terms that remain to be bounded are of the form $\|\eta_i L_i T f\|$.
As $\eta_i$ is bounded, it is enough to bound $\|L_i T f\|$. We have
\equs[equ:est2]{
\|L_i Tf\|^2 &= \scal{Tf, L_i^* L_i Tf} \le \scal{Tf, \tilde\Delta Tf} \le
\|f\|^2\;.
}
This completes the proof of the statement about the relative compactness of
$\chi$.

This implies that we can add to $H$ any function with compact support without
changing its essential spectrum (see \cite[Thm.~XIII.14]{RS}). But the
assumption we made concerning $V$ and the positivity of $\tilde \Delta$ imply
that for any constant $C$, we can raise the spectrum of $H + \chi$ above $C$ by
taking for $\chi$ a smooth function satisfying
\equ{
	\chi(x) = \left\{ \begin{array}{cc@{}l}
	C \quad& x \in \CK_C &\;,\\
	0 \quad& d(x,\CK_C) > 1&\;.
\end{array} \right.
}
Therefore, the essential spectrum of $H$ is empty and thus $H$ has compact
resolvent.
\end{proof}

It is now easy to give the

\begin{proof}[of Theorem \ref{theo:Chain}]
By Proposition \ref{prop:estG} and \ref{prop:estDeltaPrime}, we can choose a
constant $\eps$ small enough to have, for every $f \in \cOinf[\CX]$, the
estimate
\equ{
	\|\tilde\Delta^\eps f\| \le C(\|Kf\| + \|f\|) \quad\text{and}\quad  \|G^\eps
f\| \le C(\|Kf\| + \|f\|)\;.
}
We moreover define
\equ[e:defHchain]{
	H \equiv \tilde\Delta + G\;.
}
By the proof of Proposition~\ref{prop:estDeltaPrime}, we see that the assumptions of
Lemma~\ref{prop:sumEps} are satisfied. We can thus write
\equs{
\|H^\eps f\|^2 &= \scal{f, H^{2\eps}f} = \scal{f,(\tilde\Delta + G)^{2\eps} f}
\le  \scal{f, \tilde\Delta^{2\eps} f} + \scal{f, G^{2\eps} f} + C\|f\|^2 \\
	& \le \|\tilde\Delta^{\eps} f\|^2 + \|G^\eps f\|^2 + C\|f\|^2 \le C(\|Kf\| +
\|f\|)^2\;.
}
Because $G$ is confining, we can apply Proposition~\ref{prop:compRes} to see
that $H$, and therefore also $H^\eps$, have compact resolvent. Therefore
Corollary~\ref{prop:Comp} applies, showing that $K$ has compact resolvent.
\end{proof}

\begin{remark}[1]
The proof still works under slightly weaker assumptions. The coupling between
the ends of the chain and the heat baths does not have to be of the dipolar
type. It is enough for example that $F_L$ and $F_R$ belong to some $\CF_\beta$
with $\beta < n$. Moreover, the potentials $V_1$ and $V_2$ can be different for
each particle. We only have to impose that assumptions {\bf A1}--{\bf
A3} can be satisfied for every particle with the same constants
$\ell$, 
$m$ and $n$.
\end{remark}

\begin{remark}[2]
Throughout this paper, we restricted ourselves to the one-dimensional case, \ie
each particle had only one degree of freedom. It is not very hard to generalize
the results of this paper to the $d$-dimensional case. It is straightforward to
generalize assumptions {\bf A1} and {\bf A2}, where $V'$ is now a vector. In
assumption {\bf A3}, the inverse of $\V2''$ has to be read as the inverse
matrix. A matrix or vector-valued function is said to belong to $\CF_\beta$ if
each of its components belong to $\CF_\beta$.

The only point that could cause some trouble is the expression \eref{e:derV},
because the $\V2''(\tilde q_j)$ are now matrices which do not commute, so the
expression for $\d_{q_{j}} V_L^{(i)}(\tilde q)$ will show terms of the form
\equ{
	 \V2''(\tilde q_{i}) \V2''(\tilde q_{i-1})\cdot\ldots\cdot \V2'''(\tilde
q_{j}) \cdot\ldots\cdot\V2''(\tilde q_{1})\;,
}
where $\V2'''$ is a trilinear form. Such a term can we written as
\equ{
\V2''(\tilde q_{i}) \V2''(\tilde q_{i-1})\cdot\ldots\cdot \V2'''(\tilde q_{j})
{\V2''(\tilde q_{j+1})}^{-1} \cdot\ldots\cdot {\V2''(\tilde q_{i})}^{-1}
V_L^{(i)}\;.
}
If we want to get expressions similar to \eref{e:estVi1} and \eref{e:estVi2},
we have to make $|\alpha|$ very big (of the order of $N$), but this is not a
problem.
\end{remark}

\begin{remark}[3]
One important assumption was that $m > n$, in other words, the
interparticle coupling is stronger at infinity than the single
particle potential.
If
this is not satisfied, our proof does not work. There could be some physical
reason behind this. If a stationary state exists, this means that even if the
chain is in a state of very high energy, the mean time to reach a region with
low energy is finite (see \eg \cite{Ha}). But if $m<n$, the relative strength
of the coupling versus the one-body potential goes to zero at high energy. The
consequence is that there is almost no energy transmitted between particles.
Since the only points where dissipation occurs are the ends of the
chain, we see that the 
higher the energy of the chain is, the slower this energy will be dissipated.
Probably this is not sufficient to destroy the existence of a stationary state,
but it could explain why the proof does not work in this situation. It is even
possible that this phenomenon destroys the compactness of the resolvent of $K$.
\end{remark}

\mysection{The invariant measure}
\label{sec:inv}

This section is devoted to the proof of Theorem~\ref{theo:exist}.
Throughout this section, we denote by $\T^t$ the semi-group generated
by the system of stochastic differential equations
\eref{e:stochChain}. We also assume that {\bf A1}--{\bf A3} are
satisfied, so Proposition~\ref{prop:estG} and \ref{prop:estDeltaPrime}
hold, as well as Theorem~\ref{theo:Chain}. The proof of
Theorem~\ref{theo:exist} is divided into three separate propositions,
showing respectively the following properties of the invariant measure
$\mu$:
\begin{list}{$\bullet$}
{\setlength{\leftmargin}{9mm}\setlength{\topsep}{2mm}\setlength{\itemsep}{0mm}}
\item[(i)] Existence and smoothness.
\item[(ii)] Decay properties.
\item[(iii)] Uniqueness and strict positivity.
\end{list}

\begin{proposition}
If Assumptions {\bf A1}--{\bf A3} are satisfied, the Markov
process given by \eref{e:stochChain} possesses an invariant measure $\mu$. It
has a density $h$, which is a $\CC^\infty$ function on $\R^{2N+4}$.
\end{proposition}

\begin{proof}
By Theorem~\ref{theo:Chain}, we know that $K$ has compact resolvent.
This implies also the compactness of the resolvent of $L_{\CH}$ and
thus of $L_0$. Since $G$ grows algebraically at infinity, we see that the
constant function $1$ belongs to $\CH_0$. Moreover, we notice that $L_0 1 = 0$,
thus the operator $L_0$ has an eigenvalue $0$, which is isolated because of the
compactness of its resolvent. This in turn implies that $L_0^*$ also has an
isolated eigenvalue $0$. We denote the corresponding eigenvector by $g$ and
normalize it so that $\scal{g,1}_{\CH_0}=1$. Since $L_0^*$ is hypoelliptic, $g$
must be $\CC^\infty$.

Assume first that $g \ge 0$. We then define
\equ[e:defh]{
	h(p,q,r) = Z_0^{-1} g(p,q,r) e^{-2\beta_0 G}\;,
}
where $Z_0$ is the normalization constant appearing in the definition of
$\CH_0$. Set $\mu(dx) = h(x)\,dx$; we want to check that $\mu$ is the invariant
measure we are looking for. Notice that $\mu(dx)$ is a probability measure
because
\equ{
	\int \mu(dx) = Z_0^{-1}\int e^{-2\beta_0 G(x)}g(x)\,dx = \scal{g,1}_{\CH_0} =
1\;.
}
Let $A$ be a Borel set of $\R^{2N+4}$. Then the characteristic function
$\chi_A$ of $A$ belongs to $\CH_0$. We have
\equs{
	\bigl((\T^t)^* \mu\bigr)(A) &= \int \bigl(\T^t \chi_A\bigr)(x)\,\mu(dx) =
Z_0^{-1} \int e^{-2\beta_0 G(x)} g(x) \bigl(\T^t \chi_A\bigr)(x)\, dx \\
	& = Z_0^{-1} \int e^{-2\beta_0 G(x)} \bigl((\T_0^t)^* g\bigr)(x)\chi_A(x)\, dx
= \mu(A)\;,
}
thus $\mu$ is an invariant measure for the Markov process defined by
\eref{e:stochChain}.

The argument showing that it was indeed justified to assume $g$ positive can be
taken over from \cite[Prop.~3.6]{EPR}.
\end{proof}

We next turn to the decay properties of the invariant measure $h$. We first
introduce a convenient family of Hilbert spaces.

\begin{definition}
Choose $\gamma \in \R$. We define the Hilbert space $W^{(\gamma)}$  as
\equ{
	W^{(\gamma)} \equiv \Ltwo(\CX,\,G^{2\gamma}(x)\,dx) = \CD(G^\gamma)\;.
}
We will denote by $\scal{\cdot,\cdot}_{(\gamma)}$ and $\|\cdot\|_{(\gamma)}$
the corresponding scalar product and norm.
We also define
\equ{
	W^{(\infty)} \equiv \bigcap_{\gamma > 0} W^{(\gamma)}\;,
}
which is the set of all functions that decay at infinity faster than any
polynomial.
\end{definition}

We already know that $h$ is a $\CC^\infty$ function, so we want to show that it
is possible to write
\equ{
	h(p,q,r) = \tilde h(p,q,r) e^{-\beta_0 G(p,q,r)}\;,\qquad \tilde h \in
W^{(\infty)}\;.
}
The function $\tilde h$ being an eigenfunction of the operator $K$, the decay
properties of the invariant measure are a consequence of the following result.

\begin{proposition}
\label{prop:decay}
The eigenfunctions of $K$ and $K^*$ belong to $\CC^\infty(\CX) \cap
W^{(\infty)}$.
\end{proposition}

We will show Proposition~\ref{prop:decay} only for the eigenfunctions of $K$.
It is a simple exercise left to the reader to retrace the proof for the
eigenfunctions of $K^*$. We already know that $K$ and $K^*$ are hypoelliptic,
so their eigenfunctions belong to $\CC^\infty(\CX)$. It remains to be proven
that they also belong to $W^{(\infty)}$.

To prove the proposition, we will show the implication
\equ[e:implSpaces]{
f \in W^{(\gamma)} \quad\text{and}\quad K f \in W^{(\gamma)}
\quad\Rightarrow\quad f \in W^{(\gamma + \eps)}\;,
}
which immediately implies that the eigenvectors of $K$ belong to
$W^{(\infty)}$.
For this purpose, we introduce the family of operators $K_\gamma$ defined by
\equs[2]{
	K_\gamma : \;&\CD(K_\gamma) &\;\to\;& W^{(\gamma)} \\
	&\quad f & \;\mapsto\; & K f\;,
}
where $\CD(K_\gamma)$ is given by
\equ{
\CD(K_\gamma) = \{ f \in W^{(\gamma)}\;|\; K f \in W^{(\gamma)}\}\;.
}
The expression $K f$ has to be understood in the sense of distributions.

We have the following preliminary result.

\begin{lemma}
\label{lem:coregamma}
$\cOinf[\CX]$ is a core for $K_\gamma$.
\end{lemma}

\begin{proof}
The proof uses the tools developed in Appendix~\ref{App:ops} and is postponed
to Appendix~\ref{App:core}.
\end{proof}

The key lemma for the proof of Proposition~\ref{prop:decay} is the following.

\begin{lemma}
\label{lem:KDecay}
There are an $\eps>0$ and constants $C_\gamma > 0$ such that for every $\gamma
> 0$ and every $u \in \CD(K_\gamma)$, the relation
\equ[e:estKDecay]{
	\|G^\eps u\|^2_{(\gamma)} \le C_\gamma\bigl(\|K_\gamma u\|^2_{(\gamma)} +
\|u\|^2_{(\gamma)}\bigr)
}
holds.
\end{lemma}

\begin{proof}
Since we know that $\cOinf[\CX]$ is a core for $K_\gamma$, it suffices to show
\eref{e:estKDecay} for $u \in \cOinf[\CX]$.
Let $L$ be the first-order differential operator associated to a
divergence-free vector field. Then we have for $f,g \in \cOinf$,
\equs{
	\scal{Lf,g}_{(\gamma)} &= -\scal{f,L G^{2\gamma}g} = -\scal{f,G^{2\gamma} L g}
- 2\gamma\scal{f,G^{2\gamma} G^{-1}(LG)g} \\[1mm]
	&= -\scal{f,L g}_{(\gamma)} - 2\gamma\scal{f, G^{-1}(LG) g}_{(\gamma)}\;.
}
We write this symbolically as
\equ{
	L_\gamma^* = - L_\gamma - 2\gamma G^{-1}(LG)\;.
}
We can use the latter equality to show that there are constants
$c^{(1)}_\gamma$ and $c^{(2)}_\gamma$ such that
\equ{
	-\frac{(L_\gamma)^2 + (L_\gamma^*)^2}{2} = L_\gamma^* L_\gamma +
c^{(1)}_\gamma G^{-2}(LG)^2 + c^{(2)}_\gamma G^{-1}(L^2G)\;.
}
Using the explicit form of $K$, this in turn yields the useful relation
\equs[e:ReKbeta]{
	\Re\scal{u,K u}_{(\gamma)} =&\, c_L^2 \|\d_\rL u\|_{(\gamma)}^2 + c_R^2
\|\d_\rR u\|_{(\gamma)}^2 \\[1mm]
	& + a_L^2 \|(\rL - \lambda_L q_0) u\|_{(\gamma)}^2 + a_R^2 \|(\rR - \lambda_R
q_N) u\|_{(\gamma)}^2 + \scal{u,f_K u}_{(\gamma)}\;,
}
where $f_K$ is some bounded function.

We now have the tools to prove the validity of \eref{e:estKDecay}. We use
Proposition~\ref{prop:estG} to write
\equs{
	\|u\|^2_{(\gamma+\eps)} &= \|G^\eps G^\gamma u\|^2 \le C(\|K G^\gamma u\|^2 +
\|G^\gamma u\|^2) \\[1mm]
	&\le C(\|G^\gamma K u\|^2 + \|[K,G^\gamma] u\|^2 + \|G^\gamma u\|^2)\;.
}
An explicit computation yields
\equ{
	[K,G^{\gamma}]u = G^{\gamma}\bigl(f_L\,\d_\rL + f_R\,\d_\rR + f_0\bigr)u\;,
}
for some smooth bounded functions $f_L$, $f_R$ and $f_0$. We are thus able to
write
\equ[e:estuDecay]{
	\|u\|^2_{(\gamma+\eps)} \le C\bigl(\|K u\|^2_{(\gamma)} + \|u\|^2_{(\gamma)} +
c_L^2 \|\d_\rL u\|^2_{(\gamma)} + c_R^2 \|\d_\rR u\|^2_{(\gamma)}\bigr)\;.
}
Using \eref{e:ReKbeta}, we can write
\equs{
c_L^2 \|\d_\rL u\|^2_{(\gamma)} + c_R^2 \|\d_\rR u\|^2_{(\gamma)} &\le
|\Re\scal{u,Ku}_{(\gamma)}| + C\|u\|_{(\gamma)}^2 \\[1mm]
	&\le C\bigl(\|Ku\|_{(\gamma)}^2 + \|u\|_{(\gamma)}^2\bigr)\;.
}
This, together with \eref{e:estuDecay}, completes the proof of the assertion.
\end{proof}

\begin{proof}[of Proposition~\ref{prop:decay}]
Lemma~\ref{lem:KDecay} immediately shows that $\CD(K_\gamma) \subset W^{(\gamma
+ \eps)}$ for every $\gamma > 0$. This proves the assertion
\eref{e:implSpaces}.

Let $f$ be an eigenfunction of $K$. We know that $f \in \Ltwo(\CX)$ and,
because it is an eigenvector of $K$, we have $K f \in \Ltwo$. Thus, by
\eref{e:implSpaces}, $f \in W^{(\eps)}$. Of course $K f \in
W^{(\eps)}$ as well, so
$f \in W^{(2 \eps)}$. This can be continued {\it ad infinitum}, and so we have
$f \in W^{(\infty)}$, which is the desired result.
\end{proof}

Finally, we want to show the strict positivity and the uniqueness of the
invariant measure. The proof of this result will only be sketched, as it simply
retraces the proof of Theorem~3.6 in \cite{EPR2}.

\begin{proposition}
The density $h$ of the invariant measure $\mu$ is a strictly positive function.
Moreover, the invariant measure is unique.
\end{proposition}

\proclaim{Sketch of proof.}The idea is to show that the control system
associated with the stochastic differential equation \eref{e:stochChain} is
strongly completely controllable. This means that, given an initial condition
$x_0$, a time $\tau$ and an endpoint $x_\tau$, it is possible to find a
realization of the Wiener process $w$ such that $\xi(\tau;x_0,w) = x_\tau$. The
main assumption needed to show that is that the gradient of the two-body
potential is a diffeomorphism. This is ensured by assumption {\bf A3}.

The consequence is that, for every time $\tau$, every initial condition $x_0$
and every open set $U$, the transition probability $P(\tau,x_0,U)$ is strictly
positive. Because $\mu$ is invariant, we have
\equ{
	\mu(U) = \int P(t,x,U)\,\mu(dx) > 0\;.
}
This implies the strict positivity of $h$. Uniqueness follows from an
elementary ergodicity argument.
{\hfill\qed}\makeappendix{Proof of Lemma \ref{lem:power}}
\label{App:main}

Throughout this appendix, we will make use of the same notations as in
Section~\ref{sec:Hormander}, \ie $\CH = \Ltwo(\R^n)$, $\CD = \cOinf[\R^n]$ and
${\mathfrak D}$ is the set of differential operators with smooth coefficients.

Moreover, $\A$ denotes some finite subset of ${\mathfrak D}$ and is identified
with closed operators on $\CH$. The operator $\Lambda^2$ is defined as
\equ[e:defLambdaApp]{
\Lambda^2 \equiv 1 + \sum_{A\in\A}A^*A\;.
}
We will moreover assume that \hyp{1} and \hyp{3} concerning $\A$ and $\F$
holds, \ie $A,B \in \A$ and $f \in \F$ imply
\equ[e:H1]{
 [A,B] \in \CY^1(\A)\;,\quad A^* \in \CY^1(\A)\;,\quad [A,f] \in \F\;.
}

In order to prepare the proof of Lemma \ref{lem:power}, we need a few auxiliary
results.

\begin{lemma}
\label{lem:estj}
Let $\A$, $\F$, $\CD$ and $\Lambda$ be as above and assume \hyp{1} and \hyp{3}
hold. Then, if $A \in \CY^j_\F(\A)$, the operator $A\Lambda^{-j}$ is bounded.
\end{lemma}

The proof of this lemma will be a consequence of

\begin{lemma}
\label{lem:estTwo}
Let $\A$, $\F$, $\CD$ and $\Lambda$ be as above and assume \hyp{1} and \hyp{3}
hold. Then, if $A_1, A_2 \in \A$, the operators $A_1\Lambda^{-1}$ and $A_1
A_2\Lambda^{-2}$ are bounded.
\end{lemma}

\begin{proof}
Let us show first that $A_1 \Lambda^{-1}$ is bounded. Since $\CD$ is a core for
$\Lambda$, it suffices to show that there is a constant $C$ such that
\equ{
	\|A_1 f\|^2 \le C\|\Lambda f\|^2 \qquad\forall\, f\in \CD\;.
}
This is an immediate consequence of
\equ{
	\|\Lambda f\|^2 = \|f\|^2 +  \sum_{A \in \A} \|Af\|^2\;.
}
In order to show that $A_1A_2\Lambda^{-2}$ is bounded, we will show that there
are constants $\tau$ and $C$ such that
\equ[e:estTwo]{
	\|A_1A_2 f\|^2 \le C\|\Lambda^2f + (\tau - 1)f\|^2\;.
}
We can write the following equality
\equs{
	\|(\Lambda^2-1)f  + \tau f\|^2 &= \tau^2\|f\|^2 + 2\tau \sum_{A\in \A} \|A
f\|^2 + \sum_{A,B\in \A} \scal{f,A^*AB^*Bf} \\
&= \tau^2\|f\|^2 + 2\tau \sum_{A\in \A} \|A f\|^2 + \sum_{A,B\in \A}
\bigl(\|ABf\|^2 + \scal{f,[A^*A,B^*]Bf}\bigr)\;.
}
We can write the operator intervening in the last term as
\equ{
[A^*A,B^*]B = A^*[A,B^*]B + [B,A]^*AB\;.
}
Because of \hyp{1}, this implies that there are positive constants $C_{ABC}$
such that
\equs{
\|(\Lambda^2-1)f  + \tau f\|^2 \ge&\; \tau^2\|f\|^2 + 2\tau \sum_{A\in \A} \|A
f\|^2 + \sum_{B,C\in \A}\|BCf\|^2 \\
	&- \sum_{A,B,C \in \A} C_{ABC} \|Af\|\|BCf\|\;.
}
If we use now
\equ{
	2xy \le x^2s^2 + \frac{y^2}{s^2}\quad x,y \ge 0\,,\; s > 0\;,
}
we see that we can choose $\tau$ big enough to have
\equ{
\|(\Lambda^2-1)f  + \tau f\|^2 \ge \tau^2\|f\|^2 + \frac 12 \sum_{A\in \A} \|A
f\|^2 + \frac 12 \sum_{B,C\in \A}\|BCf\|^2\;.
}
This immediately implies \eref{e:estTwo}.
\end{proof}

This lemma can now be used to prove Lemma \ref{lem:estj}.

\begin{proof}[of Lemma \ref{lem:estj}]
We want to show that $A \in \CY^i_\F(\A)$ implies $A\Lambda^{-i}$ bounded. We
already treated the cases $i=1$ and $i=2$. For the other cases, we proceed by
induction. Let us fix $j>2$ and assume the assertion has been proved for $i<j$.
Then the operators of the form
\equ[e:estj]{
	A_1 A_2 \cdot\ldots\cdot A_j \Lambda^{-j}\qquad A_i \in \A\;,
}
are bounded. We distinguish two cases.
\begin{list}{}{\setlength{\leftmargin}{1.3truecm}}
\item[$\boldsymbol{j=2n}$.] We write the operator of \eref{e:estj} as
\equ{
	A_1A_2 \Lambda^{-2}\cdot\Lambda^2
A_3A_4\Lambda^{-4}\cdot\ldots\cdot\Lambda^{2n-1}A_{2n-2}A_{2n}\Lambda^{-2n}\;.
}
We show that operators of the form
\equ{
	\Lambda^{2m-2}AB\Lambda^{-2m}\quad A,B\in \A\,,\;m \le n\;,
}
are bounded. We write
\equ{
\Lambda^{2m-2}AB\Lambda^{-2m} = AB\Lambda^{-2} +
[\Lambda^{2m-2},AB]\Lambda^{-(2m-1)}\Lambda^{-1}\;.
}
The first term is bounded by Lemma \ref{lem:estTwo}. The second term is bounded
by noticing that $[\Lambda^{2m-2},AB] \in \CY^{2m-1}_\F(\A)$ and using the
induction hypothesis.
\item[$\boldsymbol{j=2n+1}$.] We write the operator of \eref{e:estj} as
\equ{
	A_1A_2 \Lambda^{-2}\cdot\Lambda^2
A_3A_4\Lambda^{-4}\cdot\ldots\cdot\Lambda^{2n}A_{2n+1}\Lambda^{-2n-1}\;.
}
The first terms are bounded exactly the same way as before. Concerning the last
term, we have
\equ{
\Lambda^{2n}A_{2n+1}\Lambda^{-2n-1} = A_{2n+1}\Lambda^{-1} +
[\Lambda^{2n},A_{2n+1}]\Lambda^{-2n}\Lambda^{-1}\;,
}
which is bounded by Lemma \ref{lem:estTwo} and the induction hypothesis,
noticing that the commutator belongs to $\CY^{2n}(\A)$.
\end{list}
This completes the proof of the lemma.
\end{proof}

We need another result from \cite{EPR}.

\begin{lemma}
Let $\{A(z)\} \subset \B(\H)$ be a family of uniformly bounded operators,
$\Lambda \ge 1$ a self-adjoint operator and let $F(\lambda,z)$ be a real,
positive bounded function. Then
\equ[e:op1]{
\left\| \int_0^\infty A(z)\, F(\Lambda,z) f\, dz\right\| \le \sup_{y \ge 0}
\|A(y)\|\|f\|\int_0^\infty \sup_{\lambda \ge 1} F(\lambda,z)\,dz\;, \qquad
\forall\, f\in\H\;.
}
If furthermore $A = A(z)$ is independent of $z$, one has the bound
\equ[e:op2]{
\left\| \int_0^\infty A\, F(\Lambda,z) f\, dz\right\| \le
\|A\|\|f\|\sup_{\lambda \ge 1} \int_0^\infty F(\lambda,z)\,dz\;, \qquad
\forall\, f\in\H\;.
}
\end{lemma}

\begin{lemma}
Let $\Lambda$, $\F$ and $\A$ be as above and assume \hyp{1} and \hyp{3} hold.
If $X \in \CY_\F^j(\A)$, then the operators
\equ{
	\Lambda^\beta X \Lambda^\gamma \quad\text{with}\quad \beta+\gamma \le -j
}
are bounded.

If $Y \in \fL$ is such that $[Y,\Lambda^2] \in \CY_\F^j(\A)$, then the
operators
\equ{
	\Lambda^\beta[\Lambda^\alpha,Y]\Lambda^\gamma \quad\text{with}\quad \alpha +
\beta + \gamma \le 2-j
}
are bounded.

If $X,Y \in \fL$ are such that
\equ{
[X,\Lambda^2] \in \CY_\F^j(\A)\;\;,\quad [Y,\Lambda^2] \in
\CY_\F^k(\A)\quad\text{and}\quad\bigl[[\Lambda^2,X],Y\bigr] \in
\CY_\F^{j+k-2}(\A)\;,
}
then the operators
\equ{
	\Lambda^\beta\bigl[[\Lambda^\alpha,X],Y\bigr]\Lambda^\gamma
\quad\text{with}\quad \alpha + \beta + \gamma \le 4-j-k
}
are bounded.
\end{lemma}

\begin{proof}
Let us prove the first assertion. The case $\gamma = 0$ is handled by noticing
that
\equ{
	\Lambda^\beta X = \Lambda^{\beta+j}\bigl(X^*\Lambda^{-j}\bigr)^*\;,
}
and that both operators of the latter product are bounded by Lemma
\ref{lem:estj}. The case $\beta=0$ is handled in the same way by considering
the adjoint.

The proof for the other cases follows exactly \cite{EPR}. We will demonstrate
the techniques involved by proving the third assertion, assuming the first two
assertions hold. The second assertion can be proved in a similar way without
using the third one.

We will first assume that $\alpha \in (-2,0)$. In this case, we can write (see
\eg \cite[\S~V.3.11]{Ka})
\equ[e:exprPower]{
\Lambda^{\alpha} = C_{\alpha} \int_0^\infty z^{\alpha/2}
(z+\Lambda^2)^{-1}\,dz\;,\qquad C_\alpha = -\frac{\sin(\pi \alpha/2)}{\pi}\;.
}
We notice moreover that it is possible to write
\equs[e:trans]{
\bigl[[(z + \Lambda^2)^{-1},X],Y\bigr] =&\; (z +
\Lambda^2)^{-1}\bigl[[\Lambda^2,X],Y\bigr] (z + \Lambda^2)^{-1} \\
&+ (z + \Lambda^2)^{-1}[\Lambda^2,X](z + \Lambda^2)^{-1}[\Lambda^2,Y](z +
\Lambda^2)^{-1} \\
&+ (z + \Lambda^2)^{-1}[\Lambda^2,Y](z + \Lambda^2)^{-1}[\Lambda^2,X](z +
\Lambda^2)^{-1}\;.
}
If we substitute the expression \eref{e:exprPower} in
$\Lambda^\beta\bigl[[\Lambda^\alpha,X],Y\bigr]\Lambda^\gamma$ and use
\eref{e:trans}, we get three terms, which we call $T_1$, $T_2$ and
$T_3$, and which
will be estimated separately.

\proclaim{Term $\boldsymbol{T_1}$.} This term is given by
\equ{
	T_1 = C_{\alpha} \int_0^\infty z^{\alpha/2} \frac{\Lambda^\beta}{z +
\Lambda^2}\bigl[[\Lambda^2,X],Y\bigr] \frac{\Lambda^\gamma}{z +
\Lambda^2}\,dz\;.
}
We define $B= \bigl[[\Lambda^2,X],Y\bigr] \in \CY^{j+k-2}(\A)$ and write
\equs{
	T_1 &= C_{\alpha} \int_0^\infty z^{\alpha/2} \Lambda^\beta B
\frac{\Lambda^\gamma}{(z + \Lambda^2)^2}\,dz + C_{\alpha} \int_0^\infty
z^{\alpha/2} \frac{\Lambda^\beta}{z + \Lambda^2}
[\Lambda^2,B]\frac{\Lambda^\gamma}{(z + \Lambda^2)^2}\,dz\\
&\equiv C_{\alpha}\bigl(T_{11} + T_{12}\bigr)\;.
}
The term $T_{11}$ is estimated by writing, for any $f \in \H$,
\equs{
	\|T_{11} f\| &= \left\|\Lambda^{\beta}B\Lambda^{2-j-k-\beta} \int_0^\infty
z^{\alpha/2} \frac{\Lambda^{\gamma + \beta + j + k - 2}}{(z +
\Lambda^2)^2}f\,dz\right\|\\
&\le \|f\|\bigl\|\Lambda^{\beta}B\Lambda^{2-j-k-\beta}\bigr\|\sup_{\lambda \ge
1} \int_0^\infty z^{\alpha/2} \frac{\lambda^{\gamma + \beta + j + k - 2}}{(z +
\lambda^2)^2}\,dz \\
&= \|f\|\bigl\|\Lambda^{\beta}B\Lambda^{2-j-k-\beta}\bigr\|\sup_{\lambda \ge 1}
\int_0^\infty s^{\alpha/2} \frac{\lambda^{\alpha + \gamma + \beta + j + k -
4}}{(s + 1)^2}\,ds\;.
}
Since the assumption yields $B \in \CY^{j+k-2}(\A)$, the norm is bounded. The
integral is also bounded because, by assumption, we have $\alpha + \gamma +
\beta \le 4 - j - k$.

To bound $T_{12}$, we observe that $[\Lambda^2, B] \in \CY^{j+k-1}(\A)$. Using
\eref{e:op2}, we find the bound
\equs{
\|T_{12}f\| &= \biggl\| \int_0^\infty z^{\alpha/2} \frac{\Lambda^\beta}{z +
\Lambda^2}
[\Lambda^2,B]\Lambda^{3-j-k-\beta}\frac{\Lambda^{\gamma+\beta+j+k-3}}{(z +
\Lambda^2)^2}f\,dz\biggr\| \\
	&\le \|f\|\; \sup_{y>0} \;\Bigl\|\frac{\Lambda^\beta}{y + \Lambda^2}
[\Lambda^2,B]\Lambda^{3-j-k-\beta}\Bigr\| \int_0^\infty z^{\alpha/2}
\sup_{\lambda \ge 1} \frac{\lambda^{\gamma+\beta+j+k-3}}{(z +
\lambda^2)^2}\,dz\;.
}
This expression is bounded when $\alpha + \beta + \gamma \le 4 - j - k$ and
$\alpha \in (-2,0)$. This can be seen by making as before the substitution $z
\mapsto \lambda^2 s$.

Before we go on, we introduce the notation $\Lambda_z \equiv (z +
\Lambda^2)^{-1}$.

\proclaim{Term $\boldsymbol{T_2}$.} This term is given by
\equ{
	T_2 = C_{\alpha} \int_0^\infty z^{\alpha/2} \frac{\Lambda^\beta}{z +
\Lambda^2}A\frac{1}{z + \Lambda^2}B\frac{\Lambda^\gamma}{z + \Lambda^2}\,dz\;,
}
where we defined
\equ{
A = [\Lambda^2,X] \qquad\text{and}\qquad B= [\Lambda^2,Y]\;.
}
Since $[\Lambda_z, B] = \Lambda_z[B,\Lambda^2]\Lambda_z$,
the term appearing under the integral can be written as
\equ{
	\Lambda^\beta \Lambda_z A \Lambda_z B \Lambda_z \Lambda^\gamma =
\Lambda^\beta \Lambda_z A B \Lambda_z^2 \Lambda^\gamma + \Lambda^\beta
\Lambda_z A \Lambda_z [B,\Lambda^2] \Lambda_z^2 \Lambda^\gamma\;.
}
According to this, the term $T_2$ is split into two terms $T_{21}$ and
$T_{22}$. We have
\equ{
\|T_{21}f\| \le \|f\| \;\sup_{y>0}\;\bigl\|\Lambda^\beta \Lambda_y A B
\Lambda^{-\beta-j-k}\bigr\|\int_0^\infty s^{\alpha/2} \sup_{\lambda \ge
1}\frac{\lambda^{\alpha+\beta+\gamma+j+k-4}}{(s+1)^2}\,ds\;.
}
The integral is bounded by hypothesis. The norm is also bounded, because $AB
\in \CY^{j+k}(\A)$. For the second term, we have
\equ{
\|T_{22}f\| \le \|f\| \;\sup_{y>0}\;\bigl\|\Lambda^\beta \Lambda_y A \Lambda_y
[\Lambda^2,B] \Lambda^{-\beta-j-k}\bigr\|\int_0^\infty s^{\alpha/2}
\sup_{\lambda \ge 1}\frac{\lambda^{\alpha+\beta+\gamma+j+k-4}}{(s+1)^2}\,ds\;.
}
This is bounded in the same fashion, noticing that
\equ{
 \sup_{y>0}\;\bigl\|\Lambda^\beta \Lambda_y A \Lambda_y [\Lambda^2,B]
\Lambda^{-\beta-j-k}\bigr\| \le \sup_{x>0}\;\bigl\|\Lambda^\beta \Lambda_x A
\Lambda^{-\beta-j}\bigr\| \sup_{y>0}\;\biggl\|\frac{\Lambda^2}{y + \Lambda^2}
\Lambda^{j+\beta-2}[\Lambda^2,B] \Lambda^{-\beta-j-k}\biggr\|\;.
}

\proclaim{Term $\boldsymbol{T_3}$.} It can be bounded in the same way as $T_2$
by symmetry.

We now have to check the assertion for the other values of $\alpha$. If $\alpha
= 0$ or $\alpha = 2$, it holds trivially. For $\alpha > 0$, we proceed by
induction, using the equality
\equs[e:equInd]{
	\bigl[[\Lambda^{\alpha+2},X],Y\bigr] &=
\Lambda^2\bigl[[\Lambda^{\alpha},X],Y\bigr] +
\Lambda^\alpha\bigl[[\Lambda^2,X],Y\bigr] + [\Lambda^\alpha,Y][\Lambda^2,X] +
[\Lambda^2,Y][\Lambda^\alpha,X]\;.
}
For $\alpha = -2$, the assertion is proved using equality \eref{e:trans} with
$z = 0$. For $\alpha < -2$, we also proceed by induction, using \eref{e:equInd}
with $2$ replaced by $-2$.
This completes the proof of Lemma \ref{lem:power}.
\end{proof}

\makeappendix{Proof of Proposition~\ref{prop:Tt}}
\label{App:ops}

\def\mL{{\tilde\CL}}

\begin{proposition}
 $\T^t$, as defined in \eref{e:defTt}, extends uniquely to a quasi-bounded
strongly continuous semi-group on $\Ltwo(\CX,\,dx)$. Its generator $L$ acts
like $\CL$ on functions in $\cOinf[\CX]$.
\end{proposition}

\begin{proof}
See the proof of Lemma~A.1 in \cite{EPR}.
\end{proof}

We now turn to the question of the domain of the generator $L$.
Recall that $\fL$ is the set of all formal expressions of the form
\equ{
	\sum_{|l| \le k} a_l(x) D^l\;,\qquad k \ge 0\;,\quad a \in \CC^\infty(\R^n)\;.
}
To any element $L \in \fL$ having the above form, we associate its formal
adjoint $L^* \in \fL$ in an obvious way.
In the sequel, the notation $\scal{f,g}$ will be used to denote the scalar
product in $\Ltwo$ if $f,g \in \Ltwo$ and the evaluation $f(g)$ if $f$ is a
distribution and $g \in \cOinf[\R^n]$. We hope this slight ambiguity will not
be too misleading.

We associate to every $L \in \fL$ the operator $T_L : \CD(T_L) \to \Ltwo(\R^n)$
by
\equ{
	\bigl(T_L f\bigr)(x) = L f(x)\quad\text{and}\quad \CD(T_L) = \{f \in
\Ltwo\;|\; Lf \in \Ltwo\}\;,
}
where $Lf$ has to be understood in the sense of distributions, \ie
\equ{
	\bigl(Lf\bigr)(g) \equiv f(L^* g)\quad\text{for all}\quad g\in \cOinf[\R^n]\;.
}
We also define the operator $S_L : \CD(S_L) \to \Ltwo(\R^n)$ by
\equ{
	S_L = \overline{T_L \upharpoonright \cOinf}\;.
}
The operators $T_L$ and $S_L$ are usually called the \emph{minimal operator}
and the \emph{maximal operator} constructed from the \emph{formal operator}
$L$.
The following result is classical, so we do not give its proof here

\begin{proposition}
\label{prop:maxmin}
For every $L \in \fL$, we have $T_L^* = S_{L^*}$ and $S_L^* = T_{L^*}$. In
particular, this shows that $T_L$ is closed.\phantom{a}\nobreak\hfill
$\qedsquare$
\end{proposition}

We prove now the quasi {\it m}-dissipativity of $S_\L$. We define
\equ{
	\mL \equiv \L - \sum_{i=1}^M \gamma_i -1 \;.
}
By definition, if $S_\mL$ is strictly {\it m}-dissipative, $S_\L$ is quasi {\it
m}-dissipative.
It is well-known that an equivalent characterization of strict {\it
m}-dissipativity is that

\begin{list}{$\bullet$}{\setlength{\leftmargin}{9mm}
\setlength{\topsep}{0mm}\setlength{\parsep}{0mm}}
\item[(a)] $S_\mL$ is strictly dissipative and
\item[(b)] $\text{Range}(S_\mL) = \CH$.
\end{list}

\begin{proposition}
\label{prop:maccr}
Assume {\bf A0} holds. Then $S_\mL$ is strictly {\it m}-dissipative.
\end{proposition}

\begin{remark}
It is clear that the statement holds if we consider the minimal operator in
$\Ltwo(\CK, dx)$, where $\CK$ is some compact domain of $\CX$. The idea is to
approximate $\CX$ by a sequence of increasing compact domains and to control
the rest terms.

This proposition fills a gap in \cite{EPR}, since the statement ``$\Re(f, L^*f)
= -\frac{1}{2}\|\sigma^T \nabla f\|^2 + (f,\div b\; f) \le B\|f\|^2$'' in the
proof of Lemma~A.1 is not justified for every $f \in \CD(L^*)$.
\end{remark}

\begin{demo}
Property (a) is immediate. By the closed-range theorem, property (b) is
equivalent to the statement

\begin{list}{$\bullet$}{\setlength{\leftmargin}{9mm}
\setlength{\topsep}{0mm}\setlength{\parsep}{0mm}}
\item[(b')] $f \in \Ltwo$ and $\mL^* f = 0$ imply $f = 0$.
\end{list}

\begin{MHwrap}{r}{5.5cm}{Func}[-5mm]
	\vspace{-5mm}
\end{MHwrap}
Assume on the contrary that there exists a non-vanishing function $f\in \Ltwo$
for which $\mL^* f = 0$ holds in the sense of distributions. Since $\mL^*$ is
hypoelliptic, $f$ must be a $\CC^\infty$ function. Let us choose some function
$\phi \in \cOinf(\R_+)$ such that $\phi(x) = 1$ if $x \in [0,1]$. We also
define
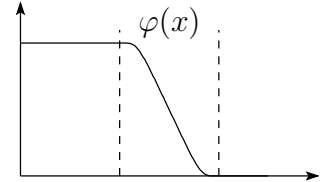
\equs[2]{
\phi_n : \;&\CX &\;\to\; &\R \\
& x &\;\mapsto\; & \phi\bigl(G(x)/n\bigr)\;.
}
By assumption, $\mL^* f = 0$, so we have
\equ{
 0 = 2\Re \scal{\phi_n f, \mL^* f} = \scal{\phi_n f, \mL^* f} + \scal{\mL^* f,
\phi_n f}\;.
}
Since $\phi_n \in \cOinf$ and all the other functions are $\CC^\infty$, we can
make all the formal manipulations we want. In particular, we have
\equ[e:altzero]{
\scal{\mL^* f, \phi_n f} = \scal{f, \mL \phi_n f} \quad \Rightarrow\quad
\scal[b]{f,(\phi_n \mL^* + \mL \phi_n) f} = 0\;.
}
Recall that $\mL$ is given by
\equs[e:genChainApp]{
	\mL &= \sum_{i=1}^M \lambda_i^2\gamma_i T_i \d_{r_i}^2 - \sum_{i=1}^M
\gamma_i\bigl(r_i - \lambda_i^2 F_i(p,q)\bigr)\d_{r_i} + X^{\HF} - \sum_{i=1}^M
r_i X^{F_i} - \sum_{i=1}^M \gamma_i -1\\
	&\equiv \sum_{i=1}^M \zeta_i \d_{r_i}^2 + Y_0 - 1\;.
}
Straightforward computation yields
\equs[e:developOp]{
	\phi_n \mL^* + \mL \phi_n &= 2\sum_{i=1}^M \zeta_i \d_{r_i}\phi_n \d_{r_i} +
\sum_{i=1}^M \zeta_i \bigl(\d_{r_i}^2 \phi_n\bigr)  + [Y_0, \phi_n] -
\phi_n\\[2mm]
	&= 2\sum_{i=1}^M \zeta_i \d_{r_i}\phi_n \d_{r_i} + \sum_{i=1}^M \zeta_i
\Bigl(\frac 1n (\d_{r_i}^2 G) \phi''(G/n) + \frac 1{n^2}(\d_{r_i} G)^2
\phi'(G/n)\Bigr)  \\[1mm]
	&\quad +  \frac 1n \sum_{i=1}^M \frac{\gamma_i}{\lambda_i^2}\bigl(r_i -
\lambda_i^2 F_i(p,q)\bigr)^2 \phi'(G/n) - \phi_n\\[2mm]
	&\equiv 2\sum_{i=1}^M \zeta_i \d_{r_i}\phi_n \d_{r_i} + \Phi_n - \phi_n\;.
}
Using {\bf A0}, we next verify that $|\Phi_n(x)| \le \tilde C$ for all $x \in
\CX$ and for all $n \ge 1$. We define
\equ{
c_1 \equiv \sup_{x \ge 0} \phi''(x)\quad\text{and}\quad c_2 \equiv \sup_{x \ge
0} x\phi'(x)\;.
}
An elementary computation shows that {\bf A0} implies that there exist
constants $c_3,\ldots, c_5 > 0$ for which
\equ{
 \bigl| \d_{r_i}^2 G(x)\bigr| \le c_3\;, \quad \bigl| \d_{r_i} G(x)\bigr|^2 \le
c_4 G(x) \;,\quad\text{and}\quad \bigl(r_i - \lambda_i^2 F_i(p,q)\bigr)^2 \le
c_5 G(p,q,r)\;.
}
We thus have
\equs{
|\Phi_n(x)| &\le \sum_{i=1}^M \Bigl( \frac{\zeta_i c_3}{n}
\bigl|\phi''(G/n)\bigr|  + \frac{\zeta_i c_4}{n} \bigl|(G/n) \phi'(G/n)\bigr| +
\frac{\gamma_i c_5}{\lambda_i^2} \bigl| (G/n) \phi'(G/n)\bigr| \Bigr)\\
	&\le \sum_{i=1}^M \Bigl( \zeta_i \frac{c_1 c_3 + c_2 c_4}{n} + \frac{\gamma_i
c_2 c_5}{\lambda_i^2}\Bigr) \le \tilde C\;,
}
as asserted. Moreover, the first part of {\bf A0} implies that there exist
constants $C, \alpha > 0$ such that
\equ[e:suppPhi]{
\supp \Phi_n \subset \{x \in \CX \;|\; \|x\|^\alpha \ge n/C\}\;.
}
Substituting \eref{e:developOp} back into \eref{e:altzero}, we get
\equ[e:altzero2]{
0 = - 2\sum_{i=1}^M \zeta_i \bigl\|\sqrt{\phi_n} \d_{r_i} f\bigr\|^2
 - \|\sqrt{\phi_n} f\bigr\|^2 + \int_\CX \Phi_n(x) |f(x)|^2\; dx\;.
}
Since $f \in \Ltwo(\CX)$, one has
\equ{
	\lim_{n \to \infty} \|\sqrt{\phi_n} f\bigr\|^2 = \|f\|^2\;.
}
Moreover, the uniform boundedness of $\Phi_n$ together with property
\eref{e:suppPhi} imply that
\equ{
	\lim_{n \to \infty} \int_\CX \Phi_n(x) |f(x)|^2\; dx = 0\;.
}
This supplies the required contradiction to \eref{e:altzero2}, thus
establishing the strict {\it m}-dissipativity of $S_\mL$.
\end{demo}

We complete now the

\begin{demo}[of Proposition~\ref{prop:Tt}]
It only remains to be proved that $L = S_\L$ and that $L^* = S_{\L^*}$.

It is clear that the generator $L$ of $\CT^t$ satisfies $S_\L \subset L$. Since
$S_\L$ is quasi {\it m}-dissipative, \ie has no proper quasi dissipative
extension, and since the generator of a quasi-bounded semi-group is always
quasi {\it m}-dissipative, we must have $L = S_\L$.

Concerning the adjoint, we have by Proposition~\ref{prop:maxmin}, $L^* =
T_{\L^*}$. It is possible to retrace the above argument for $\L^*$ to show that
$S_{\L^*}$ is quasi {\it m}-dissipative. Since $L^*$ is also quasi {\it
m}-dissipative and $S_{\L^*} \subset L^*$, we must have $L^* = S_{\L^*}$.
\end{demo}

\makeappendix{Proof of Lemma~\ref{lem:coregamma}}
\label{App:core}

Using the technique developed in Appendix~\ref{App:ops}, we can now turn to the
proof of Lemma~\ref{lem:coregamma}. Recall that $K$ is given by
\eref{e:defKchain} and that
\equ{
	W^{(\gamma)} = \Ltwo(\CX, G^{2\gamma}\,dx)\;.
}
Moreover, $K_\gamma$ is the maximal operator constructed from $K$ when
considering it as a differential operator in $W^{(\gamma)}$. We have

\begin{proposition}
$\cOinf[\CX]$ is a core for $K_\gamma$.
\end{proposition}

\begin{proof}
We introduce the unitary operator $U : W^{(\gamma)} \to \Ltwo(\CX)$ defined by
\equ{
	\bigl(U f\bigr)(x) = G^\gamma (x) f(x)\;.
}
We also define $K_\gamma^0 \equiv \overline{K_\gamma \upharpoonright
\cOinf[\CX]}$. The operators $K_\gamma$ and $K_\gamma^0$ are unitarily
equivalent to the operators $\tilde K_\gamma$ and $\tilde K_\gamma^0$
respectively by the following relations.
\[
\newdimen\UHeight
\settoheight{\UHeight}{$\scriptstyle{U}$}
\advance\UHeight 4mm
\def\myrule{\rule[-2mm]{0mm}{\the\UHeight}}
\def\lef#1{\vcenter{\llap{$\scriptstyle{#1}$}}}
\def\rig#1{\vcenter{\rlap{$\scriptstyle{#1}$}}}
\begin{array}{CCCCCCC}
\CD(K_\gamma) & \stackrel{K_\gamma}{\longrightarrow} & W^{(\gamma)} &
\qquad\qquad\qquad & \CD(K_\gamma^0) & \stackrel{K_\gamma^0}{\longrightarrow} &
W^{(\gamma)}\\[1mm]
\lef{U \myrule} \Big\downarrow \Big\uparrow \rig{\myrule U^{-1}}  & & \lef{U
\myrule}\Big\downarrow\Big\uparrow\rig{\myrule U^{-1}} && \lef{U \myrule}
\Big\downarrow \Big\uparrow \rig{\myrule U^{-1}}  &  & \lef{U
\myrule}\Big\downarrow\Big\uparrow\rig{\myrule U^{-1}}\\[2mm]
\CD(\tilde K_\gamma) & \underset{\tilde K_\gamma}{\longrightarrow} & \Ltwo(\CX)
&& \CD(\tilde K_\gamma^0) & \underset{\tilde K_\gamma^0}{\longrightarrow} &
\Ltwo(\CX) \end{array}
\]
By construction, $\tilde K_\gamma$ is maximal. Thus, by
Proposition~\ref{prop:maxmin}, its adjoint $\tilde K_\gamma^*$ is minimal. It
is immediate that the formal expressions for $\tilde K_\gamma^*$ and
$K_\gamma^0$ are given by
\equ{
	\tilde K_\gamma^* = G^{-\gamma} K^* G^\gamma \qquad\text{and}\qquad \tilde
K_\gamma^0 = G^\gamma K G^{-\gamma}\;.
}
It is now a simple exercise to retrace the proof of
Proposition~\ref{prop:maccr} to see that $\tilde K_\gamma^*$ and $\tilde
K_\gamma^0$ are both {\it m}-accretive. The remark of Section~\ref{sec:defs}
concerning the adjoints of {\it m}-accretive operators implies that $\tilde
K_\gamma$ is also {\it m}-accretive. Since $\tilde K_\gamma^0 \subset \tilde
K_\gamma$, we must have $\tilde K_\gamma^0 = \tilde K_\gamma$ and thus
$K_\gamma^0 = K_\gamma$. This proves the assertion.
\end{proof}

\markboth{\sc \refname}{\sc \refname}
\def\Rom#1{\uppercase\expandafter{\romannumeral #1}}
\providecommand{\bysame}{\leavevmode\hbox to3em{\hrulefill}\thinspace}

\end{document}